\documentclass[12pt]{article}
\pdfoutput=1
\usepackage{amsmath,amssymb,amsthm,bm,graphicx,float,array,multirow,multicol,rotfloat,caption,subcaption,hyperref,cleveref,enumerate,geometry,mathdots,adjustbox,booktabs,parskip,mathtools,tikz,tikz-cd,pdflscape,csquotes,lscape,rotating,empheq,mathrsfs,longtable}
\usetikzlibrary{decorations.pathreplacing,calligraphy}
\usepackage{stackengine} 
\usepackage{lscape} 
\usepackage{xcolor}
\usepackage[bottom]{footmisc}
\usepackage[para]{threeparttable}
\usepackage[all]{xy}
\usepackage[normalem]{ulem}
\usepackage[aligntableaux=center]{ytableau}
\usepackage[numbers,sort&compress]{natbib}
\numberwithin{equation}{section}
\numberwithin{figure}{section}
\allowdisplaybreaks
\makeatletter
\usetikzlibrary{arrows,shapes.misc,positioning,decorations.pathmorphing,decorations.markings,decorations.pathreplacing,matrix,patterns,backgrounds}

\tikzset{
scale cd/.style={every label/.append style={scale=#1}, cells={nodes={scale=#1}}}}

\tikzset{gauge/.style={rounded rectangle, draw=black!100, thick, minimum size=5mm},  gaugeD/.style={rounded rectangle, draw=black!100,double,thick,minimum size=5mm},  empty/.style={rounded rectangle, draw=white!100, thick, minimum size=5mm}, flavor/.style={rectangle, draw=black!100, thick, minimum size=5mm},flavorD/.style={rectangle, draw=black!100, double,thick, minimum size=5mm}}

\tikzset{
node/.style={circle, thick, draw=black!100,fill=white!100,  minimum size=2mm, inner sep=0pt},
sqnode/.style={rectangle
, thick, draw=black!100,fill=white!100,  minimum size=2mm, inner sep=0pt
},
sonode/.style={circle, thick, draw=black!100,fill=red!100,  minimum size=3mm, inner sep=0pt},
spnode/.style={circle, thick, draw=black!100,fill=blue!100,  minimum size=3mm, inner sep=0pt},
fnode/.style={rectangle, thick, draw=black!100,fill=white!100,  minimum size=3mm, inner sep=0pt},
tnode/.style={rounded rectangle, outer sep=0pt, thick, minimum size=5mm}
}

\tikzset{
brace/.style={
     decoration={brace, mirror},
     decorate
   }
}

\newcommand\notsoscript{\@setfontsize\notsoscript{9}{7}}

\theoremstyle{plain}
\newtheorem*{thm*}{Theorem}

\theoremstyle{definition}

\newtheorem*{defn*}{Definition}

\makeatother

\newcommand{\su}{\mathfrak{su}}

\newcommand{\EsixSU}[6]{
  \overset{\mathfrak{su}_{#1}}{2}
  \overset{\mathfrak{su}_{#2}}{2}
  \overset{
    \displaystyle \overset{\mathfrak{su}_{#6}}{2}
  }
  {
    \overset{\mathfrak{su}_{#3}}{2}
  }
  \overset{\mathfrak{su}_{#4}}{2}
  \overset{\mathfrak{su}_{#5}}{2}
}

\newcommand{\EsevenSU}[7]{
  \overset{\mathfrak{su}_{#1}}{2}
  \overset{\mathfrak{su}_{#2}}{2}
  \overset{\mathfrak{su}_{#3}}{2}
  \overset{
    \displaystyle \overset{\mathfrak{su}_{#7}}{2}
  }
  {
    \overset{\mathfrak{su}_{#4}}{2}
  }
  \overset{\mathfrak{su}_{#5}}{2}
  \overset{\mathfrak{su}_{#6}}{2}
}

\newcommand{\EeightSU}[8]{
  \overset{\mathfrak{su}_{#1}}{2}
  \overset{\mathfrak{su}_{#2}}{2}
  \overset{\mathfrak{su}_{#3}}{2}
  \overset{\mathfrak{su}_{#4}}{2}
  \overset{
    \displaystyle \overset{\mathfrak{su}_{#8}}{2}
  }
  {
    \overset{\mathfrak{su}_{#5}}{2}
  }
  \overset{\mathfrak{su}_{#6}}{2}
  \overset{\mathfrak{su}_{#7}}{2}
}

\newcommand{\Lpagenumber}{\ifdim\textwidth=\linewidth\else\bgroup
  \dimendef\margin=0 
  \ifodd\value{page}\margin=\oddsidemargin
  \else\margin=\evensidemargin
  \fi
  \raisebox{\dimexpr+2\topmargin-\headheight-\headsep-0.5\linewidth}[0pt][0pt]{%
    \rlap{\hspace{\dimexpr\margin+\textheight+3.5\footskip}%
    \llap{\rotatebox{90}{\thepage}}}}%
\egroup\fi}
\AddToHook{shipout/background}{\Lpagenumber}%

\begin{document}

\begin{titlepage}
\vspace*{-3cm} 
\begin{flushright}
{\tt DESY-24-079}\\
\end{flushright}
\begin{center}
\vspace{2cm}
{\LARGE\bfseries The Higgs Branch of 6d \boldmath{$(1,0)$} SCFTs \& LSTs \\[0.3em] with DE-type SUSY Enhancement}
\vspace{1.2cm}

{\large
Craig Lawrie and Lorenzo Mansi\\}
\vspace{.7cm}
{ Deutsches Elektronen-Synchrotron DESY,}\par
{Notkestr.~85, 22607 Hamburg, Germany}\par
\vspace{.2cm}

\vspace{.3cm}

\scalebox{1.0}{\tt craig.lawrie1729@gmail.com,
lorenzo.mansi@desy.de}\par
\vspace{1.2cm}
\textbf{Abstract}
\end{center}

We detail the Higgs branches of 6d $(1,0)$ superconformal field theories (SCFTs) and little string theories (LSTs) that exhibit supersymmetry-enhancing Higgs branch renormalization group flows to the 6d $(2,0)$ SCFTs and LSTs of type DE. Generically, such theories are geometrically engineered in F-theory via a configuration of $(-2)$-curves, arranged in an (affine) DE-type Dynkin diagram, and supporting special unitary gauge algebras; this describes the effective field theory on the tensor branch of the SCFT. For the Higgsable to D-type $(2,0)$ SCFTs/LSTs, there generically also exists a Type IIA brane description, involving a Neveu--Schwarz orientifold plane, which allows for the derivation of a magnetic quiver for the Higgs branch. These are 3d $\mathcal{N}=4$ unitary-orthosymplectic quivers whose Coulomb branch is isomorphic to the Higgs branch of the 6d theories. From this magnetic quiver, together with an extended quiver subtraction algorithm that we explain, the foliation structure of the Higgs branch as a symplectic singularity is unveiled. For this class of 6d SCFTs, we observe a simple rule, which we refer to as ``slice subtraction,'' to read off the transverse slice in the foliation from the tensor branch. Based on this slice subtraction observation, we conjecture the transverse slices in the Higgsable to E-type $(2,0)$ Hasse diagram, where the SCFTs lack any known magnetic quiver for their Higgs branches.

\vfill 
\end{titlepage}

\tableofcontents

\newpage

\section{Introduction}\label{sec:intro}

In recent years, the study of six-dimensional superconformal fields theories (SCFTs) with minimal supersymmetry has been burgeoning. While the existence of these novel quantum fields theories has been established for many years \cite{Nahm:1977tg, Witten:1995zh,Strominger:1995ac,Seiberg:1996qx}, their concrete construction and the extraction of their physical features has been rendered challenging by the absence of supersymmetry-preserving relevant or marginal deformations \cite{Cordova:2015fha,Louis:2015mka,Cordova:2016xhm,Cordova:2016emh,Buican:2016hpb}, and thus of direct bottom-up Lagrangian descriptions. 

Nevertheless, 6d SCFTs have been fruitfully explored via top-down constructions from string theory.\footnote{See \cite{Heckman:2018pqx,Heckman:2015bfa,Heckman:2018jxk,Heckman:2013pva,Argyres:2022mnu} for recent reviews of such explorations.} The maximally-supersymmetric SCFTs in 6d were first constructed \cite{Witten:1995gx} by considering the compactification of Type IIB string theory on the orbifold
\begin{equation}
    \mathbb{C}^2/\Gamma \qquad \text{ with } \qquad \Gamma \subset SU(2) \,.
\end{equation}
The finite subgroups, $\Gamma$, of $SU(2)$ organize themselves into an ADE classification, consisting of two infinite series and three sporadics:
\begin{equation}\label{eqn:6d(2,0)}
    A_{n \geq 1} \,, \quad D_{n \geq 4} \,, \quad E_6 \,, \quad E_7 \,, \quad E_8 \,,
\end{equation}
and thus, each 6d $(2,0)$ SCFT is associated with a choice of simple and simply-laced Lie group via the McKay correspondence \cite{MR604577}.

To construct 6d SCFTs with minimal supersymmetry, one can replace Type IIB, used in the construction of 6d $(2,0)$ SCFTs, with F-theory \cite{Vafa:1996xn,Morrison:1996na,Morrison:1996pp}. In \cite{Heckman:2013pva,Heckman:2015bfa}, it was argued that F-theory compactified on certain non-compact elliptically-fibered Calabi--Yau threefolds engineers 6d $(1,0)$ SCFTs, and furthermore a mechanism was provided for the construction of a vast landscape of such Calabi--Yau spaces. In particular, we need a non-compact Calabi--Yau threefold $Y$ that is elliptically-fibered over a non-compact base $B$, containing no compact complex curves, and such that the elliptic fiber over every point of $B$ is irreducible. Since we are interested in non-product SCFTs, $B$ may contain at most one singular point, $b_0$, and at most one non-minimal fiber, which must be supported over $b_0$.\footnote{An explanation of these technical conditions can be found in the review \cite{Heckman:2018jxk}.} A method to determine such elliptically-fibered Calabi--Yau threefolds was provided in \cite{Heckman:2013pva,Heckman:2015bfa}. In particular, take a non-compact elliptically-fibered Calabi--Yau threefold
\begin{equation}
    \pi \, : \, \widetilde{Y} \rightarrow \widetilde{B} \,,
\end{equation}
such that $\widetilde{B}$ has no singular points, and the fiber over each point of $B$ is irreducible and minimal. Suppose that $\widetilde{B}$ contains a network of connected compact curves, $C_i$. If there exists a contraction map 
\begin{equation}
    \rho \, : \, \widetilde{B} \rightarrow B \,,
\end{equation}
that shrinks all compact curves to zero volume such that either the image of the $C_i$ under the map is singular, or, uplifting $\rho$ to a map $\rho \, : \, \widetilde{Y} \rightarrow Y$, the singular fiber above the image of the $C_i$ is non-minimal, then the resulting $Y$ engineers a 6d $(1,0)$ SCFT. All possible $\widetilde{Y}$ can be constructed from a small set of building blocks; see \cite{Lawrie:2023uiu} for a recent review. F-theory compactified on $\widetilde{Y}$ directly gives rise to a 6d $(1,0)$ quantum field theory, not an SCFT, which is referred to as the SQFT living at the generic point of the tensor branch of the SCFT associated with $Y$. Therefore, we often refer to $\widetilde{Y}$ as the ``tensor branch geometry'' or the ``tensor branch curve configuration''. Since $\widetilde{Y}$ is non-singular, it is generally easier to extract physical properties from $\widetilde{Y}$ rather than $Y$ directly; therefore this tensor branch description is particularly powerful for determining SCFT properties which are protected under the curve contraction map.

There is an extensive family of 6d $(1,0)$ SCFTs associated with each 6d $(2,0)$ SCFT in the following manner. Let $Y$ be a non-compact elliptically-fibered Calabi--Yau threefold that engineers some 6d $(1,0)$ SCFT in F-theory, and further suppose that there exists a sequence of complex structure deformations of $Y$ such that 
\begin{equation}\label{eqn:cxdef}
    Y \quad \xrightarrow{\,\text{cx.~str.~def.}\,} \quad T^2 \times \mathbb{C}^2/\Gamma \,,
\end{equation}
where $\Gamma$ is a finite subgroup of $SU(2)$. Then the SCFT engineered by $Y$ belongs to the family of the $(2,0)$ SCFT labelled by $\Gamma$.\footnote{It is straightforward to see that if there exists a $\Gamma$ such that equation \eqref{eqn:cxdef} is satisfied, then it is unique.} Since complex structure deformations correspond physically to Higgs branch renormalization group flows, we can refer to any such 6d $(1,0)$ SCFT as ``Higgsable to the 6d $(2,0)$ SCFT of type $\Gamma$'' \cite{Ohmori:2015pua,Ohmori:2015pia}. The most famous examples are the rank $N$ $(\mathfrak{g}, \mathfrak{g})$ conformal matter theories \cite{DelZotto:2014hpa}, which are Higgsable to the 6d $(2,0)$ SCFT of type $\Gamma = A_{N-1}$. Our focus in this paper is precisely this set of 6d $(1,0)$ SCFTs where $\Gamma$ is of DE-type; such theories have received limited attention (see, e.g., \cite{Chen:2019njf,Hayashi:2015vhy}) compared to their A-type cousins.

A vital aspect of the study of any supersymmetric quantum field theory is the exploration of the moduli space of supersymmetric vacua. A 6d $(1,0)$ SCFT can have a Higgs branch which captures the half-BPS operators of the theory. There are many interesting physical questions related to the Higgs branch, for example: 
\begin{enumerate}
    \item what are the interacting fixed points on the Higgs branch? how are they related?
    \item what are the half-BPS operators belonging to the Higgs branch chiral ring?
    \item is the Higgs branch chiral ring freely generated? 
    \item what are the generators (and relations if it is not freely generated)?
\end{enumerate} 
As an eight-supercharge theory, the Higgs branch of a 6d $(1,0)$ SCFT is both a hyperk\"ahler space and a symplectic singularity \cite{Hitchin:1986ea,Beauville_2000}. A symplectic singularity possesses a natural foliation by symplectic leaves, and this induces a partial ordering on the leaves given by inclusion; from this partial ordering, we can associate a Hasse diagram, the so-called Higgs branch Hasse diagram. The physical interpretation of this Hasse diagram is that it encompasses all of the patterns of partial Higgsing of a given theory; a 6d SCFT is associated with each symplectic leaf in the foliation, and the transverse slice between two leaves captures the parameters that need to be tuned to perform a Higgs branch renormalization group flow between the two associated SCFTs \cite{Bourget:2019aer,Bourget:2023dkj}.

In terms of the elliptically-fibered Calabi--Yau threefold, $Y$, engineering the SCFT, the Higgs branch is related to the space of complex structure deformations of $Y$ \cite{Heckman:2015ola}. In particular, if there exists a complex structure deformation
\begin{equation}
    Y \quad \xrightarrow{\,\text{cx.~str.~def.}\,} \quad Y' \,,
\end{equation}
then there exists a Higgs branch renormalization flow between the SCFTs associated with $Y$ and $Y'$. However, the question of whether such complex structure deformations exist can be difficult to answer from the study of the non-singular Calabi--Yau geometries engineering the tensor branch effective field theories: $\widetilde{Y}$ and $\widetilde{Y}'$. As the geometry does not provide a particularly transparent window on the structure of the Higgs branch, especially on the vacuum expectation values that are given to trigger a particular flow, we use alternative methods to study the Higgs branch Hasse diagram in this paper, and we then compare to the expected geometric structure.

One particularly powerful method to study the structure of the Higgs branch of a 6d $(1,0)$ SCFT is the method of magnetic quivers \cite{Hanany:1996ie, Hanany:1997gh, Ferlito:2017xdq}. Let $\mathcal{T}$ be an arbitrary 6d $(1,0)$ SCFT; if we can find a 3d $\mathcal{N}=4$ Lagrangian theory $\mathcal{T}_M$ such that the Higgs branch of $\mathcal{T}$ is the same as the Coulomb branch of $\mathcal{T}_M$:
\begin{equation}\label{eqn:MQ}
    \operatorname{HB}\big[ \mathcal{T} \big] \,\,=\,\, \operatorname{CB} \big[ \mathcal{T}_M \big] \,,
\end{equation}
then we say that $\mathcal{T}_M$ is a magnetic quiver for the Higgs branch of $\mathcal{T}$.\footnote{More generally, the Higgs branch can be isomorphic to a union of Coulomb branches of magnetic quivers:
\begin{equation}
    \operatorname{HB}\big[ \mathcal{T} \big] \,\,=\,\, \bigcup_i \operatorname{CB} \big[ \mathcal{T}_M^{(i)} \big] \,.
\end{equation}
This will not be relevant for this paper, where the Higgs branch is always given by a single magnetic quiver.} Of course, for this to be a useful perspective, it is necessary to have an algorithm for the construction of a $\mathcal{T}_M$. Luckily, if $\mathcal{T}$ can be engineered via a brane system in Type IIA (or Type I') string theory, then one can pass to the magnetic phase of the brane system to obtain a quiver $\mathcal{T}_M$ satisfying equation \eqref{eqn:MQ}. This process has been carried out for certain 6d $(1,0)$ SCFTs in \cite{Cabrera:2019izd, Cabrera:2019dob, Hanany:2018uhm}.

Knowing that the Higgs branch of the 6d SCFT is isomorphic to the Coulomb branch of a Lagrangian 3d $\mathcal{N}=4$ quiver is particularly powerful, as we can then use the sophisticated tools developed to study the Coulomb branches of such quivers. For example, we can use the monopole formulae \cite{Cremonesi:2013lqa} to calculate the Coulomb branch Hilbert series, and thus understand the structure of the 6d Higgs branch chiral ring; or we can use a quiver subtraction algorithm \cite{Cabrera:2018ann} or the decay and fission algorithm \cite{Bourget:2023dkj,Bourget:2024mgn} to determine the structure of the interacting fixed points on the 6d Higgs branch. As one of the principle results of this paper, we use the Type IIA brane system description of the Higgsable to $(2,0)$ D-type SCFTs, which involves a Neveu--Schwarz orientifold \cite{Kapustin:1998fa}, to determine unitary-orthosymplectic magnetic quivers for their Higgs branches. Then, we use the quiver subtraction algorithm to study the interacting fixed points and compare them to the 6d $(1,0)$ SCFTs expected to lie along subloci of the Higgs branch.

The structure of this paper is as follows. First, in Section \ref{sec:thetheories}, we review the geometric construction of 6d $(1,0)$ SCFTs via F-theory and introduce the classes of theories that are of interest in this paper. In Section \ref{sec:Magnetic_quiver}, we determine the magnetic quivers for the Higgs branches of the $D^{\mathfrak{su}_{2k}}_N(O)$ SCFTs from the Type IIA brane engineering description involving $\operatorname{ON}^-$-planes. We perform a consistency check by comparing the unitary magnetic quivers obtained when $\Gamma = A_3$ with the unitary-orthosymplectic magnetic quivers obtained when $\Gamma = D_3$ in Section \ref{sec:A3vsD3}. With the magnetic quiver firmly in hand, we turn, in Section \ref{sec:Higgsbranch}, to the determination of the Higgs branch Hasse diagram, and we observe that the quiver subtraction algorithm leads to the same result as expected from the F-theory geometry. In Section \ref{sec:product}, we make a brief digression to discuss the complex structure deformations behind Higgsings that break an SCFT into a product of SCFTs. In Section \ref{sec:Etype}, we explore the 6d $(1,0)$ SCFTs with supersymmetry-enhancing RG flows to 6d $(2,0)$ SCFTs of E-type; due to the lack of a magnetic quiver, we study the Higgs branch purely from the 6d perspective. Next, in Section \ref{sec:LSTs}, we study the 6d $(1,0)$ little strings theories (LSTs) that have DE-type supersymmetry-enhancing Higgs branch RG flows, and determine the magnetic quivers capturing the structure of their Higgs branches. Finally, in Section \ref{sec:disc}, we summarize our results and discuss some suggested future directions.

\section{6d \texorpdfstring{\boldmath{$(1,0)$}}{(1,0)} SCFTs with SUSY-enhancing RG Flows}\label{sec:thetheories}

As we have mentioned in Section \ref{sec:intro}, 6d $(1,0)$ SCFTs can be engineered by considering F-theory compactified on certain non-compact elliptically-fibered Calabi--Yau threefolds. The atomic construction \cite{Heckman:2013pva,Heckman:2015bfa} provides a constructive algorithm to produce such threefolds. This geometric procedure has been reviewed in detail in \cite{Heckman:2018jxk}, as well as in recent work of the current authors \cite{Lawrie:2023uiu}, and so we only briefly summarize the necessary aspects of the construction here, and refer to these references for a fuller exposition. 

We begin with a non-compact elliptically-fibered Calabi--Yau threefold $\widetilde{Y}$. The base of the fibration, $\widetilde{B}$, contains a collection of intersecting smooth rational curves, $C_i$, and the intersection matrix of these compact curves is presumed to be negative-definite:
\begin{equation}\label{eqn:negdef}
    C_i \cdot C_j \,\prec\, 0 \,.
\end{equation}
Furthermore, the fibration over every point of $\widetilde{B}$ is assumed to be minimal, in the technical sense explained in \cite{Heckman:2013pva,Heckman:2015bfa}. Then, there exists a (unique) contraction map $\pi: \widetilde{Y} \rightarrow Y$, which simultaneously takes the volume of all the $C_i$ to zero, such that F-theory compactified on $Y$ leads to a 6d $(1,0)$ SCFT.\footnote{Depending on the $\widetilde{Y}$, the 6d SCFT engineered in this way may be non-interacting.} The construction of little string theories (LSTs) from F-theory \cite{Bhardwaj:2015oru} proceeds in a similar way, except that the intersection matrix of compact curves is negative-semi-definite with a single zero eigenvalue. Then, there exists a (possibly non-unique) contraction map which simultaneously shrinks all-but-one of the compact curves; F-theory compactified on the Calabi--Yau space obtained after contraction gives rise to an LST. We return to LSTs in Section \ref{sec:LSTs}.

Therefore, a 6d $(1,0)$ SCFT engineered in such a way can be encoded in the relevant data of $\widetilde{Y}$, which is simply the configuration of curves $C_i$, and the singular fibers supported over the generic points of each $C_i$. Each $C_i$ has a negative self-intersection number, $(-n)$ for some positive $n$, due to the condition in equation \eqref{eqn:negdef}. The minimal singular fiber supported over the curve $C_i$ can be captured by a simple Lie algebra, $\mathfrak{g}_i$.\footnote{There are Kodaira--Neron singular fibers that correspond to the same simple Lie algebra, e.g., $I_3$ and $IV$ fibers both correspond to $\mathfrak{su}(3)$, however, such distinctions are not generally relevant in this paper.} The intersections of the $C_i$ necessarily occur pairwise, and with intersection number $1$, for the contraction to lead to a $Y$ that gives rise to a 6d $(1,0)$ SCFT. Therefore, we can simply use the now-standard notation where we denote a curve $C_i$ of self-intersection number $(-n_i)$ and supporting singular fiber associated with the Lie algebra $\mathfrak{g}_i$ as
\begin{equation}
    \overset{\mathfrak{g}_i}{n_i} \,.
\end{equation}
Then, we can depict several such expressions adjacent to each other to indicate a non-trivial intersection of the associated curves. Such a notation captures all the salient data of the elliptically-fibered Calabi--Yau threefold $\widetilde{Y}$. For example, 
\begin{equation}
    1 \overset{\mathfrak{su}_3}{3} 1 \,,
\end{equation}
denotes three genus-zero curves, with self-intersection numbers $(-1)$, $(-3)$, and $(-1)$, that intersect in a linear chain, where only the $(-3)$-curve carries a non-trivial singular fiber,\footnote{For ease of notation, we refer to the smooth torus as the trivial singular fiber.} which is associated with the Lie algebra $\mathfrak{su}(3)$. Such data, which we refer to as a ``curve configuration,'' then specifies an interacting 6d $(1,0)$ SCFT.\footnote{In some circumstances, such as those discussed in \cite{Distler:2022yse}, additional data may be required to uniquely describe the 6d $(1,0)$ SCFT, however, for the theories discussed in this paper, this is not necessary.}

When considering F-theory compactified on $\widetilde{Y}$, as opposed to $Y$, we obtain a 6d $(1,0)$ QFT -- this is a gauge theory where the gauge algebras are specified by the singular fibers over the $C_i$. In addition to the gauge sector, there are also hypermultiplets associated with matter fields. In the cases of interest to us, the number of hypermultiplets and the representations of $\mathfrak{g}_i$ under which they transform are prescribed, via gauge-anomaly cancellation, by the choice of self-intersection number of the corresponding curve $C_i$. The resulting QFT is the effective description at the generic point of the tensor branch of the 6d $(1,0)$ SCFT associated with the Calabi--Yau geometry $Y$. 

\subsection{Higgsable to D-type \texorpdfstring{$(2,0)$}{(2,0)}}\label{sec:DDD}

The first family of theories that we explore in this paper are those that possess supersymmetry-enhancing Higgs branch renormalization group flows to the $(2,0)$ SCFT of type $D_N$. In the atomic construction, the 6d $(2,0)$ SCFT of type $D_N$ is associated with the following curve configuration:
\begin{equation}\label{eqn:TBDN}
    2\,\overset{\displaystyle 2}{2}\underbrace{\,2\,\cdots\,2\,}_{N-3} \,.
\end{equation}
I.e., a $D_N$ Dynkin diagram formed out of intersecting $(-2)$-curves, where the elliptic fiber over each $(-2)$-curve is just a smooth torus.\footnote{One can continue to $N=1$ and $N=2$, which correspond to either a single $(-2)$-curve, or to two disjoint $(-2)$-curves, respectively. Throughout this paper, we assume that $N \geq 3$ to avoid these edge cases.} 

One family of theories that are Higgsable to the 6d $(2,0)$ SCFTs of type $D_N$ are associated with the curve configurations where the $(-2)$-curves in equation \eqref{eqn:TBDN} are decorated with singular fibers associated with special unitary gauge algebras:
\begin{equation}\label{eqn:TBDNdecorated}
    \overset{\mathfrak{su}_{k_1}}{2}\,\overset{\displaystyle \overset{\mathfrak{su}_{k_2}}{2}}{\overset{\mathfrak{su}_{k_3}}{2}}\underbrace{\,\overset{\mathfrak{su}_{k_4}}{2}\,\cdots\,\overset{\mathfrak{su}_{k_N}}{2}\,}_{N-3} \,.
\end{equation}
Gauge-anomaly cancellation mandates that when an $\mathfrak{su}(k)$ gauge algebra is supported over a $(-2)$-curve there must be precisely $2k$-hypermultiplets transforming in the fundamental representation of $\mathfrak{su}(k)$.\footnote{Care must be taken when any $k = 0, 1$.} Thus, we need to have that 
\begin{equation}\label{eqn:Cartan_Gauge=flavor}
    \sum_{j=1}^N A^{ij}k_j = m_i \geq 0 \,,
\end{equation}
where $A^{ij}$ is the (positive-definite) Cartan matrix for the $D_N$ Lie algebra and all the $k_i$ and $m_i$ are non-negative. In general, this is a necessary condition, but it is not sufficient; for example, any curve configuration of the form
\begin{equation}\label{eqn:screwy}
    \overset{\mathfrak{su}_{2}}{2}\,\overset{\displaystyle \overset{\mathfrak{su}_{2}}{2}}{\overset{\mathfrak{su}_{3}}{2}}\,\overset{\mathfrak{su}_{4}}{2} \cdots \,,
\end{equation}
does not, in fact, correspond to a 6d $(1,0)$ SCFT \cite{Morrison:2016djb}, even though it may satisfy equation \eqref{eqn:Cartan_Gauge=flavor}. The $m_i$ appearing in equation \eqref{eqn:Cartan_Gauge=flavor} are referred to as ``dangling hypermultiplets'', and they are associated with $\mathfrak{su}(m_i)$ flavor algebras rotating these hypermultiplets. We can depict all this information together in one curve configuration as
\begin{equation}
    \underset{[m_1]}{\overset{\mathfrak{su}_{k_1}}{2}}\,
    \overset{\displaystyle \overset{[m_2]}{\overset{\mathfrak{su}_{k_2}}{2}}}{\underset{[m_3]}{\overset{\mathfrak{su}_{k_3}}{2}}}\underbrace{\,\underset{[m_4]}{\overset{\mathfrak{su}_{k_4}}{2}}\,\cdots\,\underset{[m_N]}{\overset{\mathfrak{su}_{k_N}}{2}}\,}_{N-3} \,,
\end{equation}
although the $m_i$ are, of course, already implicitly fixed when writing the configuration as in equation \eqref{eqn:TBDNdecorated}.\footnote{As almost all of the theories studied in detail in this paper consist of $\mathfrak{su}(k)$ algebras supported over $(-2)$-curves, we introduce a simplifying notation where we write the dual Coxeter numbers of the gauge algebras arranged in the same way as the $(-2)$-curves. For example: \begin{equation}\label{eqn:Notation_Example}
    \overset{\mathfrak{su}_3}{2} 
    \underset{[1]}{ \overset{\displaystyle \overset{\mathfrak{su}_3}{2}}{\overset{\mathfrak{su}_6}{2}}}
    \overset{\mathfrak{su}_5}{2} 
    \underset{[3]}{\overset{\mathfrak{su}_4}{2} }
     \ \ \xRightarrow[\text{denoted as}]{\text{will be}} \ \ 3 \stackunder{\stackon{6}{3}}{[1]} 5 \ 4  [3] \,.
\end{equation}
While this could, a priori, be confusing, in every case that we consider in this paper, it is unambiguous whether we are referring to a curve configuration in the usual notation, or in this simplified notation. } 

For theories described by tensor branch configurations as in equation \eqref{eqn:TBDNdecorated}, there is a straightforward approach to understanding the Higgs branch renormalization group flows between theories. Consider a pair of consistent SCFTs of the form in equation \eqref{eqn:TBDNdecorated}, with the same value of $N$, where the gauge algebras are captured by the tuples
\begin{equation}\label{eqn:RGflow}
    (k_1, k_2, \cdots, k_N) \qquad \text{ and } \qquad (k_1', k_2', \cdots, k_N') \,,
\end{equation}
respectively. Then, there exists a Higgs branch renormalization group flow from the unprimed to the primed theory iff
\begin{equation}\label{eqn:ordering}
    k_i' \leq k_i \,,
\end{equation}
for all $i$. This is equivalent to the existence of a complex structure deformation of the Calabi--Yau engineering the unprimed SCFT, $Y$, to that engineering the primed SCFT, $Y'$.

The formulation in equation \eqref{eqn:TBDNdecorated} is highly redundant: most combinations of positive integers $(k_1, \cdots, k_N)$ are ruled out as they violate the condition in equation \eqref{eqn:Cartan_Gauge=flavor}. Instead, all valid tensor branch configurations of the form in equation \eqref{eqn:TBDNdecorated} arise as elements of a family parametrized by a choice of algebra $\mathfrak{g} = \mathfrak{su}(2k)$ and a nilpotent orbit, $O$, of $\mathfrak{g}$. For fixed $N$ with $N \geq 2k + 1$,\footnote{When we refer to $N$ as ``sufficiently large'' throughout this paper, we mean that this condition is satisfied.} every combination of $(\mathfrak{g}, O)$ corresponds to an interacting SCFT, and every interacting SCFT arises via a unique pair $(\mathfrak{g}, O)$. These families are obtained by starting with a parent theory with a $\mathfrak{g}$ flavor symmetry, and the children in the family are those that can be obtained from the parent theory by giving a nilpotent vacuum expectation value to the moment map of the $\mathfrak{g}$ flavor symmetry. The parent of each family is an SCFT which we denote as
\begin{equation}
    D_N^{\mathfrak{su}_{2k}} \,.
\end{equation}
The description of these SCFTs at the generic point of their tensor branch is given by the configuration
\begin{equation}\label{eqn:TBDNsu2k}
    \overset{\mathfrak{su}_{k}}{2}\,\overset{\displaystyle \overset{\mathfrak{su}_{k}}{2}}{\overset{\mathfrak{su}_{2k}}{2}}\underbrace{\,\overset{\mathfrak{su}_{2k}}{2}\,\cdots\,\overset{\mathfrak{su}_{2k}}{2}\,}_{N-3} \,[\mathfrak{su}_{2k}] \,,
\end{equation}
where we have highlighted the existence of an $\mathfrak{su}(2k)$ flavor symmetry. 

As we have said, each such $D_N^{\mathfrak{su}_{2k}}$ SCFT forms the parent for a whole family of theories related by Higgs branch renormalization group flow via giving nilpotent vacuum expectation values to the moment map of the $\mathfrak{su}(2k)$ flavor symmetry. We label the resulting SCFTs as
\begin{equation}\label{eqn:DNO}
    D_N^{\mathfrak{su}_{2k}}(O) \,,
\end{equation}
where $O$ is a nilpotent orbit of $\mathfrak{su}(2k)$. Nilpotent orbits of $\mathfrak{su}(2k)$ are in one-to-one correspondence with integer partitions of $2k$, and the mapping between integer partitions and tensor branch configurations is given in the appendices of \cite{Baume:2023onr}. From the tensor branch description, the anomaly polynomial can be determined following the usual algorithm \cite{Ohmori:2014pca,Ohmori:2014kda,Intriligator:2014eaa,Baume:2021qho}, and thus the dimensions of the Higgs branches of each of the SCFTs $D_N^{\mathfrak{su}_{2k}}(O)$ can be obtained \cite{Ohmori:2015pua,Ohmori:2015pia,Baume:2023onr}. We find
\begin{equation}
    \operatorname{dim}(\mathcal{H}) = N + 2k^2 - \operatorname{dim}(O) \,,
\end{equation}
where $\operatorname{dim}(O)$ is the dimension of the nilpotent orbit $O$.\footnote{Note: this holds for all $k \geq 0$. When $k=0$, for notational convenience, we formally consider the existence of a single nilpotent orbit of dimension zero.} The trivial orbit has dimension zero -- consistent with it corresponding to the trivial Higgsing.

More than just the dimension of the Higgs branch, we would like to understand the \emph{structure} of the Higgs branch. This can mean several things. For example, we would like to understand the effective theory (especially when that is an interacting SCFT) which exists along the different subloci of the Higgs branch; we would like to know when there exists an RG flow from a theory on one subloci to another, the transverse slice corresponding to that transition, as well as the operator(s) to which one gives a vacuum expectation value, and how the spectrum of the theory changes under the flow. 

\begin{figure}[p]
    \centering
    \begin{subfigure}[b]{0.32\textwidth}
    \centering
    \includegraphics[scale=0.35]{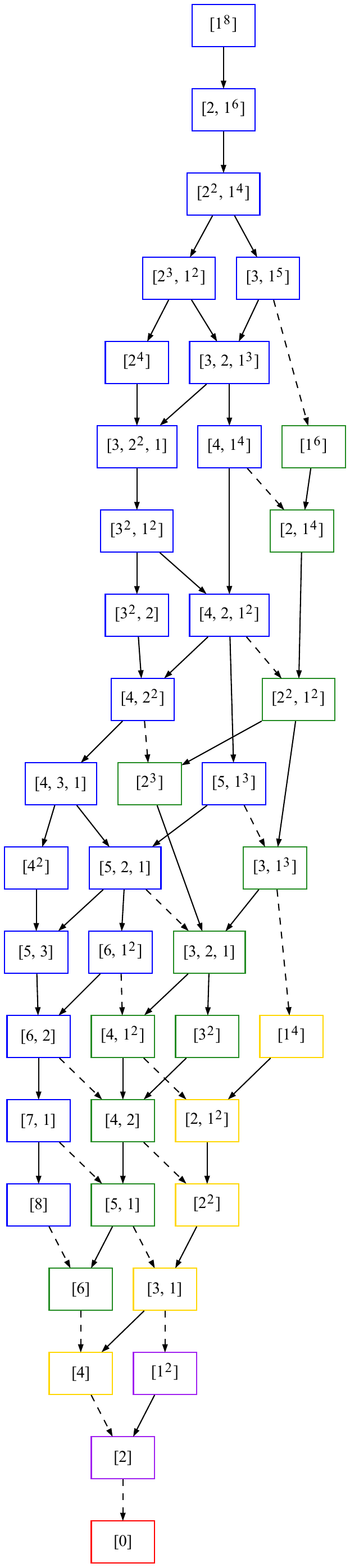}
    \caption{$N > 9$.}
    \end{subfigure}
    \begin{subfigure}[b]{0.32\textwidth}
    \centering
    \includegraphics[scale=0.35]{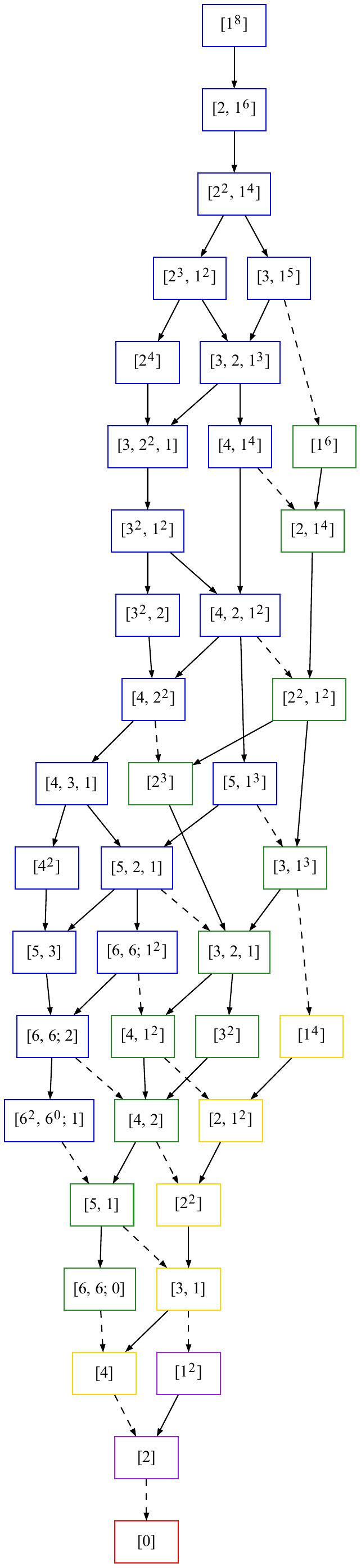}
    \caption{$N = 7$.}
    \end{subfigure}
    \begin{subfigure}[b]{0.32\textwidth}
    \centering
    \includegraphics[scale=0.35]{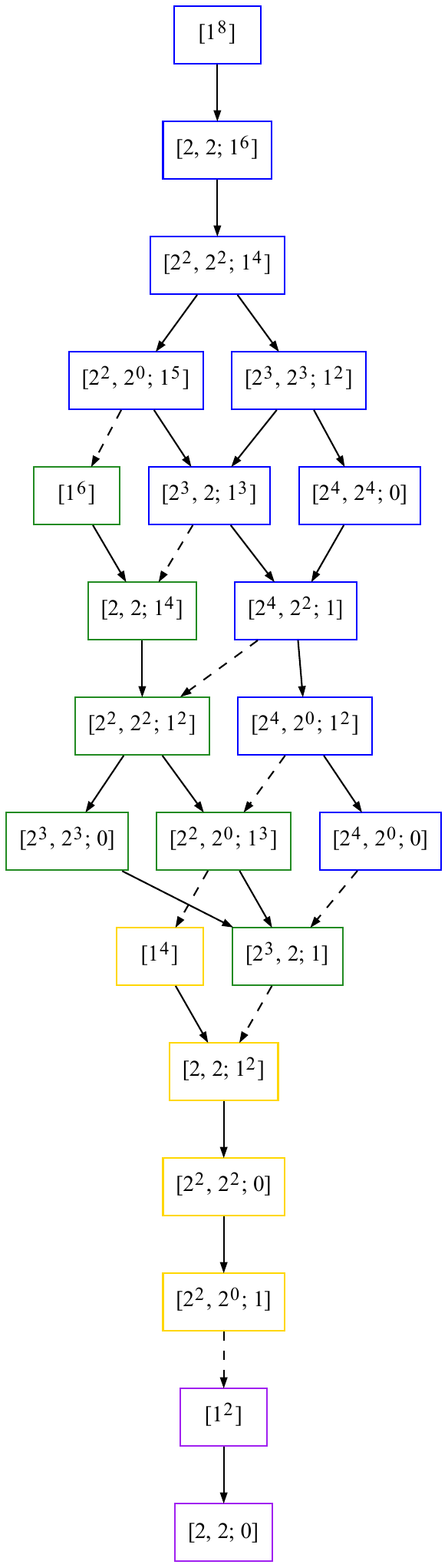}
    \caption{$N = 3$.}
    \end{subfigure}
    \caption{We depict the Hasse diagram of the theories $D_N^{\mathfrak{su}_{2\ell \leq 8}}(O)$ under the partial ordering given in equation \eqref{eqn:RGflow}. In (a), we assume that $N > 9$. We label each vertex by $O$, which we write as an integer partition of $2\ell$. Vertices bordered in blue, green, yellow, purple, and red denote $\ell = 4, 3, 2, 1$, and $0$, respectively; dashed edges indicate a change in $\ell$. We use $[0]$ for convenience to denote the 6d $(2,0)$ SCFT of type $D_N$. In (b) and (c), we depict the Hasse diagrams when $N = 7$ and $N=3$, respectively, where the labels are instead given by generalized partitions.}
    \label{fig:hasseeeee}
\end{figure}

First, it is clear from equation \eqref{eqn:RGflow} and the tensor branch description of the $D_N^{\mathfrak{su}_{2k}}(O)$ theories that there exists a Higgs branch renormalization group flow between the following SCFTs
\begin{equation}\label{eqn:rightflow}    
    D_N^{\mathfrak{su}_{2k}}(O) \quad \longrightarrow \quad D_N^{\mathfrak{su}_{2k}}(O') \,,
\end{equation}
if $O' < O$ under the partial ordering on nilpotent orbits of $\mathfrak{su}(2k)$ defined via the dominance ordering of their associated integer partitions. That is, there exists a subloci of the Higgs branch of the theory $D_N^{\mathfrak{su}_{2k}}(O)$ along which the theory $D_N^{\mathfrak{su}_{2k}}(O')$ lives, and the Higgs branch of the latter theory is contained within that of the former. The specific operators belonging to the Higgs branch chiral ring whose vacuum expectation value triggers the renormalization group flow for the class of Higgsings in equation \eqref{eqn:rightflow} are the same as those studied for $(A, A)$ conformal matter in \cite{DKL}. In addition, it is straightforward to see that one can Higgs from a family with $\mathfrak{su}(2k)$ to a family with $\mathfrak{su}(2k-2)$:
\begin{equation}
    D_N^{\mathfrak{su}_{2k}} \quad \longrightarrow \quad D_N^{\mathfrak{su}_{2k-2}} \,.
\end{equation}
More generally, it is straightforward to see, using equation \eqref{eqn:RGflow},\footnote{Without using the tensor branch configuration, we can ask abstractly when there exists a flow between theories defined by the data $(N, 2k, O)$ and $(N, 2\ell, O')$, where $O$ and $O'$ are nilpotent orbits of $\mathfrak{su}(2k)$ and $\mathfrak{su}(2\ell)$, respectively. The answer is a D-type analogue of what is called ``simultaneous two-box deletion'' in \cite{DKL}. This is triggered via giving a vacuum expectation value to the D-type analogue of an ``end-to-end'' operator, which has been studied in the A-type theories in \cite{Baume:2020ure,Razamat:2019mdt,Bergman:2020bvi,Baume:2022cot,DKL}. However, in this paper, we proceed case-by-case, and thus we do not need to explore the general behavior of such transitions.} precisely when there exists a Higgs branch renormalization group flow between theories 
\begin{equation}
    D_N^{\mathfrak{su}_{2k}}(O) \quad \longrightarrow \quad D_N^{\mathfrak{su}_{2k-2}}(O') \,.
\end{equation}
This provides a partial ordering on the set of theories of the form $D_N^{\mathfrak{su}_{2k}}(O)$, and the Hasse diagram of such theories with this partial ordering provides a subdiagram of the Hasse diagram of the Higgs branch of $D_N^{\mathfrak{su}_{2k}}$. We emphasize that this is a priori only a subdiagram, as there may be subloci of the Higgs branch where there are interacting fixed points not of the form $D_N^{\mathfrak{su}_{2k}}(O)$, or else product theories, theories with a free sector, \&c. We depict this Hasse diagram for theories $D_N^{\mathfrak{su}_{2\ell}}(O)$, for $2\ell \leq 8$, in Figure \ref{fig:hasseeeee}.

While the SCFTs associated with tensor branch configurations as in equation \eqref{eqn:TBDNdecorated} are the main subject of analysis in this paper, there are other 6d $(1,0)$ SCFTs which are Higgsable to the D-type $(2,0)$ SCFTs. We briefly review these theories here for completeness. There are 6d $(1,0)$ SCFTs that we label as 
\begin{equation}
    D_N^{\mathfrak{e}_6}(O) \,,
\end{equation}
where $O$ is a nilpotent orbit of $\mathfrak{e}_6$. These theories are so-named as they behave similarly to the theories described around equation \eqref{eqn:DNO}. When $O$ is the nilpotent orbit of dimension zero, the tensor branch configuration takes the form
\begin{equation}\label{eqn:DNE6}
    \overset{\mathfrak{su}_3}{3}1 \overset{\displaystyle \overset{\mathfrak{su}_3}{3}}{\overset{\displaystyle 1}{\overset{\mathfrak{e}_6}{6}}} 1\overset{\mathfrak{su}_3}{3}1\overbrace{\overset{\mathfrak{e}_6}{6} 1\overset{\mathfrak{su}_3}{3}1 \cdots \overset{\mathfrak{e}_6}{6} 1\overset{\mathfrak{su}_3}{3}1}^{N-3}\,[\mathfrak{e}_6] \,,
\end{equation}
where we have also depicted the $\mathfrak{e}_6$ flavor symmetry on the right. When $O$ is non-trivial, the configuration is modified following the rules described in the appendices of \cite{Baume:2023onr}. In addition, there are a small number of exceptional theories, which we denote only by their tensor branch configurations. We have the parent theories
\begin{equation}\label{eqn:DtypeEXCEPT}
    2 \overset{\displaystyle 2}{\overset{\mathfrak{su}_{2}}{2}}\underbrace{\,\overset{\mathfrak{su}_{3}}{2} \cdots \overset{\mathfrak{su}_{3}}{2} \,}_{N-3}\,[\mathfrak{su}_3] \qquad \text{ and } \qquad 
    \overset{\mathfrak{su}_2}{2}\overset{\displaystyle\overset{\mathfrak{su}_2}{2}}{\overset{\mathfrak{so}_7}{3}}1\underbrace{\overset{\mathfrak{so}_8}{4}1\cdots\overset{\mathfrak{so}_8}{4}1\,}_{N-3}\,[\mathfrak{so}_8]
    \,,
\end{equation}
together with their descendants obtained via nilpotent Higgsing of the $\mathfrak{su}(3)$ and $\mathfrak{so}(8)$ flavor symmetries, respectively. 

These four classes of theories exhaust the 6d $(1,0)$ SCFTs that are Higgsable to D-type $(2,0)$, where $N$ can be taken to be arbitrarily large. When considering SCFTs Higgsable to the D-type $(2,0)$ SCFTs with small $N$, there are more options available. 

For an SCFT $D_N^{\mathfrak{su}_{2k}}$ with $N \geq 2k + 1$, then for each nilpotent orbit $O$ of $\mathfrak{su}(2k)$ there exists a unique child theory associated with $O$, and the set of interacting non-product fixed points on the Higgs branch of $D_N^{\mathfrak{su}_{2k}}$ is saturated by the set of theories $D_N^{\mathfrak{su}_{2\ell}}(O)$, for all $\ell \leq k$ and $O$ a nilpotent orbit of $\mathfrak{su}(2\ell)$. We refer to this as the ``long quiver'' case. In contrast, for short quivers, there may be no child SCFTs associated with particular nilpotent orbits, and there may be interacting non-product fixed points on the Higgs branch which are not of the form $D_N^{\mathfrak{su}_{2\ell}}(O)$. Let $O$ be a nilpotent orbit of $\mathfrak{su}(2k)$ associated with a partition
\begin{equation}
    P = [p_1, \cdots] \,,
\end{equation}
of $2k$, written in weakly-decreasing order. It is straightforward to see that the SCFT $D_N^{\mathfrak{su}_{2k}}(O)$ is well-defined when 
\begin{equation}
    p_1 \leq N - 1 \,.
\end{equation}
One can also check that the anomaly polynomial of the $D_N^{\mathfrak{su}_{2k}}(O)$ SCFT is exactly what one would expect from the nilpotent Higgsing of the moment map operator for the $\mathfrak{su}(2k)$ flavor symmetry, as in \cite{Baume:2023onr}. Therefore, the first speciality of short quivers is that nilpotent orbits where the partition contains an element $> N-1$ do not exist.

If $N < 2k + 1$, then the interacting fixed points on the Higgs branch of $D_N^{\mathfrak{su}_{2k}}$ can be captured by $D_N^{\mathfrak{su}_{2\ell}}(O)$ where $\ell \leq k$ and $O$ is a \emph{generalized partition}. A generalized partition of $2\ell$ is either an integer partition of $2\ell$ such no element is larger than $N-2$:
\begin{equation}\label{eqn:genpartboring}
    \left[(N-2)^{m_{N-2}}, \cdots, 1^{m_{1}} \right] \qquad \text{ such that } \qquad \sum_{i=1}^{N-2} i m_i = 2\ell \,,
\end{equation}
or else it can be written as
\begin{equation}\label{eqn:genpart}
     \left[ (N-1)^{m_{N-1}^L}, (N-1)^{m_{N-1}^R} ; (N-2)^{m_{N-2}}, \cdots, 1^{m_{1}} \right] \,.
\end{equation}
Here, the exponents are non-negative integers such that there exist non-negative integers $k_i$, $k_{N-1}^{L,R}$ satisfying
\begin{equation}\label{eqn:cartanm}
  \begin{aligned}
    k_{N-1}^L = \ell &\,, \\
    2k_{N-1}^L - k_{N-2} - m_{N-1}^L  = 0 &\,, \\
    2k_{N-1}^R - k_{N-2} - m_{N-1}^R = 0 &\,, \\
    2k_{N-2} - k_{N-1}^L - k_{N-1}^R -k_{N-3} - m_{N-2} = 0 &\,, \\
    2k_{N-3} - k_{N-2} - k_{N-4} - m_{N-3} = 0 &\,, \\
    \cdots &\,, \\
    2k_1 - k_2 - m_1 = 0 &\,,
  \end{aligned}
\end{equation}
where we have assumed, without loss of generality, that $m_{N-1}^L \geq m_{N-1}^R$ and $m_{N-1}^L \neq 0$. For the generalized partition written in equation \eqref{eqn:genpartboring}, the description of the associated tensor branch is straightforward, and is written in \cite{Baume:2023onr}; in particular, it is treated as a regular partition of $2\ell$. For a generalized partition of the form in equation \eqref{eqn:genpart}, the curve configuration follows from the solutions to equation \eqref{eqn:cartanm}; it is
\begin{equation}
    \overset{\mathfrak{su}_{k_{N-1}^L}}{2}
    \overset{\displaystyle \overset{\mathfrak{su}_{k_{N-1}^R}}{2}}{\overset{\mathfrak{su}_{k_{N-2}}}{2}} \cdots \overset{\mathfrak{su}_{k_1}}{2} \,.
\end{equation}

There are simple rules for determining the partial ordering on these generalized partitions, similar in style to those given in \cite{DKL}, however, we do not go into details here as, case-by-case, it is straightforward to determine the partial ordering using equation \eqref{eqn:RGflow}. To give the reader a clear picture of the generalized partitions, we have included in Figure \ref{fig:hasseeeee} the Hasse diagram of interacting fixed points on the Higgs branch of $D_N^{\mathfrak{su}_8}$ for $N = 3, 7$. They are of the form $D_N^{\mathfrak{su}_{2\ell}}(O)$ where $\ell \leq 4$ and $O$ is a generalized partition.\footnote{The observant reader may notice that there is no $[0]$ at the bottom of Figure \ref{fig:hasseeeee} (c). This is special to $N=3$ and is related to the fact that $D_3^{\mathfrak{su}_2}([1^2])$ has enhanced $\mathfrak{su}(3)$ flavor symmetry \cite{Morrison:2016djb}, instead of $\mathfrak{su}(2)$.}

The structure of the generalized partition in equation \eqref{eqn:genpart} is particularly natural when considering Type IIA brane engineering with ON$^-$-planes, as we shall see in Section \ref{sec:Magnetic_quiver}.

\subsection{Higgsable to E-type \texorpdfstring{$(2,0)$}{(2,0)}} \label{sec:EEE}

We now turn to those 6d $(1,0)$ SCFTs that have SUSY-enhancing Higgs branch renormalization group flows to the 6d $(2,0)$ SCFTs of exceptional type. In fact, there is not a lot of variety in such 6d $(1,0)$ SCFTs. There are three infinite families which are associated with the configurations
\begin{equation}
  \begin{gathered}
    \EsixSU{k_1}{k_2}{k_3}{k_4}{k_5}{k_6} \,, \qquad \EsevenSU{k_1}{k_2}{k_3}{k_4}{k_5}{k_6}{k_7} \,, \qquad \EeightSU{k_1}{k_2}{k_3}{k_4}{k_5}{k_6}{k_7}{k_8} \,,
    \end{gathered}
\end{equation}
which have SUSY-enhancing flows to the 6d $(2,0)$ SCFTs of type $E_6$, $E_7$, and $E_8$, respectively. We refer to the SCFTs associated with these tensor branch curve configurations as
\begin{equation}\label{eqn:EtypeNot}
    E_N(k_1, \cdots, k_N) \,,
\end{equation}
for each of $N = 6, 7, 8$. As in the Higgsable to D-type case, most of the tuples $(k_1, \cdots, k_N)$ need to be discarded since they give rise to gauge-anomalous theories; only those obeying the E-type analogue of equation \eqref{eqn:Cartan_Gauge=flavor}, i.e., where the Cartan matrix is the $E_N$ Cartan matrix, realize physical 6d $(1,0)$ SCFTs. 

\begin{figure}[p]
    \centering
    \includegraphics[page=1,height=0.90\textheight]{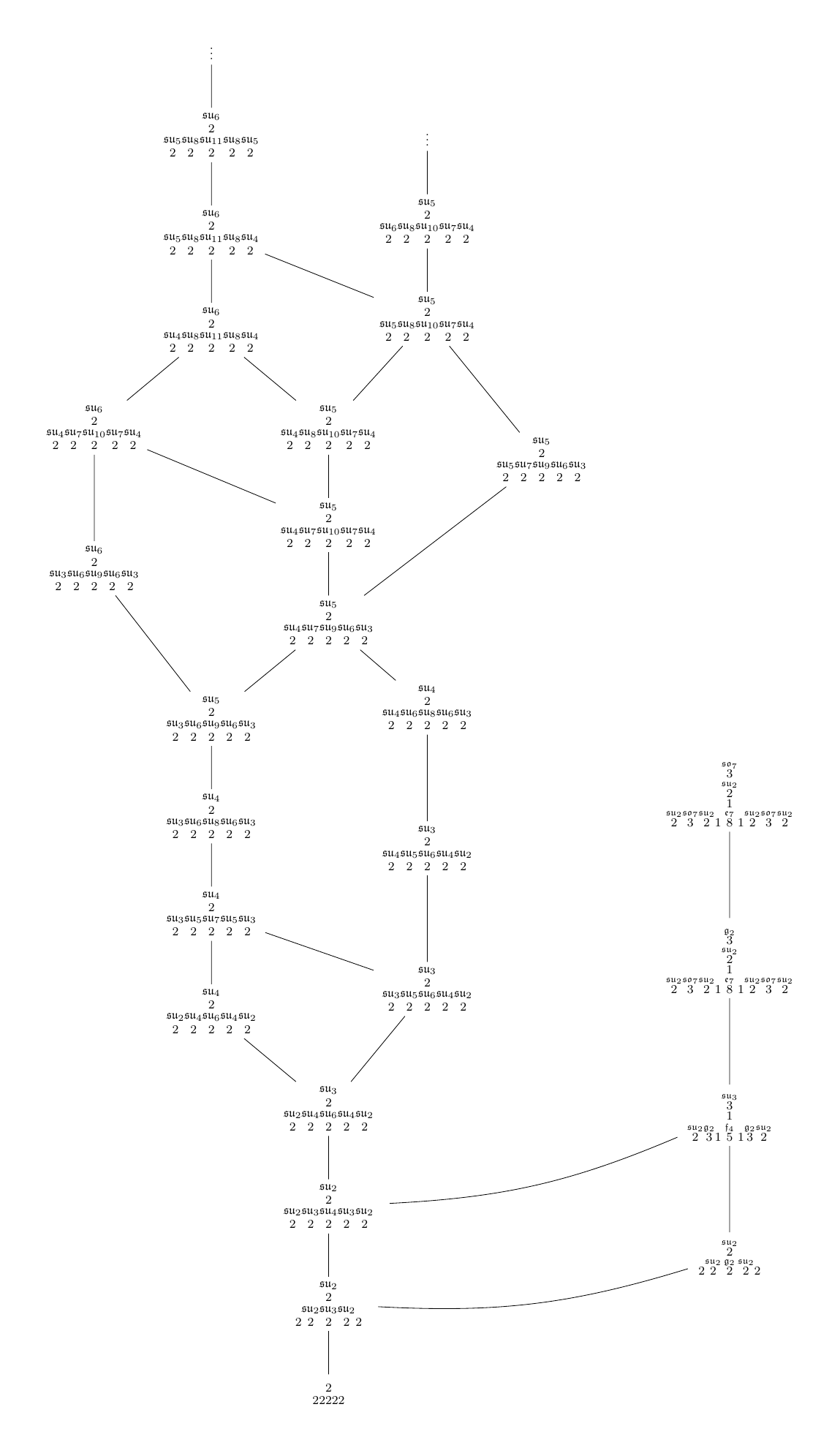}
    \caption{The Hasse diagram of the 6d $(1,0)$ SCFTs that have SUSY-enhancing Higgs branch RG flows to the 6d $(2,0)$ SCFT of type $E_6$. For symmetric configurations, we have depicted only one rather than both.}
    \label{fig:hasseE6}
\end{figure}

For each of these theories, it is straightforward to determine the dimension of the Higgs branch using the standard anomaly polynomial machinery, just as we did for the $D_N^{\mathfrak{su}_{2k}}(O)$ SCFTs in Section \ref{sec:DDD}. We find that the dimension of the Higgs branch of $E_N(k_1, \cdots, k_N)$ is
\begin{equation}\label{eqn:dimHE}
    \operatorname{dim}(\mathcal{H}) = N + \frac{1}{2} \sum_{i,j = 1}^N A^{ij} k_i k_j \,,
\end{equation}
where $A^{ij}$ is the positive-definite Cartan matrix of $E_N$. Note that this expression holds for all $k_i \geq 0$ satisfying the E-type analogue of equation \eqref{eqn:Cartan_Gauge=flavor}.

For the families in equation \eqref{eqn:EtypeNot}, the presence of a Higgs branch renormalization group flow between two theories is particularly straightforward. If 
\begin{equation}\label{eqn:EPO}
    (k_1', \cdots, k_N') < (k_1, \cdots, k_N) \,,
\end{equation}
applied element-by-element,\footnote{I.e., if $k_i' \leq k_i$ for all $i = 1, \cdots, N$, excepting the trivial case where the inequality is saturated for all $i$.} then there exists a flow between the SCFTs:
\begin{equation}
    E_N(k_1, \cdots, k_N) \quad \rightarrow \quad E_N(k_1', \cdots, k_N') \,.
\end{equation}
We have depicted a part of the Hasse diagram, under the partial ordering defined in equation \eqref{eqn:EPO}, of consistent SCFTs of this form with $N=6$ in Figure \ref{fig:hasseE6}. We further analyze the types of elementary transition in this Hasse diagram in Section \ref{sec:Etype}.

In addition to these standard families of 6d $(1,0)$ SCFTs that have supersymmetry-enhancing RG flows to the 6d $(2,0)$ SCFTs of type $E_{6,7,8}$, there are a small number of exceptional curve configurations that also realize such flows. When considering the $E_6$ theory, there are four such exceptional curve configurations, which are
\begin{equation}\label{eqn:E6exceptionals}
  \begin{gathered}
    2\overset{\mathfrak{su}_2}{2}\overset{\displaystyle \overset{\mathfrak{su}_2}{2}}{\overset{\mathfrak{g}_2}{2}}\overset{\mathfrak{su}_2}{2}2
    \,, \qquad 
    \overset{\mathfrak{su}_2}{2}\overset{\mathfrak{g}_2}{3}1\overset{\displaystyle \overset{\displaystyle \overset{\mathfrak{su}_3}{3}}{1}}{\overset{\mathfrak{f}_4}{5}}1\overset{\mathfrak{g}_2}{3}\overset{\mathfrak{su}_2}{2} \,, 
    \qquad
    \overset{\mathfrak{su}_2}{2}\overset{\mathfrak{so}_7}{3}\overset{\mathfrak{su}_2}{2}1\overset{\displaystyle \overset{\displaystyle \overset{\displaystyle \overset{\mathfrak{g}_2}{3}}{\overset{\mathfrak{su}_2}{2}}}{1}}{\overset{\mathfrak{e}_7}{8}}1\overset{\mathfrak{su}_2}{2}\overset{\mathfrak{so}_7}{3}\overset{\mathfrak{su}_2}{2} \,, 
    \qquad 
    \overset{\mathfrak{su}_2}{2}\overset{\mathfrak{so}_7}{3}\overset{\mathfrak{su}_2}{2}1\overset{\displaystyle \overset{\displaystyle \overset{\displaystyle \overset{\mathfrak{so}_7}{3}}{\overset{\mathfrak{su}_2}{2}}}{1}}{\overset{\mathfrak{e}_7}{8}}1\overset{\mathfrak{su}_2}{2}\overset{\mathfrak{so}_7}{3}\overset{\mathfrak{su}_2}{2} \,.
  \end{gathered}
\end{equation}
We have incorporated the SCFTs associated with these curve configurations into the Hasse diagram that we have drawn in Figure \ref{fig:hasseE6}. For the $E_7$ theory, there is only a single exceptional configuration:
\begin{equation}
  \begin{gathered}
    2\overset{\mathfrak{su}_2}{2}\overset{\mathfrak{g}_2}{3}1\overset{\displaystyle \overset{\displaystyle \overset{\mathfrak{su}_3}{3}}{1}}{\overset{\mathfrak{f}_4}{5}}1\overset{\mathfrak{g}_2}{3}\overset{\mathfrak{su}_2}{2} \,. 
  \end{gathered}
\end{equation}
The are no exceptional curve configurations which have Higgs branch RG flows to the 6d $(2,0)$ SCFT of type $E_8$.

\section{A Magnetic Quiver for the \texorpdfstring{\boldmath{$D_N^{\mathfrak{su}_{2k}}(O)$}}{DNsu2k(O)} SCFTs}\label{sec:Magnetic_quiver}

Now that we have enumerated the 6d $(1,0)$ SCFTs that have supersymmetry-enhancing renormalization group flows to the 6d $(2,0)$ SCFTs of DE-type, we would like to analyze the structure of the Higgs branch of these SCFTs. Conveniently, the $D_N^{\mathfrak{su}_{2k}}(O)$ theories can be obtained via a brane construction in Type IIA string theory. This allows us to obtain a magnetic quiver for the Higgs branch of $D_N^{\mathfrak{su}_{2k}}(O)$, and we can then study the structure of the Higgs branch using techniques from the study of the Coulomb branch of 3d $\mathcal{N}=4$ Lagrangian theories. This approach will provide non-trivial insights into the Higgs branch renormalization group flows which can then be generalized to the $E_N(k_1, \cdots, k_N)$ SCFTs that do not have such a brane construction.

\subsection{The Electric Type IIA Description}\label{sec:Electric_SetUp}

A construction in Type IIA string theory for D-type quiver theories has been long-known \cite{Kapustin:1998fa}. The ingredients are the usual D6-D8-NS5-branes that engineer six-dimensional theories, together with a Neveu--Schwarz orientifold plane: $\mathrm{ON}^-$. These are arranged in the ten-dimensional spacetime as depicted in Table \ref{tab:Brane_conf}. This provides us with a mechanism to engineer the 6d SQFTs that are the effective field theory living on the generic point of the tensor branch of the SCFTs of interest.

To understand the gauge theories engineered on the brane worldvolume, it is useful to pass to the T-dual Type IIB description \cite{Kapustin:1998fa}. Of primary importance is the counterpart of the Neveu--Schwarz orientifold plane. This plane, characterized by a negative NS5-charge, is inevitably paired with an NS5-brane to create an uncharged object. Consequently, its Type IIB dual is an $\mathrm{O5}^-$-plane paired with a D5-brane, such that the total Ramond--Ramond charge is vanishing. The worldvolume of this combined $\mathrm{O5}^-$-plane plus D5-brane carries an $SO(2)$ gauge theory, and D3-branes ending on this combined object can have one of two possible boundary conditions. Different stacks with different boundary conditions lead to the characteristic bifurcation in the D-type Dynkin diagram. The analogous picture for D6-branes ending on the combined ON$^-$-plane and paired NS5-brane explains the bifurcation in the 6d tensor branch description.

\begin{table}[t]
    \centering
    \begin{threeparttable}
    \begin{tabular}{cccccccccccc}
        \toprule
         & $x^0$ & $x^1$& $x^2$ & $x^3$ & $x^4$& $x^5$& $x^6$& $x^7$& $x^8$& $x^9$  \\ \midrule
         D6 & $\times$ & $\times$& $\times$& $\times$& $\times$& $\times$& $\times$& $\cdot$& $\cdot$& $\cdot$ \\
         NS5 &  $\times$ & $\times$& $\times$& $\times$& $\times$& $\times$& $\cdot$& $\cdot$& $\cdot$& $\cdot$ \\
         D8 &  $\times$ & $\times$& $\times$& $\times$& $\times$& $\times$& $\cdot$& $\times$& $\times$& $\times$ \\
         $\mathrm{ON}^-$ & $\times$ & $\times$& $\times$& $\times$& $\times$& $\times$& $\cdot$& $\cdot$& $\cdot$& $\cdot$ \\ \bottomrule
    \end{tabular}
    \end{threeparttable}
    \caption{Spacetime brane extension in Type IIA string theory for D-type quivers.}
    \label{tab:Brane_conf}
\end{table}

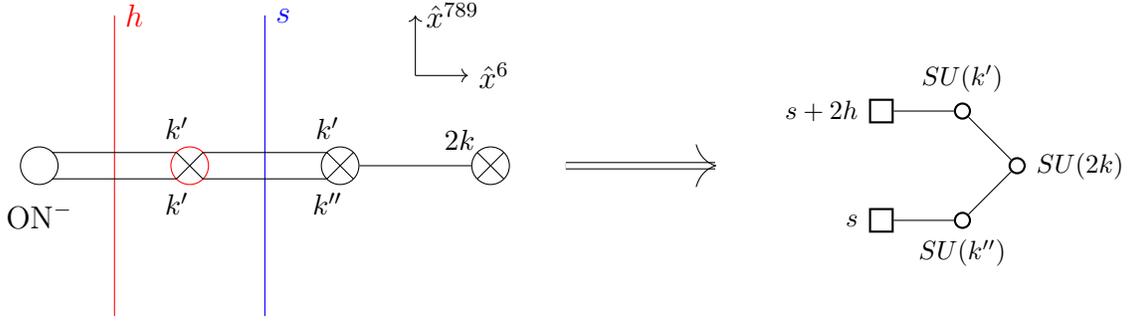
\begin{figure}[t]
    \centering
    \begin{tikzpicture}[cross/.style={path picture={ 
  \draw[black]
(path picture bounding box.south east) -- (path picture bounding box.north west) (path picture bounding box.south west) -- (path picture bounding box.north east);
}}]
            \draw[->] (3,1.2)--(3.7,1.2) node[right] () {$\hat{x}^6$};
            \draw[->] (3,1.2)--(3,2) node[right] () {$\hat{x}^{789}$};
            \node[circle, draw=black, minimum size=5mm] at (-2,0) (nodo0) {};
            \node[circle, draw=red, cross, minimum size=5mm] at (0,0) (nodo1) {};
            \node[circle, draw=black, cross, minimum size=5mm] at (2,0) (nodo2) {};
            \node[circle,draw=black, cross, minimum size=5mm] at (4,0) (nodo3) {};

            \draw[red] (-1,-2)--(-1,2) node[right] () {$h$};
            \draw[blue] (1,-2)--(1,2) node[right] () {$s$};
            \draw (-1.83,0.175)--(-0.17,0.175) node[left=3pt,above=1pt] () {\small{$k'$}};
            \draw (-1.83,-0.175)--(-0.17,-0.175) node[left=3pt,below=1pt] () {\small{$k'$}};
            \draw (0.17,0.175)--(1.83,0.175) node[left=3pt,above=1pt] () {\small{$k'$}};
            \draw (0.17,-0.175)--(1.83,-0.175) node[left=3pt,below=1pt] () {\small{$k''$}};
            
            \draw (nodo2)--(nodo3) node[above left =2.5pt] () {\small{$2k$}};
        
            \draw[->,double distance = 2pt] (5,0)--(7,0);

            \node[node, label=right:{\footnotesize $SU(2k)$}] (2k) at (11,0) {};
            \node[node, label=above:{\footnotesize $SU(k')$}] (ktop)  [above left=8mm of 2k] {};
            \node[node, label=below:{\footnotesize $SU(k'')$}] (kbot)  [below left=8mm of 2k] {};
            \node[fnode, label=left:{\footnotesize $s+2h$}] (ftop)  [left=8mm of ktop] {};
            \node[fnode, label=left:{\footnotesize $s$}] (fbot)  [left=8mm of kbot] {};
            \draw (2k)--(ktop);
            \draw (2k)--(kbot);
            \draw (ktop)--(ftop);
            \draw (kbot)--(fbot);
            
        \node[circle, minimum size=5mm,below] at (-2,0) () {$\mathrm{ON}^-$};

    \end{tikzpicture}
    \caption{Quiver obtained on the $D_3$ bifurcation of the quiver theory by placing D8-branes between the $\mathrm{ON}^-$-plane and its paired NS5-brane (in red) and between the paired NS5-brane and NS5-brane to the right (in blue). We use the standard convention, throughout this paper, of depicting NS5-branes with crossed circles, D6-branes as horizontal lines, and D8-branes as vertical lines.}
    \label{fig:flavor_branes}
\end{figure}

Focusing on the Type IIA engineering, the $\mathrm{ON}^-$ orientifold plane can interplay with D8-branes to construct bound states \cite{Hanany:1999sj}. In fact, binding D8-branes to the $\mathrm{ON}^-$-plane (this case will be drawn as D8-branes between the $\mathrm{ON}^-$ and its paired NS5-brane) provides a flavor symmetry factor only in one direction of the bifurcation. Differently, inserting D8-branes between the NS5-brane paired to the $\mathrm{ON}^-$ and the adjacent NS5-brane, the heavy branes act as a source of flavor symmetry on both branches of the bifurcation. Both these possibilities are shown in Figure \ref{fig:flavor_branes}.

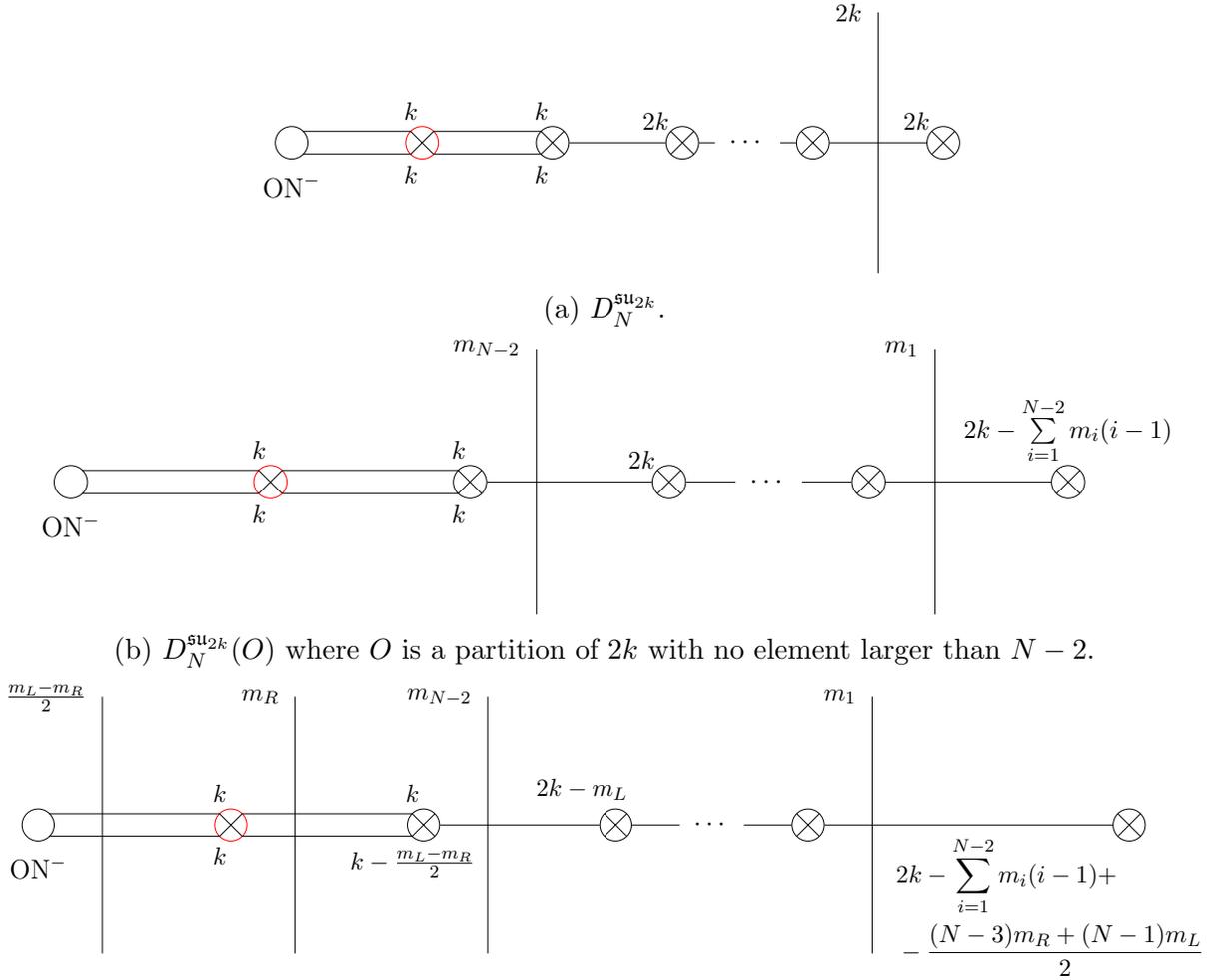
\begin{figure}[p]
    \centering
    \begin{subfigure}[b]{0.9\textwidth}
    \centering
    \resizebox{0.65\textwidth}{!}{
    \begin{tikzpicture}[cross/.style={path picture={ 
  \draw[black]
(path picture bounding box.south east) -- (path picture bounding box.north west) (path picture bounding box.south west) -- (path picture bounding box.north east);
}}]
            \node[circle, draw=black, minimum size=5mm] at (-2,0) (nodo0) {};
            \node[circle, draw=red, cross, minimum size=5mm] at (0,0) (nodo1) {};
            \node[circle, draw=black, cross, minimum size=5mm] at (2,0) (nodo2) {};
            \node[circle,draw=black, cross, minimum size=5mm] at (4,0) (nodo3) {};
            \node[circle,draw=black, cross, minimum size=5mm] at (6,0) (nodo4) {};
            \node[circle,draw=black, cross, minimum size=5mm] at (8,0) (nodo5) {};

            \draw (-1.83,0.175)--(-0.17,0.175) node[left=3pt,above=1pt] () {\small{$k$}};
            \draw (-1.83,-0.175)--(-0.17,-0.175) node[left=3pt,below=1pt] () {\small{$k$}};
            \draw (0.17,0.175)--(1.83,0.175) node[left=3pt,above=1pt] () {\small{$k$}};
            \draw (0.17,-0.175)--(1.83,-0.175) node[left=3pt,below=1pt] () {\small{$k$}};
            \draw (nodo2)--(nodo3) node[above left =2.5pt] () {\small{$2k$}};
            \draw (nodo3)--(4.5,0);
            \draw (5.5,0)--(nodo4);
            \node at (5,0) () {$\cdots$};
            \draw (nodo4)--(nodo5) node[above left =2.5pt] () {\small{$2k$}};
            \draw (7,-2)--(7,2) node[left=3pt] () {\small{$2k$}};
                   \node[circle, minimum size=5mm,below] at (-2,0) () {$\mathrm{ON}^-$};
    \end{tikzpicture}
    }
    \caption{$D_N^{\mathfrak{su}_{2k}}$.}\label{fig:D_n-Branes_a}
    \end{subfigure}
    \begin{subfigure}[b]{0.99\textwidth}
    \centering
    \resizebox{0.95\textwidth}{!}{
    \begin{tikzpicture}[cross/.style={path picture={ 
  \draw[black]
(path picture bounding box.south east) -- (path picture bounding box.north west) (path picture bounding box.south west) -- (path picture bounding box.north east);
}}]
            \node[circle, draw=black, minimum size=5mm] at (-3,0) (nodo0) {};
            \node[circle, draw=red, cross, minimum size=5mm] at (0,0) (nodo1) {};
            \node[circle, draw=black, cross, minimum size=5mm] at (3,0) (nodo2) {};
            \node[circle,draw=black, cross, minimum size=5mm] at (6,0) (nodo3) {};
            \node[circle,draw=black, cross, minimum size=5mm] at (9,0) (nodo4) {};
            \node[circle,draw=black, cross, minimum size=5mm] at (12,0) (nodo5) {};

            \draw (-2.83,0.175)--(-0.17,0.175) node[left=3pt,above=1pt] () {\small{$k$}};
            \draw (-2.83,-0.175)--(-0.17,-0.175) node[left=3pt,below=1pt] () {\small{$k$}};
            \draw (0.17,0.175)--(2.83,0.175) node[left=3pt,above=1pt] () {\small{$k$}};
            \draw (0.17,-0.175)--(2.83,-0.175) node[left=3pt,below=1pt] () {\small{$k$}};
            \draw (nodo2)--(nodo3) node[above left =2.5pt] () {\small{$2k$}};
            \draw (nodo3)--(7,0);
            \draw (8,0)--(nodo4);
            \node at (7.5,0) () {$\cdots$};
            \draw (nodo4)--(nodo5) node[label=above:{\small{$2k-\sum \limits_{i=1}^{N-2} m_i (i-1)$}}] () {};
            \draw (10,-2)--(10,2) node[left=3pt] () {\small{$m_1$}};
            \draw (4,-2)--(4,2) node[left=3pt] () {\small{$m_{N-2}$}};
                   \node[circle, minimum size=5mm,below] at (-3,0) () {$\mathrm{ON}^-$};
    \end{tikzpicture}
    }
    \caption{$D_N^{\mathfrak{su}_{2k}}(O)$ where $O$ is a partition of $2k$ with no element larger than $N - 2$.}\label{fig:D_n-Branes_b}
    \end{subfigure}
    \begin{subfigure}[b]{0.99\textwidth}
    \centering
    \resizebox{\textwidth}{!}{
    \begin{tikzpicture}[cross/.style={path picture={ 
  \draw[black]
(path picture bounding box.south east) -- (path picture bounding box.north west) (path picture bounding box.south west) -- (path picture bounding box.north east);
}}]
            \node[circle, draw=black, minimum size=5mm] at (-3,0) (nodo0) {};
            \node[circle, draw=red, cross, minimum size=5mm] at (0,0) (nodo1) {};
            \node[circle, draw=black, cross, minimum size=5mm] at (3,0) (nodo2) {};
            \node[circle,draw=black, cross, minimum size=5mm] at (6,0) (nodo3) {};
            \node[circle,draw=black, cross, minimum size=5mm] at (9,0) (nodo4) {};
            \node[circle,draw=black, cross, minimum size=5mm] at (14,0) (nodo5) {};

            \draw (-2.83,0.175)--(-0.17,0.175) node[left=3pt,above=1pt] () {\small{$k$}};
            \draw (-2.83,-0.175)--(-0.17,-0.175) node[left=3pt,below=1pt] () {\small{$k$}};
            \draw (0.17,0.175)--(2.83,0.175) node[left=3pt,above=1pt] () {\small{$k$}};
            \draw (0.17,-0.175)--(2.83,-0.175) node[left=3pt,below=1pt] () {\small{$k-\frac{m_{L}-m_R}{2}$}};
            \draw (nodo2)--(nodo3) node[label={[xshift=-15pt, yshift=3pt]\small{$2k-m_L$}}] () {};
            \draw (nodo3)--(7,0);
            \draw (8,0)--(nodo4);
            \node at (7.5,0) () {$\cdots$};
            \draw (nodo4)--(nodo5) node[label={[xshift=-35pt, yshift=-75pt] \small $\begin{gathered}
                \begin{aligned}
                     &2k-\sum \limits_{i=1}^{N-2} m_i (i-1) +\\
                     &-\frac{(N-3)m_R+(N-1)m_L}{2}
                \end{aligned}
            \end{gathered}$}] () {};
            \draw (10,-2)--(10,2) node[left=3pt] () {\small{$m_1$}};
            \draw (4,-2)--(4,2) node[left=3pt] () {\small{$m_{N-2}$}};
            \draw (1,-2)--(1,2) node[left=3pt] () {\small{$m_R$}};
            \draw (-2,-2)--(-2,2) node[left=3pt] () {\small{$\frac{m_{L}-m_R}{2}$}};
                   \node[circle, minimum size=5mm,below] at (-3,0) () {$\mathrm{ON}^-$};
    \end{tikzpicture}
    }
    \caption{$D_N^{\mathfrak{su}_{2k}}(O)$ where $O$ is a generalized partition of $2k$ of the form in equation \eqref{eqn:genpart}, under the general assumption $m_L\ge m_R$.}\label{fig:D_n-Branes_c}
    \end{subfigure}
    \caption{Type IIA brane configuration engineering the effective field theory on the tensor branch of the  $D^{\su_{2k}}_{N}\left( O \right)$ SCFTs. The number of D6-branes between each pair of NS5-branes, when not written, can be determined from the anomaly cancellation conditions.}
    \label{fig:D_n-Branes}
\end{figure}

Hence, it is straightforward to write down the brane construction for $D^{\su_{2k}}_{N}$. It is realized as follows. Consider a linear chain of $N$ NS5-branes, with an ON$^-$ on the left. In the rightmost interval between NS5-branes, we place a stack of $2k$ D8-branes. This configuration is depicted in Figure \ref{fig:D_n-Branes_a}. The generalization to $D^{\su_{2k}}_{N}(O)$, where $O$ is a partition of $2k$ with no element larger than $N-2$, is also clear: there are instead $m_i$ D8-branes in the interval between the $i$th and $(i+1)$th NS5-branes. Since $m_i = 0$ for $i \geq N-1$, the D8-branes do not lie in an interval that affect the number of D6-branes in the bifurcation. We show this schematically in Figure \ref{fig:D_n-Branes_b}.

Finally, it remains for us to write down the brane configuration associated with $D^{\su_{2k}}_{N}(O)$ where $O$ is a generalized partition given in the form written in equation \eqref{eqn:genpart}. Again, there are $N$ NS5-branes arranged in a linear chain, together with the ON$^-$-plane to the left of the left-most NS5-brane. Between the $i$th and $(i+1)$th NS5-branes, counting from the right, for $i = 1, \cdots, N-2$, there are $m_i$ D8-branes, where $m_i$ is as defined in equation \eqref{eqn:genpart}. Furthermore, we define the following two integers:
\begin{equation}
    h= \frac{\left|m_{N-1}^L-m_{N-1}^R\right|}{2} \,, \qquad s=\mathrm{min}(m_{N-1}^L,m_{N-1}^R) \,.
\end{equation} 
From equation \eqref{eqn:cartanm}, we can see that $h$ is always an integer. In the brane configuration, we have $s$ D8-branes between the $(N-1)$th and the $N$th NS5-branes, and $h$ D8-branes between the $N$th NS5-brane and the ON$^-$-plane. The number of D6-branes in each interval is fixed by the anomaly cancellation condition in equation \eqref{eqn:cartanm}, and this is analogous to setting the number of D6-branes ending on the left and on the right of an NS5-brane via the Type IIA cosmological constant whilst taking into account the presence of D8-branes \cite{Hanany:1997gh,Hanany:1997sa}. We have depicted this brane system in Figure \ref{fig:D_n-Branes_c}.

While we have given a prescription to associate a brane description to a generalized partition, it remains for us to motivate the connection to Higgsing. Each D6-brane ends on a D8-brane that can be consider as brought in from infinity on the right. The generalized partition then specifies which D6-brane emanating from which NS5-branes end on which D8-branes. The generalized nature of the partition is necessary as the D6-branes coming out of the two left-most NS5-branes are paired due to the ON$^-$. Hanany--Witten transitions can then be transformed to pull all of the D8-branes into the intervals between the necessary pair of NS5-branes, leading to the configurations that we have described in Figure \ref{fig:D_n-Branes}. We have depicted the configurations before and after the Hanany--Witten transitions for an example generalized partition in Figure \ref{fig:HananyWitten}.

\begin{figure}[p]
    \centering
    \resizebox{\textwidth}{!}{
    \begin{tikzpicture}[cross/.style={path picture={ 
  \draw[black]
(path picture bounding box.south east) -- (path picture bounding box.north west) (path picture bounding box.south west) -- (path picture bounding box.north east);
}}]

            \node[circle, draw=black, minimum size=5mm] at (-2,0) (nodo0) {};
            \node[circle, draw=black, cross, minimum size=5mm] at (0,0) (nodo1) {};
            \node[circle, draw=black, cross, minimum size=5mm] at (2,0) (nodo2) {};
            \node[circle,draw=black, cross, minimum size=5mm] at (4,0) (nodo3) {};
            \node[circle,draw=black, cross, minimum size=5mm] at (6,0) (nodo4) {};
            \node[circle,draw=black, cross, minimum size=5mm] at (8,0) (nodo5) {};
            \node[circle,draw=black, cross, minimum size=5mm] at (10,0) (nodo6) {};

            \draw (-1.83,0.175)--(-0.17,0.175) node[left=3pt,above=1pt] () {\small{$11$}};
            \draw (-1.83,-0.175)--(-0.17,-0.175) node[left=3pt,below=1pt] () {\small{$11$}};
            \draw (0.17,0.175)--(1.83,0.175) node[left=3pt,above=1pt] () {\small{$11$}};
            \draw (0.17,-0.175)--(1.83,-0.175) node[left=3pt,below=1pt] () {\small{$10$}};
            \draw (nodo2)--(nodo3) node[above left =2.5pt] () {\small{$19$}};
            \draw (nodo3)--(nodo4) node[above left =2.5pt] () {\small{$16$}};
            \draw (nodo4)--(nodo5) node[above left =2.5pt] () {\small{$11$}};
            \draw (nodo5)--(nodo6) node[above left =2.5pt] () {\small{$6$}};
            \draw (11,-2)--(11,2) node[left=3pt] () {};
            \draw (11.5,-2)--(11.5,2) node[left=3pt] () {};
            \draw (12,-2)--(12,2) node[left=3pt] () {};
            \draw (12.5,-2)--(12.5,2) node[left=3pt] () {};
            \draw (13,-2)--(13,2) node[left=3pt] () {};
            \draw (13.5,-2)--(13.5,2) node[left=3pt] () {};

            \draw (nodo6)--(11,0);
            \draw (nodo4)--(6.75,-0.5)--(11.5,-0.5);
            \draw (nodo4)--(6.5,-0.75)--(12,-0.75);
            \draw (nodo3)--(4.5,0.5)--(12.5,0.5);
            \draw (nodo2)--(2.5,-1)--(13,-1);
            \draw (nodo1)--(0.5,0.75)--(13.5,0.75);
    \end{tikzpicture}
}
    \begin{tikzpicture}[cross/.style={path picture={ 
  \draw[black]
(path picture bounding box.south east) -- (path picture bounding box.north west) (path picture bounding box.south west) -- (path picture bounding box.north east);
}}]
\node at (4,3) () {After Hanany--Witten transition:};
            \node[circle, draw=black, minimum size=5mm] at (-2,0) (nodo0) {};
            \node[circle, draw=black, cross, minimum size=5mm] at (0,0) (nodo1) {};
            \node[circle, draw=black, cross, minimum size=5mm] at (2,0) (nodo2) {};
            \node[circle,draw=black, cross, minimum size=5mm] at (4,0) (nodo3) {};
            \node[circle,draw=black, cross, minimum size=5mm] at (6,0) (nodo4) {};
            \node[circle,draw=black, cross, minimum size=5mm] at (8,0) (nodo5) {};
            \node[circle,draw=black, cross, minimum size=5mm] at (10,0) (nodo6) {};

            \draw (-1.83,0.175)--(-0.17,0.175) node[left=3pt,above=1pt] () {\small{$11$}};
            \draw (-1.83,-0.175)--(-0.17,-0.175) node[left=3pt,below=1pt] () {\small{$11$}};
            \draw (0.17,0.175)--(1.83,0.175) node[left=3pt,above=1pt] () {\small{$11$}};
            \draw (0.17,-0.175)--(1.83,-0.175) node[left=3pt,below=1pt] () {\small{$10$}};
            \draw (nodo2)--(nodo3) node[above left =2.5pt] () {\small{$19$}};
            \draw (nodo3)--(nodo4) node[above left =2.5pt] () {\small{$16$}};
            \draw (nodo4)--(nodo5) node[above left =2.5pt] () {\small{$11$}};
            \draw (nodo5)--(nodo6) node[above left =2.5pt] () {\small{$6$}};
            \draw (-1,-2)--(-1,2) node[left=3pt] () {1};
            \draw (1,-2)--(1,2) node[left=3pt] () {1};
            \draw (3,-2)--(3,2) node[left=3pt] () {1};
            \draw (5,-2)--(5,2) node[left=3pt] () {2};
            \draw (9,-2)--(9,2) node[left=3pt] () {1};
    \end{tikzpicture}

        \begin{tikzpicture}[cross/.style={path picture={ 
  \draw[black]
(path picture bounding box.south east) -- (path picture bounding box.north west) (path picture bounding box.south west) -- (path picture bounding box.north east);
}}]
\node at (0,2) () {Corresponding to the curve configuration:};
            \node at (0,0) (nodo1) {$[1] \ 10 \stackunder{\stackon{19}{\stackon{11}{[3]}}}{[1]}  \stackunder{16}{[2]} \ 11 \ 6 \ [1]$};
    \end{tikzpicture}
    
    \caption{Type IIA brane configuration associated with the tensor branch of $D_6^{\mathfrak{su}_{22}}(O)$ where the generalized partition of $22$ is given by $O=\left[5^3, 5;4, 3^2, 1\right]$. We show the configurations before and after having performed Hanany--Witten transitions to bring the D8-branes from infinity into their respective interval. At the bottom, we depict the corresponding curve configuration.}
    \label{fig:HananyWitten}
\end{figure}

\subsection{Magnetic Quivers at Finite Coupling}\label{sec:finitecoupling}

The realization via a brane description for the electric phase of the theories $D_{N}^{\su_{2k}}\left( O \right)$ is significant because it allows us to gain insights into the properties of the Higgs branch of these theories by transitioning to the magnetic phase. This approach has been partially explored in previous works such as \cite{Kapustin:1998fa,Hanany:1999sj,Ferlito:2016grh,Sperling:2021fcf,Bourget:2022tmw}. By studying the brane configurations in the magnetic phase, we can obtain valuable information about the behavior and characteristics of the Higgs branch in the corresponding electric phase. This provides a powerful tool for understanding the dynamics and properties of these six-dimensional theories from a three-dimensional perspective.

The general procedure for transitioning from an electric brane system to a magnetic one in a Type IIA D6-D8-NS5 configuration involves two stages. In the first stage, D8-branes are pulled from infinity and used to suspend between them the available D6-branes. In the second stage, the NS5-branes located in the segment between the $\mathrm{ON}^-$-plane and the first (counting from the left) D8-brane are put at different positions on the $x^6$ axis. The result of the process is illustrated in Figure \ref{fig:magnetic_brane_configuration_D_su2k,N} for the $D^{\su_{2k}}_{N}\left( O \right)$ theory, where $O$ is the nilpotent orbit corresponding to the partition $[1^{2k}]$.

\begin{figure}[t]
    \centering
    \begin{tikzpicture}[cross/.style={path picture={ 
  \draw[black]
(path picture bounding box.south east) -- (path picture bounding box.north west) (path picture bounding box.south west) -- (path picture bounding box.north east);
}}]
          \node[circle, draw=black, minimum size=5mm] at (0,0) (nodo1) {};
            \node[circle, draw=black, cross, minimum size=5mm] at (1,1) (nodo2) {};
            \node[circle,draw=black, cross, minimum size=5mm] at (3,1) (nodo4) {};
\draw [ultra thick, decorate,
    decoration = {calligraphic brace}] (1,1.5) --  (3,1.5) node[midway,above] () {$N$};

            \draw (5,-2)--(5,2) node[midway] (d8-1) {};
            \draw (7,-2)--(7,2) node[midway] (d8-2) {};
            \draw (9,-2)--(9,2) node[midway] (d8-3) {};
            \draw (11,-2)--(11,2) node[midway] (d8-4) {};
            \node at (2,1) () {$\cdots$};
            \node at (8,0) () {$\cdots$};

            \draw (5,0)--(7,0) node[above=5pt,left] () {\small{$2k-1$}};
            \draw (9,0)--(11,0) node[above=5pt,left] () {\small{$1$}};
            \draw (nodo1)--(5,0) node[above=5pt,left] () {\small{$2k$}};       
    \end{tikzpicture}
    \caption{Magnetic brane system for the tensor branch of the $D^{\su_{2k}}_{N}\left( [1^{2k}] \right)$ SCFT.}
    \label{fig:magnetic_brane_configuration_D_su2k,N}
\end{figure} 

In the brane configuration, be it electric or magnetic, the distance in the $x^6$ direction between adjacent NS5-branes determines the gauge couplings of the corresponding gauge nodes and, consequently, the volumes of the associated curves in the F-theory description. Similarly, the $x^6$-distance between the $\mathrm{ON}^-$-plane and the leftmost NS5-brane determines the gauge coupling of one of the nodes in the fork. By separating the NS5-branes from the $\mathrm{ON}^-$-plane and the leftmost D6-brane in space, a finite coupling six-dimensional $\mathcal{N}=(1,0)$ theory is realized. This is then the magnetic description of the $(1,0)$ SQFT that exists at the generic point of the tensor branch. For the superconformal field theory (SCFT) to emerge no scales can be present, otherwise, conformality would be ruined, and thus all the curves have to be shrunk to zero volume; equivalently, all gauge couplings must be taken to infinity.
It is a special feature of the magnetic phase that this infinite-coupling limit can be taken in a controlled manner, and thus information about the SCFT itself is obtained, see Section \ref{sec:infcoupling}.

The rules to read a magnetic quiver from the associated brane system are given in \cite{Cabrera:2019izd, Cabrera:2019dob,Sperling:2021fcf}: a stack of $k$ D6-branes contributes as a $\mathfrak{u}(k)$ gauge node whereas a stack of $n$ NS5-branes is responsible for a $\mathfrak{u}(n)$ node together with an adjoint hypermultiplet, the $\mathrm{ON}^-$-plane is responsible for the projection $\mathfrak{u}(2k) \rightarrow \mathfrak{usp}(2k)$ associated with D6-branes crossing it and $\mathfrak{u}(n)+ \mathrm{adjoint} \rightarrow \mathfrak{so}(2n) + \mathrm{antisymmetric}$ associated with a stack of $k$ NS5-branes on top of the orientifold. 

The characteristic of this finite coupling magnetic quiver is the presence of a bouquet of $\mathfrak{u}(1) = \mathfrak{so}(2)$ nodes coming from the spatially separated NS5-branes. To illustrate this with an example, the magnetic quiver for the tensor branch SQFT of $D_N^{\mathfrak{su}_{2k}}$ is
\begin{equation}\label{quiver:Magnetic_Dn_fin}
\mathcal{Q}_{fin} \,\, = \qquad 
      \begin{gathered}
    \begin{tikzpicture}
      \node[node,  label=below:{\footnotesize $1$}] (A1) {};
      \node[node, label=below:{\footnotesize $2$}] (A2) [left=8mm of A1] {};
      \node[tnode] (T1) [left=8mm of A2] {$\cdots$};
      \node[node, label=below:{\footnotesize $2k-2$}] (A3) [left=8mm of T1] {};
      \node[node, label=below:{\footnotesize $2k-1$}] (A4) [left=8mm of A3] {};
    \node[node, label=below:{\footnotesize $2k$},fill=blue] (Usp) [left=8mm of A4] {};
      \node[node, label=above:{\footnotesize $1$},draw=red] (Nu1) [yshift=1.0cm, left=3mm of Usp] {};
      \node[node, label=above:{\footnotesize $1$},draw=red] (Nu2) [yshift=1.0cm, right=3mm of Usp] {};
      \node[tnode] (T2) [above=6.5mm of Usp] {\color{red}$\cdots$};
    
      \draw (A1.west) -- (A2.east);
      \draw (A2.west) -- (T1.east);
      \draw (T1.west) -- (A3.east);
      \draw (A3.west) -- (A4.east);
      \draw (A4.west) -- (Usp.east);    
      \draw (Usp) -- (Nu1);
    \draw (Usp) -- (Nu2);

  \coordinate[above=0.6cm of Nu1.north] (aux1);
  \coordinate[above=0.6cm of Nu2.north] (aux2);
                \draw [thick, decorate,
    decoration = {calligraphic brace}] (aux1) --  (aux2) node[midway,above=1mm] () {\footnotesize $N$ nodes};
   
    \end{tikzpicture} 
  \end{gathered} \,.
\end{equation}
The $\mathfrak{usp}(2k)$ node has been drawn as a filled blue circle following the conventions of \cite{Sperling:2021fcf} whilst $\mathfrak{u}(1)$s in the bouquet have been drawn as an empty red circle anticipating the fact that when the NS5-branes are put on top of the $\mathrm{ON}^-$-plane they give rise to an $\mathfrak{so}$ gauge node that, according to the aforementioned conventions, is drawn as a filled red circle.

In this way, a 3d $\mathcal{N}=4$ quiver gauge theory can be obtained for each $D_N^{\mathfrak{su}_{2k}}(O)$, where $O$ is an arbitrary generalized partition. This is a magnetic quiver for the Higgs branch of the SQFT living at the generic point of the tensor branch of the associated SCFT. In Section \ref{sec:infcoupling}, we discuss the extension required to understand the Higgs branch of the SCFT at the origin of the tensor branch; however, we can already answer questions about the operators belonging to the Higgs branch of the tensor branch SQFT from this magnetic quiver perspective.

One of the most immediate pieces of information that can be extracted from a magnetic quiver is the global symmetry, to wit, the presence of 1/2-BPS moment map operators in the Higgs branch chiral ring. Of course, this can be directly determined by computing the leading order terms in the (refined) Coulomb branch Hilbert series of the magnetic quiver. Instead, a balance algorithm was introduced in \cite{Cabrera:2019izd} which allows for a straightforward identification of the global symmetry in certain cases. In many of our cases, the balance algorithm only provides a subset of the full global symmetry \cite{Gledhill:2021cbe}, and more refined computations are necessary. We compute the global symmetry and determine the Coulomb branch Hilbert series for several explicit examples in Section \ref{sec:A3vsD3}.

\subsection{Magnetic Quivers at Infinite Coupling}\label{sec:infcoupling}

In six dimensions, the Higgs branch can be thought of as the fibers of a fibration over the tensor branch of a given theory \cite{Bourget:2022tmw}. Over the generic point of the tensor branch, the fiber is simply the Coulomb branch of the finite-coupling magnetic quiver discussed in Section \ref{sec:finitecoupling}. Moving around on the tensor branch corresponds to adjusting the various distances in the $x^6$ direction between the NS5-branes. Special subloci of the tensor branch occur when some number of these distance moduli are taken to zero; typically the fiber, and thus the Higgs branch, is modified along these subloci as some tensionful strings become tensionless. These are referred to as the phases of the theory.

In addition to the finite-coupling phase associated with the generic point of the tensor branch, two interesting phases are the ``discretely-gauged'' phase and the ``infinite-coupling'' phase. The former occurs when the distances between all of the $N$ NS5-branes are taken to zero, but the distance between this NS5-brane stack and the ON$^-$-plane is non-zero. The infinite-coupling phase occurs when all NS5-branes are atop the ON$^-$-plane; in this latter configuration all of the scales have been removed from the theory and we are at the origin of the tensor branch where the SCFT is located. 

The discretely-gauged phase is so-named due to the observation of the phenomenon of discrete gauging as described in \cite{Hanany:2018vph,Bourget:2022tmw,Hanany:2023uzn} along this subloci of the tensor branch. The magnetic quiver for the Higgs branch in this phase, $\mathcal{Q}_{dg}$, is related to $\mathcal{Q}_{fin}$ by a discrete gauging of the $\mathfrak{S}_N$ symmetry that permutes the $\mathfrak{u}(1)$ nodes of the bouquet. This replaces the bouquet with a $\mathfrak{u}(N)$ gauge node together with an adjoint hypermultiplet. At the level of the Hilbert series, the discrete-gauging action is rephrased in the following identity:
\begin{equation}
\operatorname{HS}(\mathcal{Q}_{dg}) = \operatorname{HS}(\mathcal{Q}_{fin} )/\mathfrak{S}_N\,.
\end{equation}

When tuning the $x^6$ moduli to reach the infinite-coupling phase, a new phenomenon is found that results in a $\mathbb{Z}_2$ projection of the discretely-gauged phase. As a result of the orientifold projection rules mentioned before, the $\mathfrak{u}(N)$ gauge algebra coming from the discrete gauging of the permutation symmetry on the $N$ NS5-branes, and similarly, the coupled adjoint hypermultiplet, are projected onto an $\mathfrak{so}(2N)$ gauge algebra paired with a matter hypermultiplet in the rank two antisymmetric representation. Therefore, the SCFT $D_{N}^{\su_{2k}}$ has the following magnetic quiver for the Higgs branch:
\begin{equation}\label{quiver:Magnetic_Dn_inf}
      \mathcal{Q}_{\infty}\,\,=\qquad\begin{gathered}
    \begin{tikzpicture}
      \node[node,  label=below:{\footnotesize $1$}] (A1) {};
      \node[node, label=below:{\footnotesize $2$}] (A2) [left=8mm of A1] {};
      \node[tnode] (T1) [left=8mm of A2] {$\cdots$};
      \node[node, label=below:{\footnotesize $2k-2$}] (A3) [left=8mm of T1] {};
      \node[node, label=below:{\footnotesize $2k-1$}] (A4) [left=8mm of A3] {};
    \node[node, label=below:{\footnotesize $2k$},fill=blue] (Usp) [left=8mm of A4] {};
      \node[node, label=right:{\footnotesize $2N$},fill=red] (Nu) [above =6mm of Usp] {};

      \draw (A1.west) -- (A2.east);
      \draw (A2.west) -- (T1.east);
      \draw (T1.west) -- (A3.east);
      \draw (A3.west) -- (A4.east);
      \draw (A4.west) -- (Usp.east);    
      \draw (Usp.north) -- (Nu.south);

    \draw (Nu) to[out=130, in=410, looseness=12] (Nu);   
    \end{tikzpicture} 
  \end{gathered}  \quad .
\end{equation}
The Coulomb branch Hilbert series of $\mathcal{Q}_\infty$, which counts the 1/2-BPS operators belonging to the Higgs branch chiral ring of the 6d SCFT, is related to the Hilbert series of the aforementioned discretely-gauged phase as: 
\begin{equation}
    \operatorname{HS}(\mathcal{Q}_{\infty}) = \operatorname{HS}(\mathcal{Q}_{dg}) / \mathbb{Z}_2  \,.
\end{equation}
For completeness, we now write the magnetic quiver for the Higgs branch of an arbitrary phase of the SCFT $D_{N}^{\su_{2k}}$. The $N$ NS5-branes can be arranged in stacks. Let $m_i$ denote the number of NS5-brane stacks containing $i$ branes, which are not sitting on top of the ON$^-$. We also assume that there are $m_{\hat{1}}$ NS5-branes on top of the ON$^-$. In this generic phase, the magnetic quiver for the Higgs branch takes the following form:
\begin{equation}\label{quiver:Generic_phase}
      \begin{gathered}
    \begin{tikzpicture}
      \node[node,  label=below:{\footnotesize $1$}] (A1) {};
      \node[node, label=below:{\footnotesize $2$}] (A2) [left=8mm of A1] {};
      \node[tnode] (T1) [left=8mm of A2] {$\cdots$};
      \node[node, label=below:{\footnotesize $2k-2$}] (A3) [left=8mm of T1] {};
      \node[node, label=below:{\footnotesize $2k-1$}] (A4) [left=8mm of A3] {};
    \node[node, label=below:{\footnotesize $2k$},fill=blue] (Usp) [left=8mm of A4] {};
      \node[node, label=left:{\footnotesize $2m_{\hat{1}}$},draw=red,fill=red] (Nu) [yshift=1.0cm, left=20mm of Usp] {};
    
      \node[node, label=left:{\footnotesize $n$},draw=red] (Nu1n) [yshift=1.0cm, left=10mm of Usp] {};
      \node[node, label=right:{\footnotesize $n$},draw=red] (Nu2n) [yshift=1.0cm, left=-1mm of Usp] {};
      \node[tnode] (T2n) [yshift=1.0cm, left=1mm of Usp] {\color{red} \footnotesize $\cdots$};

            \node[node, label=above:{\footnotesize $1$},draw=red] (Nu11) [yshift=1.0cm, right=12mm of Usp] {};
      \node[node, label=above:{\footnotesize $1$},draw=red] (Nu21) [yshift=1.0cm, right=24mm of Usp] {};
      \node[tnode] (T21) [yshift=1.0cm, right=15mm of Usp] {\color{red}\footnotesize $\cdots$};

      \node[tnode] (T21) [yshift=1.0cm, right=3mm of Usp] {\footnotesize $\cdots$};
    
      \draw (A1.west) -- (A2.east);
      \draw (A2.west) -- (T1.east);
      \draw (T1.west) -- (A3.east);
      \draw (A3.west) -- (A4.east);
      \draw (A4.west) -- (Usp.east);    
      \draw (Usp) -- (Nu1n);
    \draw (Usp) -- (Nu2n);
          \draw (Usp) -- (Nu11);
    \draw (Usp) -- (Nu21);
    \draw (Usp) -- (Nu);

    \draw (Nu1n) to[out=130, in=410, looseness=12] (Nu1n); 
    \draw (Nu2n) to[out=130, in=410, looseness=12] (Nu2n);
    \draw (Nu) to[out=130, in=410, looseness=12] (Nu);

  \coordinate[above=5mm of Nu1n.north] (aux1n);
  \coordinate[above=5mm of Nu2n.north] (aux2n);
                \draw [thick, decorate,
    decoration = {calligraphic brace}] (aux1n) --  (aux2n) node[midway,above] () {\footnotesize$m_n$};
      \coordinate[above=0.5cm of Nu11.north] (aux11);
  \coordinate[above=0.5cm of Nu21.north] (aux21);
                \draw [thick, decorate,
    decoration = {calligraphic brace}] (aux11) --  (aux21) node[midway,above] () {\footnotesize$m_1$};
   
    \end{tikzpicture} 
  \end{gathered} \quad .
\end{equation}
While we have only discussed the $D_{N}^{\su_{2k}}$ theories explicitly in this section, the same considerations about phases, discrete gauging, and $\mathbb{Z}_2$ quotients apply for every theory $D_{N}^{\su_{2k}}\left( O \right)$ where $O$ is a generalized partition. In particular, we can determine the infinite coupling magnetic quiver in each case, which provides the Higgs branch for each SCFT $D_{N}^{\su_{2k}}\left( O \right)$.

\section{Exploiting the \texorpdfstring{\boldmath{$A_3$}}{A3} \texorpdfstring{\boldmath{$\cong$}}{cong} \texorpdfstring{\boldmath{$D_3$}}{D3} Isomorphism}\label{sec:A3vsD3}

In the previous section, we determined the magnetic quiver for the Higgs branch of the $D_{N}^{\mathfrak{su}_{2k}}(O)$ SCFTs from the brane engineering of the theory in Type IIA string theory. In fact, we have determined the magnetic quiver for the Higgs branch emanating from each sublocus of the tensor branch, not just for the SCFT at the origin. Generically, this is a unitary-orthosymplectic quiver; however, when $N=3$ we can use the isomorphism between $\mathfrak{so}(6)$ and $\mathfrak{su}(4)$ to write a dual unitary quiver. In this section, we briefly use this isomorphism to provide a cross-check on the unitary-orthosymplectic quivers that we have derived.

At the generic point of the tensor branch, the SCFT $D_3^{\mathfrak{su}_{2k}}$ is described via the curve configuration:
\begin{equation}
    \overset{\mathfrak{su}_k}{2}\overset{\displaystyle \overset{\mathfrak{su}_{k}}{2}}{ \overset{\mathfrak{su}_{2k}}{2}} \, [\mathfrak{su}_{2k}] \,.
\end{equation}
Rearranging, we can write this curve configuration as
\begin{equation}
    \overset{\mathfrak{su}_k}{2}\underset{[\mathfrak{su}_{2k}]}{ \overset{\mathfrak{su}_{2k}}{2}}\overset{\mathfrak{su}_k}{2} \,,
\end{equation}
which is nothing other than the description at the generic point of the tensor branch of rank four $(\mathfrak{su}(2k), \mathfrak{su}(2k))$ conformal matter, Higgsed on both the left and the right by the nilpotent orbit associated with the partition $[2^k]$. We refer to this theory as $A_{3}^{\mathfrak{su}_{2k}}([2^k], [2^k])$. The Type IIA brane engineering of this latter theory, and thus the magnetic quiver for the Higgs branch, was studied in \cite{Hanany:2018vph}, where it arises through D6-NS5-D8-branes, however without the presence of any ON$^-$-plane.

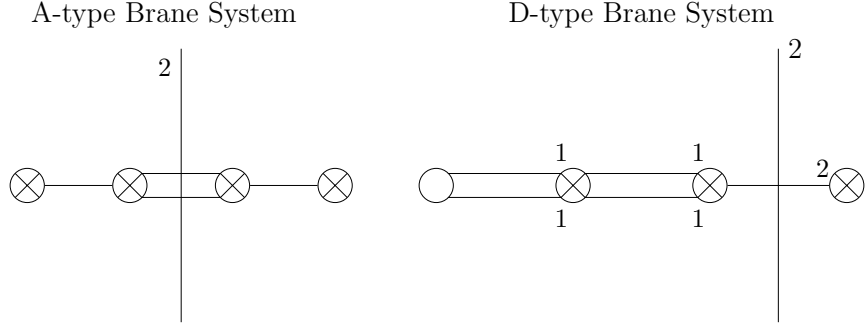
\begin{figure}[t]
    \centering
    \resizebox{0.7\textwidth}{!}{
 \begin{tikzpicture}[cross/.style={path picture={ 
  \draw[black]
(path picture bounding box.south east) -- (path picture bounding box.north west) (path picture bounding box.south west) -- (path picture bounding box.north east);
}}]
\node at (2,2.5) () {A-type Brane System};
            \node[circle, draw=black, cross, minimum size=5mm] at (0,0) (nodo1) {};
            \node[circle,draw=black, cross, minimum size=5mm] at (1.5,0) (nodo2) {};
            \node[circle, draw=black, cross, minimum size=5mm] at (3,0) (nodo3) {};
            \node[circle,draw=black, cross, minimum size=5mm] at (4.5,0) (nodo4) {};
            
            \draw (2.25,-2)--(2.25,2) node[anchor=north east] (d8-2) {\small{ $2$}};
            \draw (nodo1)--(nodo2);
            \draw (1.67,0.175)--(2.83,0.175) ;
            \draw (1.67,-0.175)--(2.83,-0.175);
            \draw (nodo3)--(nodo4);

    \end{tikzpicture} \qquad
      \begin{tikzpicture}[cross/.style={path picture={ 
  \draw[black]
(path picture bounding box.south east) -- (path picture bounding box.north west) (path picture bounding box.south west) -- (path picture bounding box.north east);
}}]
\node at (1,2.5) () {D-type Brane System};
            \node[circle, draw=black, minimum size=5mm] at (-2,0) (nodo0) {};
            \node[circle, draw=black, cross, minimum size=5mm] at (0,0) (nodo1) {};
            \node[circle, draw=black, cross, minimum size=5mm] at (2,0) (nodo2) {};
            \node[circle,draw=black, cross, minimum size=5mm] at (4,0) (nodo3) {};

            \draw[] (3,-2)--(3,2) node[right] () {$2$};
            \draw (-1.83,0.175)--(-0.17,0.175) node[left=3pt,above=1pt] () {\small{$1$}};
            \draw (-1.83,-0.175)--(-0.17,-0.175) node[left=3pt,below=1pt] () {\small{$1$}};
            \draw (0.17,0.175)--(1.83,0.175) node[left=3pt,above=1pt] () {\small{$1$}};
            \draw (0.17,-0.175)--(1.83,-0.175) node[left=3pt,below=1pt] () {\small{$1$}};
            
            \draw (nodo2)--(nodo3);
            \node at (3.65,0.25) () {\small{$2$}};

    \end{tikzpicture}   
    }
    \caption{Two different Type IIA brane systems engineering the $2\overset{\mathfrak{su}_2}{2}2$ theory at the generic point of the tensor branch.}
    \label{fig:A3vsD3branesystem}
\end{figure}

We consider only the case of $k=1$; this is sufficient to demonstrate the subtleties of the various phases,\footnote{For the generic case the phase structure is identical.} and is particularly interesting as the structure of the non-Abelian flavor symmetry is special \cite{Baume:2021qho,Heckman:2015bfa,Morrison:2016djb,Ohmori:2015pia}. That is, we wish to compare the magnetic quivers of
\begin{equation}
    2\,\underset{[\mathfrak{su}_3]}{\overset{\mathfrak{su}_2}{2}}\,2 \qquad \text{ and } \qquad 2\,\overset{\displaystyle 2}{\overset{\mathfrak{su}_2}{2}}\,[\mathfrak{su}_3] \,,
\end{equation}
where we can already see that the flavor symmetry is enhanced from the naive $\mathfrak{su}(2)$ to an $\mathfrak{su}(3)$. The two different Type IIA brane systems engineering these theories, at the generic point of the tensor branch, are depicted in Figure \ref{fig:A3vsD3branesystem}. Using the $A_3$-type brane description, we have determined the magnetic quivers for each phase, and furthermore, we have determined the Coulomb branch Hilbert series using the standard methods \cite{Cremonesi:2013lqa}. These quivers and their Coulomb branch Hilbert series appeared already in previous studies \cite{Hanany:2018vph}.

We can then perform the same analysis for the phases of the $D_3$-type brane description. A priori, there appear to be seven phases, as opposed to five, however, there is a redundancy in the $D_3$-type brane description -- if only a single NS5-brane is placed on top of the ON$^-$-plane, then the resulting quiver is the same as if that NS5-brane was placed elsewhere. Removing this redundancy, we find the five phases listed in Table \ref{tab:D3_results}. Again, we can compute the Coulomb symmetry of the resulting magnetic quivers and their Coulomb branch Hilbert series; these quantities precisely match those from the unitary magnetic quivers listed in Table \ref{tab:A3_results}.

\begin{landscape}
\pagestyle{empty}
\begin{table}[p]
    \centering
\resizebox{1.3\textwidth}{!}{
    \begin{threeparttable}
        \begin{tabular}{c c c c}\toprule
         Brane System & Magnetic Quiver & Global Symmetry & Coulomb Branch Hilbert Series \\ \midrule
           $\begin{gathered}
             \begin{tikzpicture}[cross/.style={path picture={ 
  \draw[black]
(path picture bounding box.south east) -- (path picture bounding box.north west) (path picture bounding box.south west) -- (path picture bounding box.north east);
}}]
            \node[circle, draw=black, minimum size=5mm] at (-2,0) (nodo0) {};
            \node[circle, draw=black, cross, minimum size=5mm] at (-1.5,1) (nodo1) {};
            \node[circle, draw=black, cross, minimum size=5mm] at (-0.9,1) (nodo2) {};
            \node[circle,draw=black, cross, minimum size=5mm] at (-0.3,1) (nodo3) {};

            \draw (0.2,-2)--(0.2,2);
            \draw (2,-2)--(2,2);
            \draw (nodo0)--(0,0);
            \draw (0,0)--(2,0);
            \node at (-0.15,0.25) () {\small{2}};

    \end{tikzpicture}
     \begin{tikzpicture}[cross/.style={path picture={ 
  \draw[black]
(path picture bounding box.south east) -- (path picture bounding box.north west) (path picture bounding box.south west) -- (path picture bounding box.north east);
}}] \node at (-2.6,0) () {$\cong$};
            \node[circle, draw=black, minimum size=5mm] at (-2,0) (nodo0) {};
            \node[circle, draw=black, cross, minimum size=5mm] at (-2,1) (nodo1) {};
            \node[circle, draw=black, cross, minimum size=5mm] at (-1.34,1) (nodo2) {};
            \node[circle,draw=black, cross, minimum size=5mm] at (-0.66,1) (nodo3) {};

            \draw (0.2,-2)--(0.2,2);
            \draw (2,-2)--(2,2);
            \draw (nodo0)--(0,0);
            \draw (0,0)--(2,0);
            \node at (-0.15,0.25) () {\small{2}};

    \end{tikzpicture} 
         \end{gathered}$  &    $\begin{gathered}
    \begin{tikzpicture}
      \node[node, minimum size=5mm, label=below:{\footnotesize $1$}] (A1) {};
      \node[node, minimum size=5mm,label=below:{\footnotesize $2$},fill=blue] (A2) [left=8mm of A1] {};
      \node[node, minimum size=5mm,label=above:{\footnotesize $1$},draw=red] (Nu) [above=8mm of A2] {};
      \node[node, minimum size=5mm,label=above:{\footnotesize $1$},draw=red] (Nun) [yshift=13.5mm,left=4mm of A2] {};
        \node[node, minimum size=5mm,label=above:{\footnotesize $1$},draw=red] (Nunn) [yshift=13.5mm,right=4mm of A2] {};
    
      \draw (A1.west) -- (A2.east);
      \draw (A2.north) -- (Nu.south);
      \draw (A2) -- (Nun);
        \draw (A2) -- (Nunn);
    \end{tikzpicture}
   \begin{tikzpicture} 
      \node[node, minimum size=5mm, label=below:{\footnotesize $1$}] (A1) {};
      \node[node, minimum size=5mm,label=below:{\footnotesize $2$},fill=blue] (A2) [left=8mm of A1] {};
      \node[node, minimum size=5mm,label=above:{\footnotesize $1$},draw=red] (Nu) [above=8mm of A2] {};
      \node[node, minimum size=5mm,label=above:{\footnotesize $2$},fill=red] (Nun) [yshift=13.5mm,left=4mm of A2] {};
        \node[node, minimum size=5mm,label=above:{\footnotesize $1$},draw=red] (Nunn) [yshift=13.5mm,right=4mm of A2] {};

\node[tnode] (T21) [yshift=1.0cm, left=14mm of A2] {$\cong$};
    
      \draw (A1.west) -- (A2.east);
      \draw (A2.north) -- (Nu.south);
      \draw (A2) -- (Nun);
        \draw (A2) -- (Nunn);
    \end{tikzpicture} \end{gathered}$ & $\mathfrak{so}(8)$ &  \Large $\frac{\left(t^2+1\right) \left(t^8+17 t^6+48 t^4+17 t^2+1\right)}{\left(1-t^2\right)^{10}}$\\ 
                $\begin{gathered}
                 
             \begin{tikzpicture}[cross/.style={path picture={ 
  \draw[black]
(path picture bounding box.south east) -- (path picture bounding box.north west) (path picture bounding box.south west) -- (path picture bounding box.north east);
}}]
            \node[circle, draw=black, minimum size=5mm] at (-2,0) (nodo0) {};
            \node[circle, draw=black, cross, minimum size=5mm] at (-1.34,1) (nodo1) {};
            \node[circle, draw=black, cross, minimum size=5mm] at (-1.34,1.5) (nodo2) {};
            \node[circle,draw=black, cross, minimum size=5mm] at (-0.66,1) (nodo3) {};

            \draw (0.2,-2)--(0.2,2);
            \draw (2,-2)--(2,2);
            \draw (nodo0)--(0,0);
            \draw (0,0)--(2,0);
            \node at (-0.15,0.25) () {\small{2}};

    \end{tikzpicture}     \begin{tikzpicture}[cross/.style={path picture={ 
  \draw[black]
(path picture bounding box.south east) -- (path picture bounding box.north west) (path picture bounding box.south west) -- (path picture bounding box.north east);
}}] \node at (-2.6,0) () {$\cong$};
            \node[circle, draw=black, minimum size=5mm] at (-2,0) (nodo0) {};
            \node[circle, draw=black, cross, minimum size=5mm] at (-2,1) (nodo1) {};
            \node[circle, draw=black, cross, minimum size=5mm] at (-1,1) (nodo2) {};
            \node[circle,draw=black, cross, minimum size=5mm] at (-1,1.5) (nodo3) {};

            \draw (0.2,-2)--(0.2,2);
            \draw (2,-2)--(2,2);
            \draw (nodo0)--(0,0);
            \draw (0,0)--(2,0);
            \node at (-0.15,0.25) () {\small{2}};

    \end{tikzpicture} \end{gathered}$ &     $\begin{gathered}
    \begin{tikzpicture}
      \node[node, minimum size=5mm, label=below:{\footnotesize $1$}] (A1) {};
      \node[node, minimum size=5mm,label=below:{\footnotesize $2$},fill=blue] (A2) [left=8mm of A1] {};
      \node[node, minimum size=5mm,label=right:{\footnotesize $2$},draw=red] (Nu) [above=8mm of A2] {};
      \node[node, minimum size=5mm,label=left:{\footnotesize $1$},draw=red] (Nun) [yshift=13.5mm,left=4mm of A2] {};
    
      \draw (A1.west) -- (A2.east);
      \draw (A2.north) -- (Nu.south);
      \draw (A2) -- (Nun);
    \draw (Nu) to[out=130, in=410, looseness=12] (Nu); 

    \end{tikzpicture}
   \begin{tikzpicture} 
        \node[node, minimum size=5mm, label=below:{\footnotesize $1$}] (A1) {};
      \node[node, minimum size=5mm,label=below:{\footnotesize $2$},fill=blue] (A2) [left=8mm of A1] {};
      \node[node, minimum size=5mm,label=right:{\footnotesize $2$},draw=red] (Nu) [above=8mm of A2] {};
      \node[node, minimum size=5mm,label=left:{\footnotesize $2$},fill=red] (Nun) [yshift=13.5mm,left=4mm of A2] {};

      \node[tnode] (T21) [yshift=1.0cm, left=14mm of A2] {$\cong$};
    
      \draw (A1.west) -- (A2.east);
      \draw (A2.north) -- (Nu.south);
      \draw (A2) -- (Nun);
    \draw (Nu) to[out=130, in=410, looseness=12] (Nu); 
    \end{tikzpicture} \end{gathered}$   & $\mathfrak{so}(7)$ & \Large  $\frac{\left(t^2+1\right) \left(t^8+10 t^6+20 t^4+10 t^2+1\right)}{\left(1-t^2\right)^{10}}$\\ 

            $\begin{gathered}
                   \begin{tikzpicture}[cross/.style={path picture={ 
  \draw[black]
(path picture bounding box.south east) -- (path picture bounding box.north west) (path picture bounding box.south west) -- (path picture bounding box.north east);
}}] 
            \node[circle, draw=black, minimum size=5mm] at (-2,0) (nodo0) {};
            \node[circle, draw=black, cross, minimum size=5mm] at (-2,1) (nodo1) {};
            \node[circle, draw=black, cross, minimum size=5mm] at (-1,1) (nodo2) {};
            \node[circle,draw=black, cross, minimum size=5mm] at (-2,1.5) (nodo3) {};

            \draw (0.2,-2)--(0.2,2);
            \draw (2,-2)--(2,2);
            \draw (nodo0)--(0,0);
            \draw (0,0)--(2,0);
            \node at (-0.15,0.25) () {\small{2}};

    \end{tikzpicture} 
             \end{gathered}$ &  $\begin{gathered}
    \begin{tikzpicture}
      \node[node, minimum size=5mm, label=below:{\footnotesize $1$}] (A1) {};
      \node[node, minimum size=5mm,label=below:{\footnotesize $2$},fill=blue] (A2) [left=8mm of A1] {};
      \node[node, minimum size=5mm,label=right:{\footnotesize $4$},fill=red] (Nu) [above=8mm of A2] {};
      \node[node, minimum size=5mm,label=left:{\footnotesize $1$},draw=red] (Nun) [yshift=13.5mm,left=4mm of A2] {};
    
      \draw (A1.west) -- (A2.east);
      \draw (A2.north) -- (Nu.south);
      \draw (A2) -- (Nun);
    \draw (Nu) to[out=130, in=410, looseness=12] (Nu); 

    \end{tikzpicture}
  \end{gathered}$ & $\mathfrak{su}(4)$ & \Large  $\frac{t^{20}+10 t^{18}+55 t^{16}+150 t^{14}+288 t^{12}+336 t^{10}+288 t^8+150 t^6+55 t^4+10 t^2+1}{\left(1-t^2\right)^{10} \left(t^2+1\right)^5}$ \\ 

     $\begin{gathered}
                       \begin{tikzpicture}[cross/.style={path picture={ 
  \draw[black]
(path picture bounding box.south east) -- (path picture bounding box.north west) (path picture bounding box.south west) -- (path picture bounding box.north east);
}}] 
            \node[circle, draw=black, minimum size=5mm] at (-2,0) (nodo0) {};
            \node[circle, draw=black, cross, minimum size=5mm] at (-1,0.5) (nodo1) {};
            \node[circle, draw=black, cross, minimum size=5mm] at (-1,1) (nodo2) {};
            \node[circle,draw=black, cross, minimum size=5mm] at (-1,1.5) (nodo3) {};

            \draw (0.2,-2)--(0.2,2);
            \draw (2,-2)--(2,2);
            \draw (nodo0)--(0,0);
            \draw (0,0)--(2,0);
            \node at (-0.15,0.25) () {\small{2}};

    \end{tikzpicture} 
                 \end{gathered}$ &  $\begin{gathered}
    \begin{tikzpicture}
      \node[node, minimum size=5mm, label=below:{\footnotesize $1$}] (A1) {};
      \node[node, minimum size=5mm,label=below:{\footnotesize $2$},fill=blue] (A2) [left=8mm of A1] {};
      \node[node, minimum size=5mm,label=left:{\footnotesize $3$},draw=red] (Nu) [above=8mm of A2] {};
    
      \draw (A1.west) -- (A2.east);
      \draw (A2.north) -- (Nu.south);
    \draw (Nu) to[out=130, in=410, looseness=12] (Nu); 

    \end{tikzpicture}
  \end{gathered}$ & $\mathfrak{g}_2$ & \Large  $\frac{\left(t^2+1\right) \left(t^8+3 t^6+6 t^4+3 t^2+1\right)}{\left(1-t^2\right)^{10}}$ \\ 

     $\begin{gathered}
                     \begin{tikzpicture}[cross/.style={path picture={ 
  \draw[black]
(path picture bounding box.south east) -- (path picture bounding box.north west) (path picture bounding box.south west) -- (path picture bounding box.north east);
}}] 
            \node[circle, draw=black, minimum size=5mm] at (-2,0) (nodo0) {};
            \node[circle, draw=black, cross, minimum size=5mm] at (-2,0.5) (nodo1) {};
            \node[circle, draw=black, cross, minimum size=5mm] at (-2,1) (nodo2) {};
            \node[circle,draw=black, cross, minimum size=5mm] at (-2,1.5) (nodo3) {};

            \draw (0.2,-2)--(0.2,2);
            \draw (2,-2)--(2,2);
            \draw (nodo0)--(0,0);
            \draw (0,0)--(2,0);
            \node at (-0.15,0.25) () {\small{2}};

    \end{tikzpicture} \end{gathered}$  &  $\begin{gathered}
    \begin{tikzpicture}
      \node[node, minimum size=5mm, label=below:{\footnotesize $1$}] (A1) {};
      \node[node, minimum size=5mm,label=below:{\footnotesize $2$},fill=blue] (A2) [left=8mm of A1] {};
      \node[node, minimum size=5mm,label=left:{\footnotesize $6$},fill=red] (Nu) [above=8mm of A2] {};
    
      \draw (A1.west) -- (A2.east);
      \draw (A2.north) -- (Nu.south);
    \draw (Nu) to[out=130, in=410, looseness=12] (Nu); 

    \end{tikzpicture}
  \end{gathered}$ & $\mathfrak{su}(3)$ &  \Large $\frac{t^{20}+3 t^{18}+13 t^{16}+25 t^{14}+46 t^{12}+48 t^{10}+46 t^8+25 t^6+13 t^4+3 t^2+1}{\left(1-t^2\right)^5 \left(1-t^4\right)^5}$ \\ \bottomrule

    \end{tabular}
    \end{threeparttable}
}
    \caption{The D-type construction of the phases of the SCFT associated with the tensor branch configuration $2\overset{\mathfrak{su}_2}{2}2$, arising from D6-NS5-D8-branes together with the ON$^-$-plane in Type IIA. We list the magnetic quiver for each phase and determine the Coulomb symmetry and the Coulomb branch Hilbert series.}
    \label{tab:D3_results}
\end{table}

\begin{table}[p]
    \centering
    \resizebox{1.1\textwidth}{!}{
    \begin{threeparttable}
        \begin{tabular}{c c c c}\toprule
         Brane System & Magnetic Quiver & Global Symmetry & Coulomb Branch Hilbert Series \\ \midrule
         & & & \\
         $\begin{gathered}
             \begin{tikzpicture}[cross/.style={path picture={ 
  \draw[black]
(path picture bounding box.south east) -- (path picture bounding box.north west) (path picture bounding box.south west) -- (path picture bounding box.north east);
}}] 
            \node[circle, draw=black, cross, minimum size=5mm] at (2.85,1) (nodo1) {};
            \node[circle,draw=black, cross, minimum size=5mm] at (3.45,1) (nodo2) {};
            \node[circle, draw=black, cross, minimum size=5mm] at (4.05,1) (nodo3) {};
            \node[circle,draw=black, cross, minimum size=5mm] at (4.65,1) (nodo4) {};

            \draw (2.5,-1)--(2.5,2) node[midway] (d8-2) {};
            \draw (5,-1)--(5,2) node[midway] (d8-3) {};

            \draw (2.5,0.2)--(5,0.2);
            \draw (2.5,-0.2)--(5,-0.2);

    \end{tikzpicture} \end{gathered}$ & $    \begin{gathered}
    \begin{tikzpicture}
      \node[node, minimum size=5mm, label=below:{\footnotesize $2$}] (A1) {};
      \node[node, minimum size=5mm,label=below:{\footnotesize $1$}] (A2) [below left=8mm of A1] {};
      \node[node, minimum size=5mm,label=below:{\footnotesize $1$}] (A3) [below right=8mm of A1] {};
      \node[node, minimum size=5mm,label=below:{\footnotesize $1$}] (A4) [above left=8mm of A1] {};
      \node[node, minimum size=5mm,label=below:{\footnotesize $1$}] (A5) [above right=8mm of A1] {};
    
      \draw (A1) -- (A2);
      \draw (A1) -- (A3);
      \draw (A1) -- (A4);
      \draw (A1) -- (A5);

    \end{tikzpicture}
  \end{gathered}$ & $\mathfrak{so}(8)$ & {\Large $\frac{\left(t^2+1\right) \left(t^8+17 t^6+48 t^4+17 t^2+1\right)}{\left(1-t^2\right)^{10}}$ } \\ 
             $\begin{gathered}
                 
            \begin{tikzpicture}[cross/.style={path picture={ 
  \draw[black]
(path picture bounding box.south east) -- (path picture bounding box.north west) (path picture bounding box.south west) -- (path picture bounding box.north east);
}}] 
            \node[circle, draw=black, cross, minimum size=5mm] at (3.125,1) (nodo1) {};
            \node[circle,draw=black, cross, minimum size=5mm] at (3.75,1) (nodo2) {};
            \node[circle, draw=black, cross, minimum size=5mm] at (3.75,1.5) (nodo3) {};
            \node[circle,draw=black, cross, minimum size=5mm] at (4.375,1) (nodo4) {};

            \draw (2.5,-1)--(2.5,2) node[midway] (d8-2) {};
            \draw (5,-1)--(5,2) node[midway] (d8-3) {};

            \draw (2.5,0.2)--(5,0.2);
            \draw (2.5,-0.2)--(5,-0.2);

    \end{tikzpicture}  \end{gathered}$ &   $\begin{gathered}
    \begin{tikzpicture}
      \node[node, minimum size=5mm, label=below:{\footnotesize $1$}] (A1) {};
      \node[node, minimum size=5mm,label=below:{\footnotesize $2$}] (A2) [left=8mm of A1] {};
      \node[node, minimum size=5mm,label=below:{\footnotesize $1$}] (A3) [left=8mm of A2] {};
      \node[node, minimum size=5mm,label=left:{\footnotesize $2$}] (Nu) [above=8mm of A2] {};
    
      \draw (A1.west) -- (A2.east);
      \draw (A2.north) -- (Nu.south);
      \draw (A2.west) -- (A3.east);
    \draw (Nu) to[out=130, in=410, looseness=12] (Nu); 

    \end{tikzpicture}
  \end{gathered}$& $\mathfrak{so}(7)$ & {\Large $\frac{\left(t^2+1\right) \left(t^8+10 t^6+20 t^4+10 t^2+1\right)}{\left(1-t^2\right)^{10}}$ }\\ 
             $\begin{gathered}
             \begin{tikzpicture}[cross/.style={path picture={ 
  \draw[black]
(path picture bounding box.south east) -- (path picture bounding box.north west) (path picture bounding box.south west) -- (path picture bounding box.north east);
}}] 
            \node[circle, draw=black, cross, minimum size=5mm] at (3.33,1) (nodo1) {};
            \node[circle,draw=black, cross, minimum size=5mm] at (3.33,1.5) (nodo2) {};
            \node[circle, draw=black, cross, minimum size=5mm] at (4.16,1) (nodo3) {};
            \node[circle,draw=black, cross, minimum size=5mm] at (4.16,1.5) (nodo4) {};

            \draw (2.5,-1)--(2.5,2) node[midway] (d8-2) {};
            \draw (5,-1)--(5,2) node[midway] (d8-3) {};

            \draw (2.5,0.2)--(5,0.2);
            \draw (2.5,-0.2)--(5,-0.2);

    \end{tikzpicture} \end{gathered}$&    $ \begin{gathered}
    \begin{tikzpicture}
      \node[node, minimum size=5mm, label=below:{\footnotesize $2$}] (A1) {};
      \node[node, minimum size=5mm,label=below:{\footnotesize $2$}] (A2) [left=8mm of A1] {};
      \node[node, minimum size=5mm,label=below:{\footnotesize $2$}] (A3) [left=8mm of A2] {};
    
      \draw (A1.west) -- (A2.east);
    \draw (A2.west) -- (A3.east);
    \draw (A1) to[out=130, in=410, looseness=12] (A1);
    \draw (A3) to[out=130, in=410, looseness=12] (A3);

    \end{tikzpicture}
  \end{gathered}$& $\mathfrak{su}(4)$ & {\Large $\frac{t^{20}+10 t^{18}+55 t^{16}+150 t^{14}+288 t^{12}+336 t^{10}+288 t^8+150 t^6+55 t^4+10 t^2+1}{\left(1-t^2\right)^{10} \left(t^2+1\right)^5}$ }\\ 
             $\begin{gathered}
                 \begin{tikzpicture}[cross/.style={path picture={ 
  \draw[black]
(path picture bounding box.south east) -- (path picture bounding box.north west) (path picture bounding box.south west) -- (path picture bounding box.north east);
}}] 
            \node[circle, draw=black, cross, minimum size=5mm] at (3.33,1) (nodo1) {};
            \node[circle,draw=black, cross, minimum size=5mm] at (3.33,1.5) (nodo2) {};
            \node[circle, draw=black, cross, minimum size=5mm] at (4.16,1) (nodo3) {};
            \node[circle,draw=black, cross, minimum size=5mm] at (3.33,0.5) (nodo4) {};

            \draw (2.5,-1)--(2.5,2) node[midway] (d8-2) {};
            \draw (5,-1)--(5,2) node[midway] (d8-3) {};

            \draw (2.5,-0.3)--(5,-0.3);
            \draw (2.5,-0.7)--(5,-0.7);

    \end{tikzpicture}
    \end{gathered}$&     $      \begin{gathered}
    \begin{tikzpicture}
      \node[node, minimum size=5mm, label=below:{\footnotesize $1$}] (A1) {};
      \node[node, minimum size=5mm,label=below:{\footnotesize $2$}] (A2) [left=8mm of A1] {};
      \node[node, minimum size=5mm,label=left:{\footnotesize $3$}] (Nu) [above=8mm of A2] {};
    
      \draw (A1.west) -- (A2.east);
      \draw (A2.north) -- (Nu.south);
    \draw (Nu) to[out=130, in=410, looseness=12] (Nu);

    \end{tikzpicture}
  \end{gathered}$ & $\mathfrak{g}_2$ & {\Large $\frac{\left(t^2+1\right) \left(t^8+3 t^6+6 t^4+3 t^2+1\right)}{\left(1-t^2\right)^{10}}$}\\ 
             $\begin{gathered}
                 
             \begin{tikzpicture}[cross/.style={path picture={ 
  \draw[black]
(path picture bounding box.south east) -- (path picture bounding box.north west) (path picture bounding box.south west) -- (path picture bounding box.north east);
}}] 
            \node[circle, draw=black, cross, minimum size=5mm] at (3.75,1) (nodo1) {};
            \node[circle,draw=black, cross, minimum size=5mm] at (3.75,1.5) (nodo2) {};
            \node[circle, draw=black, cross, minimum size=5mm] at (3.75,0) (nodo3) {};
            \node[circle,draw=black, cross, minimum size=5mm] at (3.75,0.5) (nodo4) {};

            \draw (2.5,-1.5)--(2.5,2) node[midway] (d8-2) {};
            \draw (5,-1.5)--(5,2) node[midway] (d8-3) {};

            \draw (2.5,-0.8)--(5,-0.8);
            \draw (2.5,-1.2)--(5,-1.2);

    \end{tikzpicture} \end{gathered}$&   $\begin{gathered}
    \begin{tikzpicture}
      \node[node, minimum size=5mm,label=below:{\footnotesize $2$}] (A2) [left=8mm of A1] {};
      \node[node, minimum size=5mm,label=left:{\footnotesize $4$}] (Nu) [above=8mm of A2] {};
    
      \draw (A2.north) -- (Nu.south);
    \draw (Nu) to[out=130, in=410, looseness=12] (Nu); 

    \end{tikzpicture}
  \end{gathered}$& $\mathfrak{su}(3)$ & {\Large $\frac{t^{20}+3 t^{18}+13 t^{16}+25 t^{14}+46 t^{12}+48 t^{10}+46 t^8+25 t^6+13 t^4+3 t^2+1}{\left(1-t^2\right)^5 \left(1-t^4\right)^5}$}\\ \bottomrule
    \end{tabular}
        \end{threeparttable}
    }
    \caption{The Type IIA brane description, the magnetic quiver, and the Coulomb branch Hilbert series together with global symmetry for the various phases in the A-type description of the theories associated with the tensor branch configuration $2\overset{\mathfrak{su}_2}{2}2$ \cite{Hanany:2018vph}. The last entry is the magnetic quiver for the Higgs branch of the 6d SCFT.}
    \label{tab:A3_results}
\end{table}
\end{landscape}

\section{Higgs Branch RG Flow for the \texorpdfstring{\boldmath{$D_N^{\mathfrak{su}_{2k}}(O)$}}{DNsu2k(O)} SCFTs}\label{sec:Higgsbranch}

The Coulomb branches of the magnetic quivers that we have just determined are isomorphic to the Higgs branches of the 6d $(1,0)$ SCFTs that we are interested in. However, we have not yet answered any of the questions posed in Section \ref{sec:intro} about the Higgs branch, except in the abstract. We would now like to explore how to extract some of these properties from the magnetic quivers. For example, to understand the operators/generators/relations of the Higgs branch chiral ring, we can use the monopole formula to compute the Coulomb branch Hilbert series of the magnetic quiver. In this section, we utilize the magnetic quiver to enumerate the interacting SCFT fixed points that arise on different subloci of the Higgs branch and provide some understanding of the operators which trigger the RG flow between a pair of theories on nested subloci. 

\subsection{Quiver Subtraction for Unitary-orthosymplectic Quivers}\label{sec:Quiver_Subtraction}

The quiver subtraction algorithm, first formulated in \cite{Cabrera:2018ann} and then extended in \cite{Bourget:2022ehw} to also include instanton moduli spaces, provides a means to read the transverse slice between two 3d $\mathcal{N}=4$ unitary gauge theories related by a Coulomb branch RG flow. That is, given a unitary quiver theory $\mathcal{Q}$ for which we can turn on some Coulomb branch moduli such that under an RG flow it reaches a quiver theory $\mathcal{Q}'$, the subtraction $\mathcal{Q}-\mathcal{Q}'$ outputs another unitary quiver $\mathcal{S}$, whose Coulomb branch moduli space gives exactly the transverse slice that connects the two theories; in a more physical language, $\mathcal{S}$ gives information about the moduli we had to tune to trigger the flow from $\mathcal{Q}$ to $\mathcal{Q}'$.

The algorithm stemmed from observations made in the context of brane dynamics in \cite{Cabrera:2016vvv}. There, borrowing the definition of the theories $\mathcal{Q}$ and $\mathcal{Q}'$ of the last paragraph, and observing the Type IIB D3-D5-NS5-brane system realisation of $\mathcal{Q}$ and $\mathcal{Q}'$, assuming that is exists, it was possible to associate with the Higgsing process and the corresponding transverse slice, a  ``move'' in the brane system. In a subsequent work \cite{Cabrera:2017njm}, the same authors explored with the same analysis pure 3d $\mathcal{N}=4$ orthosymplectic quiver gauge theories, but the observations made in this case culminated only in a partial quiver subtraction algorithm \cite{Cabrera:2019dob} for orthosymplectic quiver theories.

In this section, we pursue the brane dynamics approach to craft a unitary-orthosymplectic quiver subtraction algorithm that can be applied to the magnetic quivers for the Higgs branch that we wish to study. We consider the following procedure.
\begin{enumerate}
    \item Engineer all the magnetic brane systems and magnetic quivers associated with the Higgsed phases of $D^{\su_{2k}}_{N}\left( O \right)$ theories, for arbitrary generalized partitions $O$.
    \item Look at the magnetic brane system of two theories we believe to be connected by an elementary RG flow according to the ordering in equation \eqref{eqn:ordering}, and try to fit this flow to a known transition identified in \cite{Cabrera:2016vvv,Cabrera:2017njm} from a brane dynamics perspective.
    \item When, after such a transition, the brane system obtained is exactly one of the theories we want to flow to, take the quivers of the theories at the two ends of the flow and work out the rules for a quiver subtraction algorithm.
\end{enumerate}

    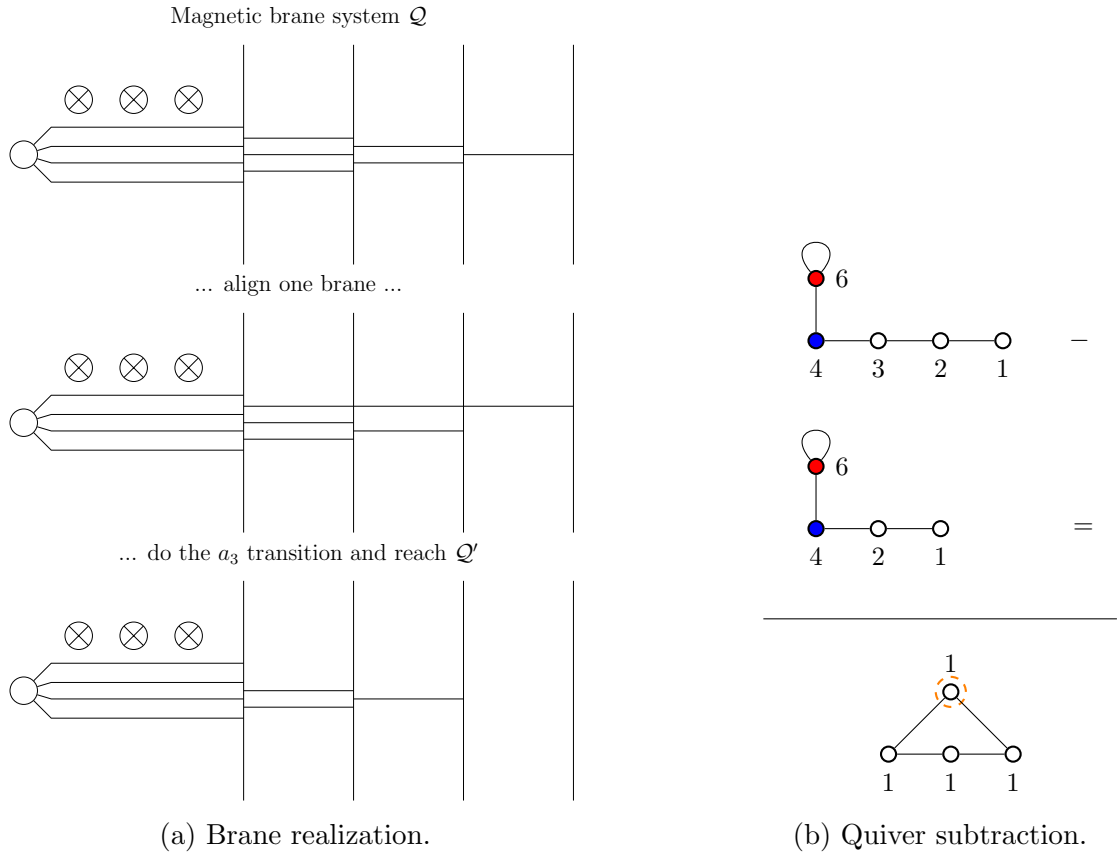
\begin{figure}[t]
\begin{subfigure}[b]{0.47\textwidth}
\resizebox{\textwidth}{!}{
    \begin{tikzpicture}[cross/.style={path picture={ 
  \draw[black]
(path picture bounding box.south east) -- (path picture bounding box.north west) (path picture bounding box.south west) -- (path picture bounding box.north east);
}}] \node[] at (5,2.5) () {Magnetic brane system $\mathcal{Q}$};
          \node[circle, draw=black, minimum size=5mm] at (0,0) (nodo1) {};
            \node[circle, draw=black, cross, minimum size=5mm] at (1,1) (nodo2) {};
            \node[circle, draw=black, cross, minimum size=5mm] at (2,1) (nodo2) {};
            \node[circle,draw=black, cross, minimum size=5mm] at (3,1) (nodo4) {};

            \draw (4,-2)--(4,2) node[midway] (d8-1) {};
            \draw (6,-2)--(6,2) node[midway] (d8-2) {};
            \draw (8,-2)--(8,2) node[midway] (d8-3) {};
            \draw (10,-2)--(10,2) node[midway] (d8-4) {};

            \draw (nodo1)--(0.5,0.5)--(4,0.5) node[above=5pt,left] () {};
            \draw (nodo1)--(0.5,0.15)--(4,0.15) node[above=5pt,left] () {}; 
            \draw (nodo1)--(0.5,-0.15)--(4,-0.15) node[above=5pt,left] () {}; 
            \draw (nodo1)--(0.5,-0.5)--(4,-0.5) node[above=5pt,left] () {};

            \draw (4,0)--(6,0);
            \draw (4,0.3)--(6,0.3);
            \draw (4,-0.3)--(6,-0.3);

            \draw (6,0.15)--(8,0.15);
            \draw (6,-0.15)--(8,-0.15);
            
            \draw (8,0)--(10,0);
    \end{tikzpicture}
    }
    \resizebox{\textwidth}{!}{
        \begin{tikzpicture}[cross/.style={path picture={ 
  \draw[black]
(path picture bounding box.south east) -- (path picture bounding box.north west) (path picture bounding box.south west) -- (path picture bounding box.north east);
}}] \node[] at (5,2.5) () {... align one brane ...};
          \node[circle, draw=black, minimum size=5mm] at (0,0) (nodo1) {};
            \node[circle, draw=black, cross, minimum size=5mm] at (1,1) (nodo2) {};
            \node[circle, draw=black, cross, minimum size=5mm] at (2,1) (nodo2) {};
            \node[circle,draw=black, cross, minimum size=5mm] at (3,1) (nodo4) {};

            \draw (4,-2)--(4,2) node[midway] (d8-1) {};
            \draw (6,-2)--(6,2) node[midway] (d8-2) {};
            \draw (8,-2)--(8,2) node[midway] (d8-3) {};
            \draw (10,-2)--(10,2) node[midway] (d8-4) {};

            \draw (nodo1)--(0.5,0.5)--(4,0.5) node[above=5pt,left] () {};
            \draw (nodo1)--(0.5,0.15)--(4,0.15) node[above=5pt,left] () {}; 
            \draw (nodo1)--(0.5,-0.15)--(4,-0.15) node[above=5pt,left] () {}; 
            \draw (nodo1)--(0.5,-0.5)--(4,-0.5) node[above=5pt,left] () {};

            \draw (4,0)--(6,0);
            \draw (4,0.3)--(6,0.3);
            \draw (4,-0.3)--(6,-0.3);

            \draw (6,0.3)--(8,0.3);
            \draw (6,-0.15)--(8,-0.15);
            
            \draw (8,0.3)--(10,0.3);
    \end{tikzpicture}
}
\resizebox{\textwidth}{!}{
        \begin{tikzpicture}[cross/.style={path picture={ 
  \draw[black]
(path picture bounding box.south east) -- (path picture bounding box.north west) (path picture bounding box.south west) -- (path picture bounding box.north east);
}}] \node[] at (5,2.5) () {... do the $a_3$ transition and reach $\mathcal{Q}'$};
          \node[circle, draw=black, minimum size=5mm] at (0,0) (nodo1) {};
            \node[circle, draw=black, cross, minimum size=5mm] at (1,1) (nodo2) {};
            \node[circle, draw=black, cross, minimum size=5mm] at (2,1) (nodo2) {};
            \node[circle,draw=black, cross, minimum size=5mm] at (3,1) (nodo4) {};

            \draw (4,-2)--(4,2) node[midway] (d8-1) {};
            \draw (6,-2)--(6,2) node[midway] (d8-2) {};
            \draw (8,-2)--(8,2) node[midway] (d8-3) {};
            \draw (10,-2)--(10,2) node[midway] (d8-4) {};

            \draw (nodo1)--(0.5,0.5)--(4,0.5) node[above=5pt,left] () {};
            \draw (nodo1)--(0.5,0.15)--(4,0.15) node[above=5pt,left] () {}; 
            \draw (nodo1)--(0.5,-0.15)--(4,-0.15) node[above=5pt,left] () {}; 
            \draw (nodo1)--(0.5,-0.5)--(4,-0.5) node[above=5pt,left] () {};

            \draw (4,0)--(6,0);

            \draw (4,-0.3)--(6,-0.3);

            \draw (6,-0.15)--(8,-0.15);

    \end{tikzpicture}
    }
        \caption{Brane realization.}
        \label{fig:a3_branes}
        \end{subfigure}
        \hspace{0.5cm}
\begin{subfigure}[b]{0.47\textwidth}
    \centering
    \begin{tikzpicture}
            \node[node,draw=black,fill=blue, label=below:{\footnotesize $4$}] at (0,0) (1) {};
            \node[node,draw=black, label=below:{\footnotesize $3$}] [right=6mm of 1] (2) {};
            \node[node,draw=black, label=below:{\footnotesize $2$}] [right=6mm of 2] (3) {};
            \node[node,draw=black, label=below:{\footnotesize $1$}] [right=6mm of 3] (4) {};
            \node[node,draw,fill=red, label=right:{\footnotesize $6$}] [above=6mm of 1] (b1) {};

            \draw (1)--(b1);
            \draw (1)--(2);
            \draw (2)--(3);
            \draw (3)--(4);

            \draw (b1) to[out=130, in=410, looseness=12] (b1);

            \node[tnode] [right=6mm of 4] {\footnotesize $-$};

            \node[node,draw=black,fill=blue, label=below:{\footnotesize $4$}] at (0,-2.5) (1) {};
            \node[node,draw=black, label=below:{\footnotesize $2$}] [right=6mm of 1] (2) {};
            \node[node,draw=black, label=below:{\footnotesize $1$}] [right=6mm of 2] (3) {};

            \node[node,draw,fill=red, label=right:{\footnotesize $6$}] [above=6mm of 1] (b1) {};

            \draw (1)--(b1);
            \draw (1)--(2);
            \draw (2)--(3);

            \draw (b1) to[out=130, in=410, looseness=12] (b1);

            \node[tnode] [right=15mm of 3] {\footnotesize $=$};

            \draw (-0.7,-3.7)--(4.0,-3.7);

            \node[tnode] at (0,-5.5) (1) {};
            \node[node,draw=black, label=below:{\footnotesize $1$}] [right=6mm of 1] (2) {};
            \node[node,draw=black, label=below:{\footnotesize $1$}] [right=6mm of 2] (3) {};
            \node[node,draw=black, label=below:{\footnotesize $1$}] [right=6mm of 3] (4) {};
            \node[node,draw=orange,dashed, minimum size=4mm] [above=5mm of 3] () {};
            \node[node,draw=black, label=above:{\footnotesize $1$}] [above=6mm of 3] (5) {};

            \draw (2)--(5);
            \draw (2)--(3);
            \draw (3)--(4);
            \draw (4)--(5);
    \end{tikzpicture}
    \caption{Quiver subtraction.}
    \label{fig:a3_quiver}
    \end{subfigure}
    \caption{An example of an $a_3$ transition via brane dynamics and quiver subtraction transitioning from the $D_{3}^{\su_4}([1^4])$ theory to the $D_{3}^{\su_4}([2,1^2])$ theory. The rebalancing node introduced in the subtraction has been drawn with an orange dashed border.}
\end{figure}

    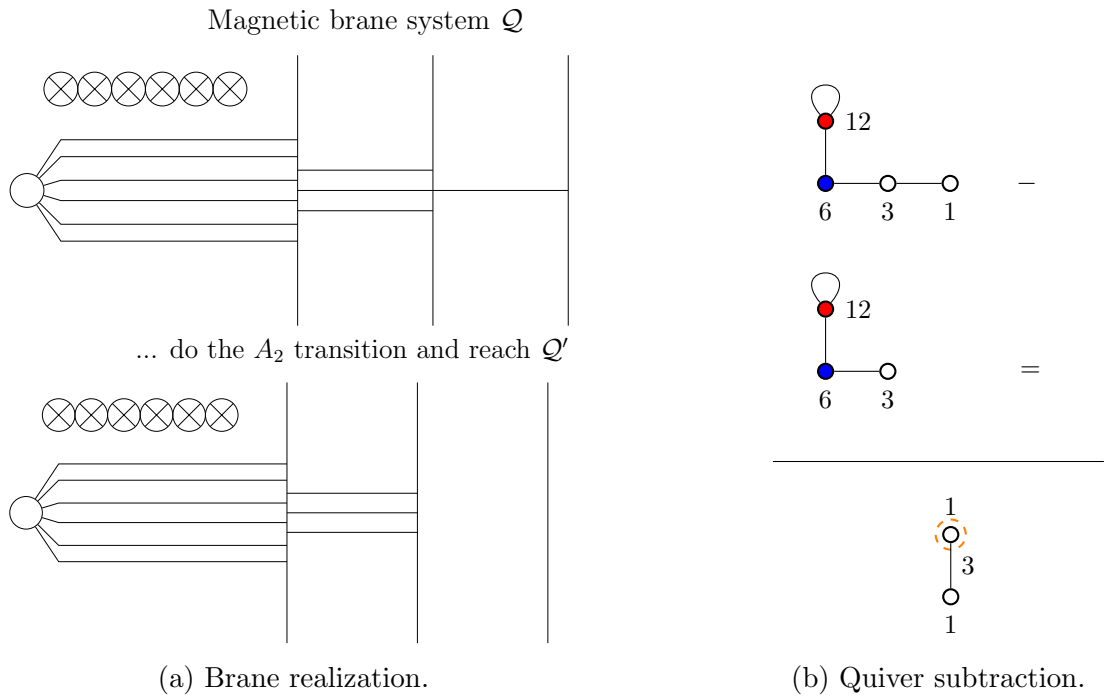
\begin{figure}[t]
        \centering
        \begin{subfigure}[b]{0.47\textwidth}
        \resizebox{\textwidth}{!}{
    \begin{tikzpicture}[cross/.style={path picture={ 
  \draw[black]
(path picture bounding box.south east) -- (path picture bounding box.north west) (path picture bounding box.south west) -- (path picture bounding box.north east);
}}] \node[] at (5,2.5) () {Magnetic brane system $\mathcal{Q}$};
          \node[circle, draw=black, minimum size=5mm] at (0,0) (nodo1) {};
          \node[circle, draw=black, cross, minimum size=5mm] at (0.5,1.5) (nodo2) {};
            \node[circle, draw=black, cross, minimum size=5mm] at (1,1.5) (nodo2) {};
            \node[circle, draw=black, cross, minimum size=5mm] at (1.5,1.5) (nodo2) {};
            \node[circle, draw=black, cross, minimum size=5mm] at (2,1.5) (nodo2) {};
            \node[circle, draw=black, cross, minimum size=5mm] at (2.5,1.5) (nodo2) {};
            \node[circle,draw=black, cross, minimum size=5mm] at (3,1.5) (nodo4) {};

            \draw (4,-2)--(4,2) node[midway] (d8-1) {};
            \draw (6,-2)--(6,2) node[midway] (d8-2) {};
            \draw (8,-2)--(8,2) node[midway] (d8-3) {};

            \draw (nodo1)--(0.5,0.75)--(4,0.75) node[above=5pt,left] () {};
            \draw (nodo1)--(0.5,0.5)--(4,0.5) node[above=5pt,left] () {};
            \draw (nodo1)--(0.5,0.15)--(4,0.15) node[above=5pt,left] () {}; 
            \draw (nodo1)--(0.5,-0.15)--(4,-0.15) node[above=5pt,left] () {}; 
            \draw (nodo1)--(0.5,-0.5)--(4,-0.5) node[above=5pt,left] () {};
            \draw (nodo1)--(0.5,-0.75)--(4,-0.75) node[above=5pt,left] () {};

            \draw (4,0)--(6,0);
            \draw (4,0.3)--(6,0.3);
            \draw (4,-0.3)--(6,-0.3);

            \draw (6,0)--(8,0);

    \end{tikzpicture}
}
\resizebox{\textwidth}{!}{
        \begin{tikzpicture}[cross/.style={path picture={ 
  \draw[black]
(path picture bounding box.south east) -- (path picture bounding box.north west) (path picture bounding box.south west) -- (path picture bounding box.north east);
}}] \node[] at (5,2.5) () {... do the $A_2$ transition and reach $\mathcal{Q}'$};
          \node[circle, draw=black, minimum size=5mm] at (0,0) (nodo1) {};
          \node[circle, draw=black, cross, minimum size=5mm] at (0.5,1.5) (nodo2) {};
            \node[circle, draw=black, cross, minimum size=5mm] at (1,1.5) (nodo2) {};
            \node[circle, draw=black, cross, minimum size=5mm] at (1.5,1.5) (nodo2) {};
            \node[circle, draw=black, cross, minimum size=5mm] at (2,1.5) (nodo2) {};
            \node[circle, draw=black, cross, minimum size=5mm] at (2.5,1.5) (nodo2) {};
            \node[circle,draw=black, cross, minimum size=5mm] at (3,1.5) (nodo4) {};

            \draw (4,-2)--(4,2) node[midway] (d8-1) {};
            \draw (6,-2)--(6,2) node[midway] (d8-2) {};
            \draw (8,-2)--(8,2) node[midway] (d8-3) {};

            \draw (nodo1)--(0.5,0.75)--(4,0.75) node[above=5pt,left] () {};
            \draw (nodo1)--(0.5,0.5)--(4,0.5) node[above=5pt,left] () {};
            \draw (nodo1)--(0.5,0.15)--(4,0.15) node[above=5pt,left] () {}; 
            \draw (nodo1)--(0.5,-0.15)--(4,-0.15) node[above=5pt,left] () {}; 
            \draw (nodo1)--(0.5,-0.5)--(4,-0.5) node[above=5pt,left] () {};
            \draw (nodo1)--(0.5,-0.75)--(4,-0.75) node[above=5pt,left] () {};

            \draw (4,0)--(6,0);
            \draw (4,0.3)--(6,0.3);
            \draw (4,-0.3)--(6,-0.3);

    \end{tikzpicture}
    }
        \caption{Brane realization.}
        \label{fig:A2_branes}
\end{subfigure}
\hspace{0.5cm}
\begin{subfigure}[b]{0.47\textwidth}
    \centering
    \begin{tikzpicture}
            \node[node,draw=black,fill=blue, label=below:{\footnotesize $6$}] at (0,0) (1) {};
            \node[node,draw=black, label=below:{\footnotesize $3$}] [right=6mm of 1] (2) {};
            \node[node,draw=black, label=below:{\footnotesize $1$}] [right=6mm of 2] (3) {};

            \node[node,draw,fill=red, label=right:{\footnotesize $12$}] [above=6mm of 1] (b1) {};

            \draw (1)--(b1);
            \draw (1)--(2);
            \draw (2)--(3);

            \draw (b1) to[out=130, in=410, looseness=12] (b1);

            \node[tnode] [right=6mm of 3] {\footnotesize $-$};

            \node[node,draw=black,fill=blue, label=below:{\footnotesize $6$}] at (0,-2.5) (1) {};
            \node[node,draw=black, label=below:{\footnotesize $3$}] [right=6mm of 1] (2) {};

            \node[node,draw,fill=red, label=right:{\footnotesize $12$}] [above=6mm of 1] (b1) {};

            \draw (1)--(b1);
            \draw (1)--(2);

            \draw (b1) to[out=130, in=410, looseness=12] (b1);

            \node[tnode] [right=15mm of 2] {\footnotesize $=$};

            \draw (-0.7,-3.7)--(3.7,-3.7);

            \node[tnode] at (0,-5.5) (1) {};
            \node[tnode] [right=3mm of 1] (2) {};
            \node[node,draw=black, label=below:{\footnotesize $1$}] [right=5mm of 2] (3) {};
            \node[node,draw=orange,dashed, minimum size=4mm] [above=5mm of 3] () {};
            \node[node,draw=black, label=above:{\footnotesize $1$}] [above=6mm of 3] (5) {};
            
            \draw (3)--(5) node[midway,right] () {\footnotesize $3$};

    \end{tikzpicture}
    \caption{Quiver subtraction.}
    \end{subfigure}
    \caption{An example of an $A_2$ transition via brane dynamics and quiver subtraction taking us from the $\mathcal{D}^{\su_6}_{6}([3,2,1])$ theory to the $\mathcal{D}^{\su_6}_{6}([3^2])$ theory. As before, the $U(1)$ rebalancing node introduced into $\mathcal{S}$ has been drawn with a dashed orange border.}
    \label{fig:A2_quiver}
\end{figure}

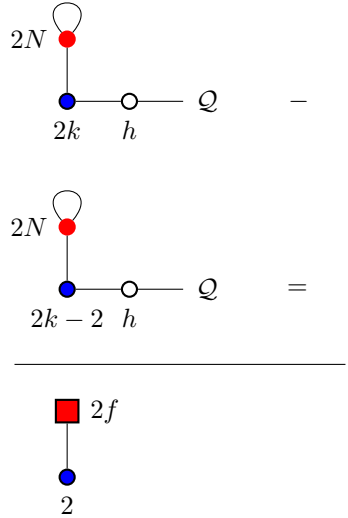
\begin{figure}[t]
    \centering
    \begin{tikzpicture}
            \node[node,draw,fill=blue, label=below:{\footnotesize $2k$}] at (0,0) (1) {};
            \node[node,draw=black, label=below:{\footnotesize $h$}] [right=6mm of 1] (2) {};
            \node[node,draw=red,fill=red, label=left:{\footnotesize $2N$}] [above=6mm of 1] (b1) {}; 
            \node[tnode] [right=6mm of 2] (qq) {{\footnotesize $\mathcal{Q}$}};

            \draw (1)--(b1);
            \draw (1)--(2);
            \draw (2)--(qq);

            \draw (b1) to[out=130, in=410, looseness=12] (b1);

            \node[tnode] [right=6mm of qq] () {\footnotesize $-$};

            \node[node,draw,fill=blue, label=below:{\footnotesize $2k-2$}] at (0,-2.5) (1) {};
            \node[node,draw=black, label=below:{\footnotesize $h$}] [right=6mm of 1] (2) {};
            \node[node,draw=red,fill=red, label=left:{\footnotesize $2N$}] [above=6mm of 1] (b1) {};
            
            \node[tnode] [right=6mm of 2] (qq) {\footnotesize $\mathcal{Q}$};

            \draw (1)--(b1);
            \draw (1)--(2);
            \draw (2)--(qq);

            \draw (b1) to[out=130, in=410, looseness=12] (b1);

            \node[tnode] [right=6mm of qq] () {\footnotesize $=$};

            \draw (-0.7,-3.5)--(3.7,-3.5);

            \node[node,draw,fill=blue, label=below:{\footnotesize $2$}] at (0,-5) (2) {};
            \node[fnode,draw,fill=red, label=right:{\footnotesize $2f$}] [above=6mm of 2] (5) {};
 
            \draw (2)--(5);
    \end{tikzpicture}
    \caption{Quiver subtraction operation on $\mathcal{Q}$ and $\mathcal{Q}'$ that results in a $D_f$ transition. The quantity $f$ is defined in equation \eqref{eqn:DN-flavors}.}
    \label{fig:DN-subtraction}
\end{figure}

Proceeding in this way, we find that in the brane system description, three kinds of transitions occur, namely the $a_n$ minimal singularity, and the $A_N$ and $D_N$ Kleinian singularities. In the brane systems, we find that they occur as follows.
\begin{itemize}
    \item \textbf{$a_n$ transitions:} are realized by aligning a single D6-brane between $n+1$ D8-branes and then pulling the merged D6-brane transversely to infinity, see Figure \ref{fig:a3_branes} for an example.
    \item \textbf{$A_N$ transitions:} occur when only a single D6-brane that sits between D8-branes is pulled away transversely to infinity, and in the intervals adjacent to the boundary D8-branes there are precisely $N+1$ D6-branes exceeding the number necessary for balance. For an example see Figure \ref{fig:A2_branes}.
    \item \textbf{$D_N$ transitions:} these one-dimensional transitions are realized by pulling away a D6-brane stretching between the ON$^-$-plane and the D8-brane next to it and are possible only when the adjacent D8 interval hosts D6-branes with balance of at least $2$. This minimal move requires the simultaneous pulling of the image D6-brane arising from the ON$^-$ projection as the orientifold plane induces an $SO(2)$ charge on the D6-branes that needs to be neutralized.
\end{itemize}

The comparison of the magnetic quiver associated with theories connected by the aforementioned brane system transitions leads to the proposal of the following quiver subtraction rules for unitary-orthosymplectic quiver theories.\footnote{We emphasize that these rules are not proposed for arbitrary unitary-orthosymplectic quivers, but for those of the form we consider in this paper.}
\begin{itemize}
    \item[$*$)] Consider all $\mathfrak{so}(2N)$ gauge nodes with matter in the rank-two antisymmetric representation as flavor nodes.\footnote{We remark that point $*)$ is necessary since there has been no understanding of an analogous concept to ``decoration'' \cite{Bourget:2022ehw} in the case of non-unitary gauge nodes.}
    \item[a)] The quiver theory $\mathcal{Q}'$ can be subtracted from the theory $\mathcal{Q}$ if there exists an alignment of the two theories such that the gauge nodes in $\mathcal{Q}'$ correspond to gauge nodes in $\mathcal{Q}$ with same type of gauge algebras but ranks not greater than that of $\mathcal{Q}$.
    \item[b)] The subtraction $\mathcal{Q}-\mathcal{Q}'$ is defined as the quiver gauge theory $\mathcal{S}$ having same links and nodes gauge algebra as $\mathcal{Q}$, but with gauge group rank given by the difference between that of the node in $\mathcal{Q}$ and the corresponding one in $\mathcal{Q}'$.
    \item[c)] The quiver theory $\mathcal{S}$ must be rebalanced according to the following procedure. 
        \begin{itemize}
            \item Call $N$ the rank of the flavor node obtained from point $*)$, $k$ the rank of the connected symplectic gauge node, and $h$ the total rank of the unitary nodes connected to the symplectic node. If after the subtraction that very same symplectic gauge node has non-zero rank then its flavor symmetry node changes to an $SO(2f)$ node with:
                \begin{equation}\label{eqn:DN-flavors}
                    2f =\begin{cases}2N+h-2k & 2N+h=0 \text{ mod } 2\,,\\
                    2N+h-2k+1 & 2N+h=1 \text{ mod } 2\,.\\
                    \end{cases}
                \end{equation}
            \item If, after the subtraction, the quiver $\mathcal{S}$ splits into a fully orthosymplectic part and a unitary part, the unitary part must be rebalanced via the introduction of the usual $U(1)$ node as in the standard quiver subtraction algorithm \cite{Cabrera:2018ann}.
         \end{itemize} 
\end{itemize}
We depicted in Figures \ref{fig:a3_quiver}, \ref{fig:A2_branes}, and \ref{fig:DN-subtraction} some examples of the subtraction algorithm applied to each transition encountered in the Higgsing of the $D_{N}^{\su_{2k}}\left( O \right)$ theories.

A much more direct approach that still manages to reproduce the Higgs branch Hasse diagram of $D^{\su_{2k}}_{N}\left( O \right)$ involves the application of an extended version of the decay and fission algorithm \cite{Bourget:2023dkj,Bourget:2024mgn} to unitary-orthosymplectic quivers.\footnote{This extension is valid only for the magnetic quivers that are the subject of this paper. Even a first extension of the decay and fission algorithm to the case of a purely orthosymplectic quiver is deemed to be challenging since ``bad'' magnetic quivers appear in Higgsed phases of theories associated with ``good'' magnetic quivers \cite{Hanany:2022itc}. Such ubiquitous pathologies are absent from the quivers we study here.} The approach is the same as adopted in the original paper and consists of writing all the possible good, in the sense of \cite{Gaiotto:2008ak}, quiver gauge theories with the same shape and type of gauge algebras as the considered theory, but with not greater gauge rank for each node. First, discard the equivalent theories, and then connect them according to whether, via this very same algorithm, a theory with smaller gauge group ranks can be obtained from a higher dimensional one. This procedure realizes the same Hasse diagram as the quiver subtraction algorithm that we have just delineated.

\subsection{The Higgs Branch}\label{sec:Higgs_generalized_partitions}

Now that we have utilized the brane system both to derive the magnetic quivers for the Higgs branches of the 6d $(1,0)$ SCFTs Higgsable to the D-type $(2,0)$ SCFTs, and the quiver subtraction algorithm for such magnetic quivers/brane systems, we are ready to explicate the structure of the Higgs branch. Since it is a straightforward application of the algorithm of Section \ref{sec:Quiver_Subtraction} to the magnetic quivers for the Higgs branch derived in Section \ref{sec:Magnetic_quiver}, in this section we simply present some explicit examples.

\begin{figure}[p]
\begin{subfigure}[b]{0.47\textwidth}
        \centering
    \includegraphics[page=4,width=\textwidth]{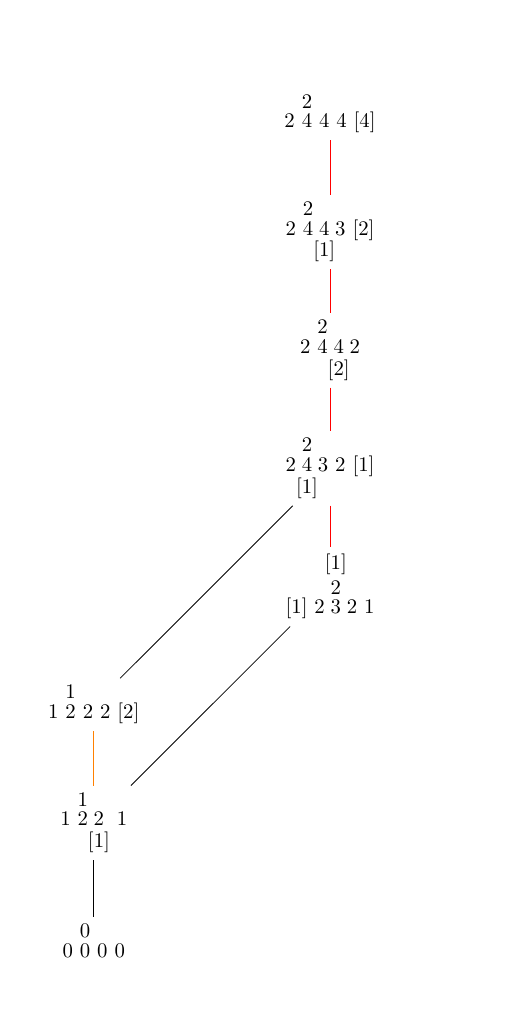}
    \caption{Hasse diagram from the tensor branch.}
    \label{fig:ProvisionalHasseD_5su_4}
\end{subfigure}
\hspace{0.5cm}
\begin{subfigure}[b]{0.47\textwidth}
        \centering
    \includegraphics[page=3,height=0.8\textheight]{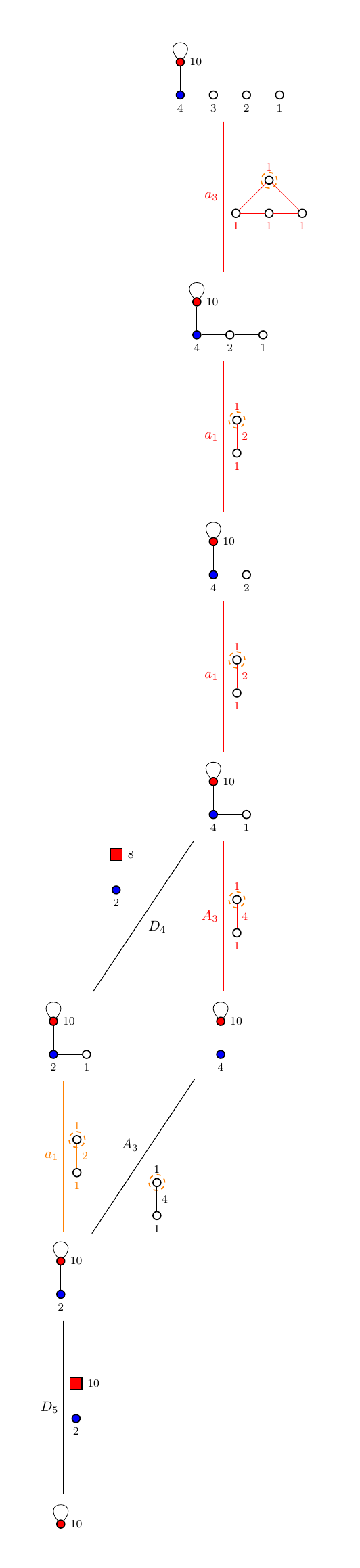}
    \caption{Hasse diagram from quiver subtraction.}
    \label{fig:QSHasseD_5su_4}
\end{subfigure}
\caption{In Figure (a) is depicted the Hasse diagram for a $D_{5}^{\su_{4}}([1^4])$ theory where each vertex is represented by its tensor branch geometry. In Figure (b)
the same Hasse diagram is proposed, but vertices show the corresponding magnetic quiver, and slices are identified via the quiver resulting from quiver subtraction of connected theories, depicted together with the standard nomenclature.}
\end{figure}

We begin by considering the $D_{5}^{\su_{4}}([1^4])$ SCFT. We determine each of the tensor branches satisfying equation \eqref{eqn:Cartan_Gauge=flavor} and such that the ranks of the gauge algebras are less than or equal to that of the gauge algebras for the tensor branch of the original theory. That is, we determine all consistent tensor branch configurations satisfying equation \eqref{eqn:RGflow}, and construct a provisional Hasse diagram using the partial ordering defined in equation \eqref{eqn:ordering}. We have depicted this in Figure \ref{fig:ProvisionalHasseD_5su_4}. The next step is to determine the nature of the slices connecting neighbouring theories, this feat can be accomplished by extracting the magnetic quiver for all the theories in the provisional Hasse and subtracting, according to the rules explained in Section \ref{sec:Quiver_Subtraction}, adjacent theories. Each subtraction will produce a 3d quiver whose Coulomb branch moduli space determines the slice nature. Therefore, from this procedure, it is possible to label each edge in the Hasse diagram, as shown in Figure \ref{fig:QSHasseD_5su_4}.

\begin{landscape}
\pagestyle{empty}
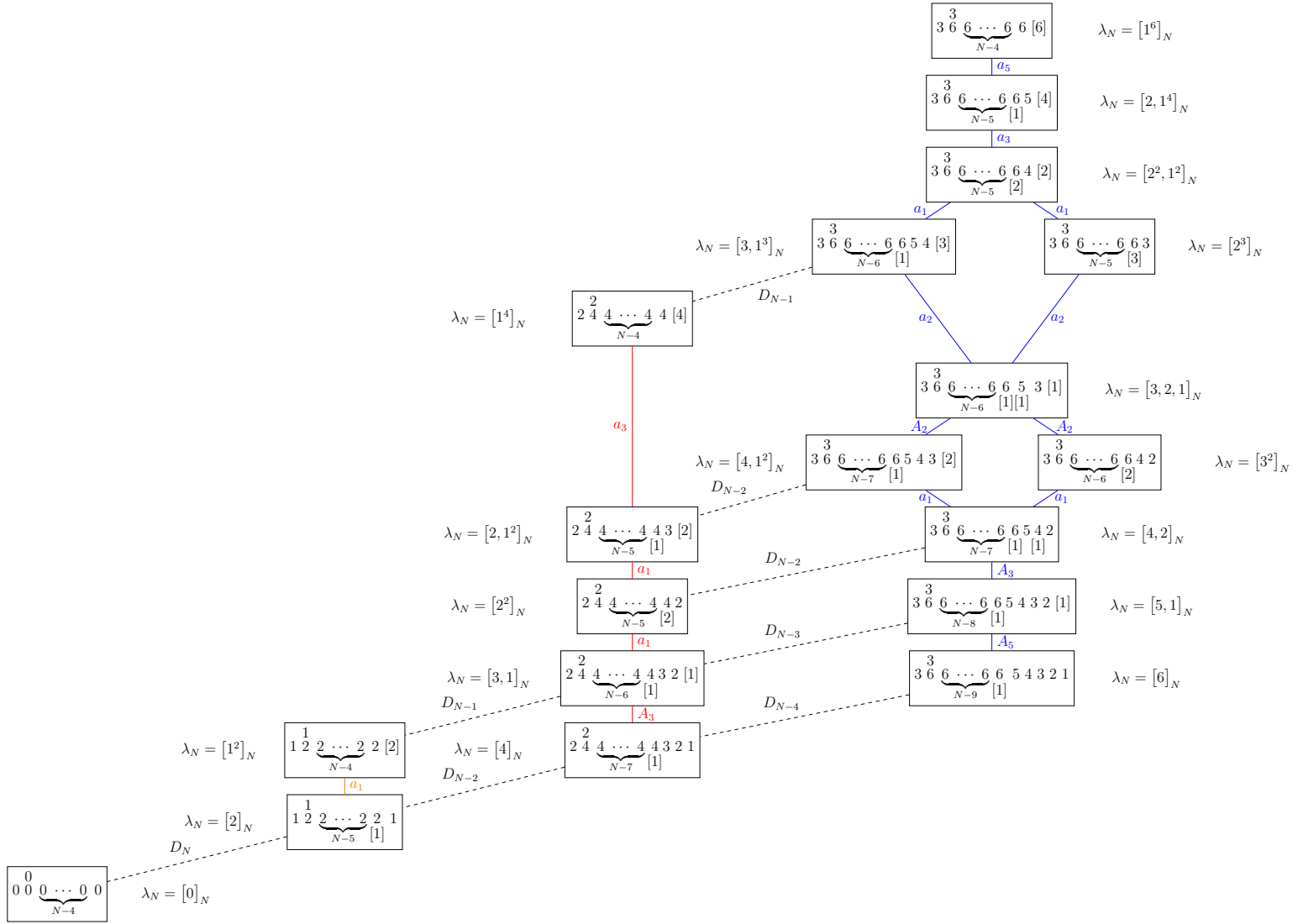
\begin{figure}[p]
    \centering
    \resizebox{!}{\textheight-2.05cm}{
    \begin{tikzpicture}
    
    \node[rectangle, draw] at (2,0) (nodo1) {$3 \ \stackon{6}{3} \ \underbrace{6 \ \cdots \ 6}_{N-4} \ 6 \ [6]$};
    \node[] at (6,0) () {$\lambda_N=\big [ 1^6  \big]_N$};
    \node[rectangle, draw] at (2,-2) (nodo2) {$3 \ \stackon{6}{3} \ \underbrace{6 \ \cdots \ 6}_{N-5} \stackunder{6}{[1]}  5 \ [4]$};
    \node[] at (6.25,-2) () {$\lambda_N=\big [ 2,1^4  \big]_N$};
    \draw[blue] (nodo1)--(nodo2) node[midway,right] {$a_5$};

    \node[rectangle, draw] at (2,-4) (nodo3) {$3 \ \stackon{6}{3} \ \underbrace{6 \ \cdots \ 6}_{N-5} \stackunder{6}{[2]}  4 \ [2]$};
    \node[] at (6.4,-4) () {$\lambda_N=\big [ 2^2,1^2  \big]_N$};
    \draw[blue] (nodo2)--(nodo3) node[midway,right] {$a_3$};

    \node[rectangle, draw] at (5,-6) (nodo4-1) {$3 \ \stackon{6}{3} \ \underbrace{6 \ \cdots \ 6}_{N-5} \stackunder{6}{[3]}  3 $};
    \node[] at (8.5,-6) () {$\lambda_N=\big [ 2^3  \big]_N$};
    \node[rectangle, draw] at (-1,-6) (nodo4-2) {$3 \ \stackon{6}{3} \ \underbrace{6  \ \cdots \ 6}_{N-6} \stackunder{6}{[1]} 5 \ 4 \ [3]$};
    \node[] at (-5,-6) () {$\lambda_N=\big [ 3,1^3  \big]_N$};
    \draw[blue] (nodo3)--(nodo4-1) node[midway,right=4pt] {$a_1$};
    \draw[blue] (nodo3)--(nodo4-2) node[midway,left=4pt] {$a_1$};

    \node[rectangle, draw] at (2,-10) (nodo5-1) {$3 \ \stackon{6}{3} \ \underbrace{6  \ \cdots \ 6}_{N-6} \stackunder{6}{[1]} \stackunder{5}{[1]} \ 3 \ [1]$};
    \node[] at (6.5,-10) () {$\lambda_N=\big [ 3,2,1  \big]_N$};
    \node[rectangle, draw] at (-8,-8) (nodo5-2) {$2 \ \stackon{4}{2} \ \underbrace{4 \ \cdots \  4}_{N-4} \ 4 \ [4]$};
    \node[] at (-12,-8) () {$\lambda_N=\big [ 1^4  \big]_N$};
    \draw[blue] (nodo4-1)--(nodo5-1) node[midway,right] {$a_2$};
    \draw[blue] (nodo4-2)--(nodo5-1) node[midway,left] {$a_2$};
    \draw[dashed] (nodo4-2)--(nodo5-2) node[midway,below right] {$D_{N-1}$};

    \node[rectangle, draw] at (-1,-12) (nodo6-1) {$3 \ \stackon{6}{3} \ \underbrace{6 \ \cdots \ 6}_{N-7} \stackunder{6}{[1]}  5 \ 4 \ 3 \ [2]$};
    \node[] at (-5,-12) () {$\lambda_N=\big [ 4, 1^2  \big]_N$};
    \node[rectangle, draw] at (5,-12) (nodo6-2) {$3 \ \stackon{6}{3} \ \underbrace{6 \ \cdots \ 6}_{N-6} \stackunder{6}{[2]}   4 \ 2$};
    \node[] at (9.25,-12) () {$\lambda_N=\big [ 3^2  \big]_N$};
    \draw (nodo5-1)[blue]--(nodo6-1) node[midway,left=4pt] {$A_2$};
    \draw (nodo5-1)[blue]--(nodo6-2) node[midway,right=4pt] {$A_2$};

    \node[rectangle, draw] at (-8,-14) (nodo7-1) {$2 \ \stackon{4}{2} \ \underbrace{4 \ \cdots \  4}_{N-5}  \stackunder{4}{[1]}  3 \ [2]$};
    \node[] at (-12,-14) () {$\lambda_N=\big [ 2,1^2  \big]_N$};
    \node[rectangle, draw] at (2,-14) (nodo7-2) {$3 \ \stackon{6}{3} \ \underbrace{6 \ \cdots \ 6}_{N-7} \stackunder{6}{[1]}  5  \stackunder{4}{[1]} 2$};
    \node[] at (6.2,-14) () {$\lambda_N=\big [ 4,2  \big]_N$};
    \draw[blue] (nodo6-1)--(nodo7-2) node[midway,left] {$a_1$};
    \draw[blue] (nodo6-2)--(nodo7-2) node[midway,right=3pt] {$a_1$};
    \draw[dashed] (nodo6-1)--(nodo7-1) node[midway,above left] {$D_{N-2}$};
    \draw[red] (nodo5-2)--(nodo7-1) node[midway,left] {$a_3$};

    \node[rectangle, draw] at (-8,-16) (nodo8-1) {$2 \ \stackon{4}{2} \ \underbrace{4  \ \cdots \  4}_{N-5} \stackunder{4}{[2]}  2$};
    \node[] at (-12,-16) () {$\lambda_N=\big [ 2^2  \big]_N$};
    \node[rectangle, draw] at (2,-16) (nodo8-2) {$ 3 \ \stackon{6}{3} \ \underbrace{6 \ \cdots \ 6}_{N-8} \stackunder{6}{[1]}  5 \ 4 \ 3 \ 2 \ [1]$};
    \node[] at (6.45,-16) () {$\lambda_N=\big [ 5,1  \big]_N$};
    \draw[red] (nodo7-1)--(nodo8-1) node[midway,right] {$a_1$};
    \draw[dashed] (nodo7-2)--(nodo8-1) node[midway,above left] {$D_{N-2}$};
    \draw[blue] (nodo7-2)--(nodo8-2) node[midway,right] {$A_3$};
    
    \node[rectangle, draw] at (-8,-18) (nodo9-1) {$2 \ \stackon{4}{2} \ \underbrace{4 \ \cdots \  4}_{N-6} \stackunder{4}{[1]} 3 \  2 \ [1]$};
    \node[] at (-12,-18) () {$\lambda_N=\big [ 3,1  \big]_N$};
    \node[rectangle, draw] at (2,-18) (nodo9-2) {$ 3 \ \stackon{6}{3} \ \underbrace{6 \ \cdots \ 6}_{N-9} \stackunder{6}{[1]} \ 5 \ 4 \ 3 \ 2 \ 1$};
    \node[] at (6.3,-18) () {$\lambda_N=\big [ 6 \big]_N$};
    \draw[red] (nodo8-1)--(nodo9-1) node[midway,right] {$a_1$};
    \draw[dashed] (nodo8-2)--(nodo9-1) node[midway,above left] {$D_{N-3}$};
    \draw[blue] (nodo8-2)--(nodo9-2) node[midway,right] {$A_5$};

       \node[rectangle, draw] at (-8,-20) (nodo10-1) {$2 \ \stackon{4}{2} \ \underbrace{4 \ \cdots \  4}_{N-7} \stackunder{4}{[1]} 3 \  2 \ 1$};
       \node[] at (-12,-20) () {$\lambda_N=\big [ 4 \big]_N$};
    \node[rectangle, draw] at (-16,-20) (nodo10-2) {$ 1  \  \stackon{2}{1} \ \underbrace{2 \ \cdots \  2}_{N-4} \  2 \ [2] $};
    \node[] at (-19.5,-20) () {$\lambda_N=\big [ 1^2  \big]_N$};
    \draw[red] (nodo9-1)--(nodo10-1) node[midway,right] {$A_3$};
    \draw[dashed] (nodo9-1)--(nodo10-2) node[midway,above left] {$D_{N-1}$};
    \draw[dashed] (nodo9-2)--(nodo10-1) node[midway,above left] {$D_{N-4}$};

    \node[rectangle, draw] at (-16,-22) (nodo11) {$1 \   \stackon{2}{1} \ \underbrace{2 \ \cdots \  2}_{N-5} \stackunder{2}{[1]} \  1 $};
    \node[] at (-19.5,-22) () {$\lambda_N=\big [ 2  \big]_N$};
    \draw[dashed] (nodo10-1)--(nodo11) node[midway,above left] {$D_{N-2}$};
    \draw[orange] (nodo10-2)--(nodo11) node[midway,right] {$a_1$};

    \node[rectangle, draw] at (-24,-24) (nodo12) {$0 \ \stackon{0}{0} \ \underbrace{0 \ \cdots \ 0}_{N-4} \  0 $};
    \node[] at (-20.7,-24) () {$\lambda_N=\big [ 0  \big]_N$};
    \draw[dashed] (nodo11)--(nodo12) node[midway,above left] {$D_N$};

    \end{tikzpicture}
    }
    \caption{Hasse diagram for the long $D_{N\geq 9}^{\su_6}$ theory. At each node, we depict the tensor branch curve configuration. Blue slices refer to the nilpotent cone of $\su(6)$, whereas red is used for the nilpotent cone of $\su(4)$, and orange for $\su(2)$. Dashed lines denote D-type transitions that link cones of different algebras together.}
    \label{fig:Hasse-D_Nsu6_Long}
\end{figure}
\end{landscape}

\begin{figure}[p]
    \centering
    \resizebox{!}{\textheight-3cm}{
    \begin{tikzpicture}
    
    \node[rectangle, draw] at (0,0) (nodo1) {$3 \ \stackon{6}{3} \ 6 \ 6 \ [6]$};
    \node[right=40pt] at (0,0) () {$\lambda_5=\big[1^6\big]_5$};
    \node[rectangle, draw] at (0,-2) (nodo2) {$3 \ \stackon{6}{3}  \stackunder{6}{[1]}  5 \ [4]$};
    \node[right=40pt] at (0,-2) () {$\lambda_5=\big[2^1,1^4\big]_5$};
    \draw[blue] (nodo1)--(nodo2) node[midway,right] {$a_5$};

    \node[rectangle, draw] at (0,-4) (nodo3) {$3 \ \stackon{6}{3}  \stackunder{6}{[2]}  4 \ [2]$};
    \node[right=40pt] at (0,-4) () {$\lambda_5=\big[2^2,1^2\big]_5$};
    \draw[blue] (nodo2)--(nodo3) node[midway,right] {$a_3$};

    \node[rectangle, draw] at (2,-6) (nodo4-1) {$3 \ \stackon{6}{3}  \stackunder{6}{[3]}  3 $};
    \node[right=40pt] at (2,-6) () {$\lambda_5=\big[2^3\big]_5$};
    \node[rectangle, draw] at (-2,-6) (nodo4-2) {$3 \ \stackunder{\stackon{6}{3}}{[1]}   5 \ 4 \ [3]$};
    \node[left=40pt] at (-2,-6) () {$\lambda_5=\big[3^1,1^3\big]_5$};
    \draw[blue] (nodo3)--(nodo4-1) node[midway,right] {$a_1$};
    \draw[blue] (nodo3)--(nodo4-2) node[midway,left] {$a_1$};

    \node[rectangle, draw] at (0,-10) (nodo5-1) {$3 \ \stackunder{\stackon{6}{3}}{[1]}   \stackunder{5}{[1]}  3 \ [1]$};
    \node[right=40pt] at (0,-10) () {$\lambda_5=\big[3^1,2^1,1^1\big]_5$};
    \node[rectangle, draw] at (-4,-8) (nodo5-2) {$2 \ \stackon{4}{2} \  4 \ 4 \ [4]$};
    \node[left=40pt] at (-4,-8) () {$\lambda_5=\big[1^4\big]_5$};
    \draw[blue] (nodo4-1)--(nodo5-1) node[midway,right] {$a_2$};
    \draw[blue] (nodo4-2)--(nodo5-1) node[midway,left] {$a_2$};
    \draw (nodo4-2)--(nodo5-2) node[midway,left=3pt] {$D_4$};

    \node[rectangle, draw] at (-2,-12) (nodo6-1) {$[1] \ 3 \ \stackon{5}{\stackon{3}{[1]}}   4 \ 3 \ [2]$};
    \node[left=60pt] at (-2,-12) () {$\lambda_5=\big[4^1,4^1;1^2\big]_5$};
    \node[rectangle, draw] at (2,-12) (nodo6-2) {$3 \ \stackunder{\stackon{6}{3}}{[2]}   4 \ 2$};
    \node[right=40pt] at (2,-12) () {$\lambda_5=\big[3^2\big]_5$};
    \draw (nodo5-1)--(nodo6-1) node[midway,right] {$A_2$};
    \draw (nodo5-1)[blue]--(nodo6-2) node[midway,left] {$A_2$};

    \node[rectangle, draw] at (-4,-14) (nodo7-1) {$2 \ \stackon{4}{2}   \stackunder{4}{[1]}  3 \ [2]$};
    \node[left=40pt] at (-4,-14) () {$\lambda_5=\big[2,1^2\big]_5$};
    \node[rectangle, draw] at (0,-14) (nodo7-2) {$[1] \ 3 \ \stackon{5}{\stackon{3}{[1]}} \stackunder{4}{[1]} 2$};
    \node[right=40pt] at (0,-14) () {$\lambda_5=\big[4^1,4^1;2^1\big]_5$};
    \draw (nodo6-1)--(nodo7-2) node[midway,right] {$a_1$};
    \draw (nodo6-2)--(nodo7-2) node[midway,left] {$a_1$};
    \draw (nodo6-1)--(nodo7-1) node[midway,left=3pt] {$A_3$};
    \draw[red] (nodo5-2)--(nodo7-1) node[midway,left] {$a_3$};

    \node[rectangle, draw] at (-4,-16) (nodo8-1) {$2 \ \stackon{4}{2}   \stackunder{4}{[2]}  2$};
    \node[left=40pt] at (-4,-16) () {$\lambda_5=\big[2^2\big]$};
    \node[rectangle, draw] at (4.5,-16) (nodo8-3) {$ 2 \ \stackon{4}{\stackon{3}{[2]}} 3 \ 2 \ [1]$};
    \node[right=40pt] at (4.5,-16) () {$\lambda_5=\big[4^2,4^0;1\big]_5$};
    \node[rectangle, draw] at (-1.5,-16) (nodo8-2) {$ [2] \ 3 \ \stackon{4}{2} \ 3 \ 2 \ [1]$};
    \node[right=40pt] at (-1.5,-16) () {$\lambda_5=\big[4^0,4^2;1\big]_5$};
    \draw[red] (nodo7-1)--(nodo8-1) node[midway,right] {$a_1$};
    \draw (nodo7-2)--(nodo8-1) node[midway,left=5pt] {$A_3$};
    \draw (nodo7-2)--(nodo8-3) node[midway,right] {$A_3$};
    \draw (nodo7-2)--(nodo8-2) node[midway,right] {$A_3$};
    
    \node[rectangle, draw] at (-4,-18) (nodo9-1) {$2  \stackunder{\stackon{4}{2}}{[1]}   3 \  2 \ [1]$};
    \node[left=40pt] at (-4,-18) () {$\lambda_5=\big[3^1,1^1\big]_5$};
    \draw[red] (nodo8-1)--(nodo9-1) node[midway,right] {$a_1$};
    \draw (nodo8-2)--(nodo9-1) node[midway,right=2pt] {$a_1$};
    \draw (nodo8-3)--(nodo9-1) node[midway,right=5pt,below=2pt] {$a_1$};
    
    \node[rectangle, draw] at (-4,-20) (nodo10-1) {$[1] \ 2   \stackon{3}{\stackon{2}{[1]}}   2 \  1 $};
    \node[rectangle, draw] at (-8,-20) (nodo10-2) {$ 1  \  \stackon{2}{1} \  2 \  2 \ [2] $};
    \node[right=40pt] at (-4,-20) () {$\lambda_5=\big[4^1,4^1;0\big]_5$};
    \node[left=40pt] at (-8,-20) () {$\lambda_5=\big[1^2 \big]_5$};
    \draw[red] (nodo9-1)--(nodo10-1) node[midway,right] {$A_3$};
    \draw (nodo9-1)--(nodo10-2) node[midway,above left] {$D_4$};

    \node[rectangle, draw] at (-8,-22) (nodo11) {$1 \   \stackon{2}{1} \stackunder{2}{[1]} \  1 $};
    \node[right=40pt] at (-8,-22) () {$\lambda_5=\big[2^1\big]_5$};
    \draw (nodo10-1)--(nodo11) node[midway,above left] {$A_3$};
    \draw[orange] (nodo10-2)--(nodo11) node[midway,right] {$a_1$};

    \node[rectangle, draw] at (-8,-24) (nodo12) {$0 \ \stackon{0}{0} \ 0 \  0 $};
    \node[right=40pt] at (-8,-24) () {$\lambda_5=\big[0\big]_5$};
    \draw (nodo11)--(nodo12) node[midway,right] {$D_5$};

    \end{tikzpicture}
    }
    \caption{Hasse diagram for the $D^{\su_6}_5$ theory. At each vertex, the tensor branch curve configuration is drawn. Blue transitions are used to identify slices in the nilpotent cone of $\mathfrak{su}(6)$, whereas red is used for the nilpotent cone of $\mathfrak{su}(4)$. We note that there are also black edges which do not correspond to any nilpotent cone.}
    \label{fig:Hasse-D5-[6]-partition}
\end{figure}

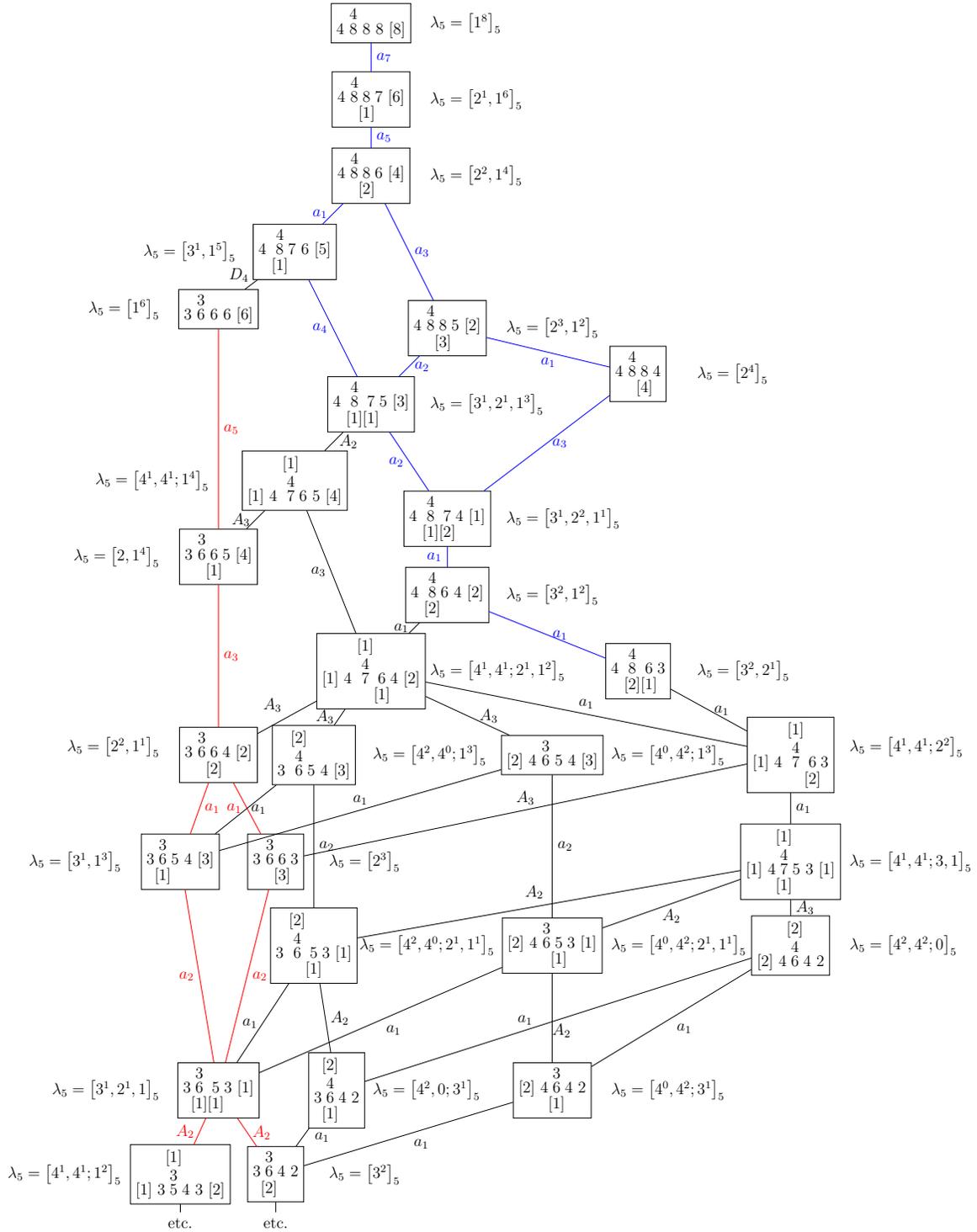
\begin{figure}[p]
    \centering
    \resizebox{!}{\textheight-2.5cm}{
    \begin{tikzpicture}
    
    \node[rectangle, draw] at (0,0) (nodo1) {$4 \ \stackon{8}{4} \ 8 \ 8 \ [8]$};
    \node[right=40pt] at (0,0) () {$\lambda_5=\big[1^8\big]_5$};
    \node[rectangle, draw] at (0,-2) (nodo2) {$4 \ \stackon{8}{4}  \stackunder{8}{[1]}  7 \ [6]$};
    \node[right=40pt] at (0,-2) () {$\lambda_5=\big[2^1,1^6\big]_5$};
    \draw[blue] (nodo1)--(nodo2) node[midway,right] {$a_7$};

    \node[rectangle, draw] at (0,-4) (nodo3) {$4 \ \stackon{8}{4}  \stackunder{8}{[2]}  6 \ [4]$};
    \node[right=40pt] at (0,-4) () {$\lambda_5=\big[2^2,1^4\big]_5$};
    \draw[blue] (nodo2)--(nodo3) node[midway,right] {$a_5$};

    \node[rectangle, draw] at (2,-8) (nodo4-1) {$4 \ \stackon{8}{4}  \stackunder{8}{[3]}  5 \ [2] $};
    \node[right=40pt] at (2,-8) () {$\lambda_5=\big[2^3,1^2\big]_5$};
    \node[rectangle, draw] at (-2,-6) (nodo4-2) {$4 \ \stackunder{\stackon{8}{4}}{[1]}   7 \ 6 \ [5]$};
    \node[left=40pt] at (-2,-6) () {$\lambda_5=\big[3^1,1^5\big]_5$};
    \draw[blue] (nodo3)--(nodo4-1) node[midway,right] {$a_3$};
    \draw[blue] (nodo3)--(nodo4-2) node[midway,left] {$a_1$};

    \node[rectangle, draw] at (0,-10) (nodo5-1) {$4 \ \stackunder{\stackon{8}{4}}{[1]}   \stackunder{7}{[1]}  5 \ [3]$};
    \node[right=40pt] at (0,-10) {$\lambda_5=\big[3^1,2^1,1^3\big]_5$};
    \node[rectangle, draw] at (7,-9.2) (nodo5-2) {$4 \ \stackon{8}{4}   \stackunder{8}{[4]}  4 $};
    \node[right=40pt] at (7,-9.2) () {$\lambda_5=\big[2^4\big]_5$};
    \node[rectangle, draw] at (-4,-7.5) (nodo5-3) {$3 \ \stackon{6}{3} \  6 \ 6 \ [6]$};
    \node[left=40pt] at (-4,-7.5) () {$\lambda_5=\big[1^6\big]_5$};
    \draw[blue] (nodo4-1)--(nodo5-1) node[midway,right] {$a_2$};
    \draw[blue] (nodo4-2)--(nodo5-1) node[midway,left] {$a_4$};
    \draw[blue] (nodo4-1)--(nodo5-2) node[midway,left,below] {$a_1$};
    \draw (nodo4-2)--(nodo5-3) node[midway,above left=-3pt] {$D_4$};

    \node[rectangle, draw] at (-2,-12) (nodo6-1) {$[1] \ 4 \ \stackon{7}{\stackon{4}{[1]}}   6 \ 5\ [4]$};
    \node[left=60pt] at (-2,-12) () {$\lambda_5=\big[4^1,4^1;1^4\big]_5$};
    \node[rectangle, draw] at (2,-13) (nodo6-2) {$4 \ \stackunder{\stackon{8}{4}}{[1]}  \stackunder{7}{[2]} 4 \ [1]$};
    \node[right=40pt] at (2,-13) () {$\lambda_5=\big[3^1,2^2,1^1\big]_5$};
    \draw (nodo5-1)--(nodo6-1) node[midway,right] {$A_2$};
    \draw (nodo5-1)[blue]--(nodo6-2) node[midway,left] {$a_2$};
    \draw (nodo5-2)[blue]--(nodo6-2) node[midway,right] {$a_3$};

    \node[rectangle, draw] at (-4,-14) (nodo7-1) {$3 \ \stackon{6}{3}   \stackunder{6}{[1]}  5 \ [4]$};
    \node[left=40pt] at (-4,-14) () {$\lambda_5=\big[2,1^4\big]_5$};
    \node[rectangle, draw] at (2,-15) (nodo7-2) {$4 \ \stackunder{\stackon{8}{4}}{[2]}  6 \ 4 \ [2]$};
    \node[right=40pt] at (2,-15) () {$\lambda_5=\big[3^2,1^2\big]_5$};
    \draw[red] (nodo5-3)--(nodo7-1) node[midway,right] {$a_5$};
    \draw (nodo6-1)--(nodo7-1) node[midway,left] {$A_3$};
    \draw (nodo6-2)[blue]--(nodo7-2) node[midway,left] {$a_1$};

    \node[rectangle, draw] at (0,-17) (nodo8-1) {$[1] \ 4 \ \stackon{7}{\stackon{4}{[1]}}   \stackunder{6}{[1]}  4\ [2]$};
    \node[right=40pt] at (0,-17) () {$\lambda_5=\big[4^1,4^1;2^1,1^2\big]_5$};
    \node[rectangle, draw] at (7,-17) (nodo8-2) {$4 \ \stackunder{\stackon{8}{4}}{[2]}  \stackunder{6}{[1]}  3 $};
    \node[right=40pt] at (7,-17) () {$\lambda_5=\big[3^2,2^1\big]_5$};
    \draw (nodo6-1)--(nodo8-1) node[midway,left] {$a_3$};
    \draw (nodo7-2)--(nodo8-1) node[midway,left] {$a_1$};   
    \draw (nodo7-2)[blue]--(nodo8-2) node[midway,right] {$a_1$};   

    \node[rectangle, draw] at (-4,-19.2) (nodo9-1) {$3 \ \stackon{6}{3}   \stackunder{6}{[2]}  4 \ [2]$};
    \node[left=40pt] at (-4,-19) () {$\lambda_5=\big[2^2,1^1 \big]_5$};
    \node[rectangle, draw] at (-1.5,-19.2) (nodo9-2) {$  3 \ \stackon{6}{\stackon{4}{[2]}}   5 \  4\ [3]$};
    \node[right=40pt] at (-1.5,-19.2) () {$\lambda_5=\big[4^2,4^0;1^3 \big]_5$};
    \node[rectangle, draw] at (4.75,-19.2) (nodo9-3) {$ [2] \ 4 \ \stackon{6}{3} \  5 \  4\ [3]$};
    \node[right=40pt] at (4.75,-19.2) () {$\lambda_5=\big[4^0,4^2;1^3 \big]_5$};
    \node[rectangle, draw] at (11,-19.2) (nodo9-4) {$[1] \ 4 \ \stackon{7}{\stackon{4}{[1]}}   \stackunder{6}{[2]}  3$};
    \node[right=40pt] at (11,-19) () {$\lambda_5=\big[4^1,4^1;2^2\big]_5$};
    \draw (nodo8-1)--(nodo9-1) node[midway,above left] {$A_3$};
    \draw (nodo8-1)--(nodo9-2) node[midway,left] {$A_3$};   
    \draw (nodo8-1)--(nodo9-3) node[midway,right=5pt] {$A_3$};   
    \draw (nodo8-1)--(nodo9-4) node[midway,right,above] {$a_1$};  
    \draw (nodo8-2)--(nodo9-4) node[midway,right] {$a_1$};
    \draw (nodo7-1)[red]--(nodo9-1) node[midway,right] {$a_3$};  

    \node[rectangle, draw] at (-5,-22) (nodo10-2) {$3  \stackunder{\stackon{6}{3}}{[1]} 5 \ 4 \ [3]$};
    \node[left=40pt] at (-5,-22) () {$\lambda_5=\big[3^1,1^3 \big]_5$};
    \node[rectangle, draw] at (-2.5,-22) (nodo10-1) {$3 \ \stackon{6}{3}   \stackunder{6}{[3]}  3 $};
    \node[right=50pt] at (-3,-22) () {$\lambda_5=\big[2^3 \big]_5$};
    \node[rectangle, draw] at (11,-22) (nodo10-3) {$[1] \ 4  \stackunder{\stackon{7}{\stackon{4}{[1]}}}{[1]}5 \ 3 \ [1]$};
    \node[right=40pt] at (11,-22) () {$\lambda_5=\big[4^1,4^1;3,1 \big]_5$};
    \draw (nodo9-1)[red]--(nodo10-1) node[midway,left] {$a_1$};
    \draw (nodo9-1)[red]--(nodo10-2) node[midway,right] {$a_1$};   
    \draw (nodo9-2)--(nodo10-2) node[midway,right] {$a_1$};   
    \draw (nodo9-3)--(nodo10-2) node[midway,right,above] {$a_1$};  
    \draw (nodo9-4)--(nodo10-1) node[midway,right, above] {$A_3$};
    \draw (nodo9-4)--(nodo10-3) node[midway,right] {$a_1$};

    \node[rectangle, draw] at (-1.5,-24.2) (nodo11-1) {$  3 \ \stackon{6}{\stackon{4}{[2]}}   \stackunder{5}{[1]}  3\ [1]$};
    \node[right=30pt] at (-1.5,-24.2) () {$\lambda_5=\big[4^2,4^0;2^1,1^1 \big]_5$};
    \node[rectangle, draw] at (4.75,-24.2) (nodo11-2) {$ [2] \ 4 \ \stackon{6}{3} \stackunder{5}{[1]}  3\ [1]$};
    \node[right=40pt] at (4.75,-24.2) () {$\lambda_5=\big[4^0,4^2;2^1,1^1 \big]_5$};
    \node[rectangle, draw] at (11,-24.2) (nodo11-3) {$[2] \ 4 \stackon{6}{\stackon{4}{[2]}} 4 \ 2 $};
    \node[right=40pt] at (11,-24.2) () {$\lambda_5=\big[4^2,4^2;0 \big]_5$};
    \draw (nodo9-2)--(nodo11-1) node[midway,right] {$a_2$};
    \draw (nodo9-3)--(nodo11-2) node[midway,right] {$a_2$};  
    \draw (nodo10-3)--(nodo11-2) node[midway,right,below] {$A_2$};
    \draw (nodo10-3)--(nodo11-1) node[midway,above left] {$A_2$};
    \draw (nodo10-3)--(nodo11-3) node[midway,right] {$A_3$};

    \node[rectangle, draw] at (-4,-28) (nodo12-1) {$3  \stackunder{\stackon{6}{3}}{[1]} \stackunder{5}{[1]}  3 \ [1]$};
    \node[left=40pt] at (-4,-28) () {$\lambda_5=\big[3^1,2^1,1 \big]_5$};
    \node[rectangle, draw] at (-0.9,-28) (nodo12-2) {$3 \stackunder{\stackon{6}{\stackon{4}{[2]}}}{[1]} 4 \ 2 $};
    \node[right=40pt] at (-1.5,-28) () {$\lambda_5=\big[4^2,0;3^1 \big]_5$};
    \node[rectangle, draw] at (4.75,-28) (nodo12-3) {$[2] \ 4 \stackunder{\stackon{6}{3}}{[1]} 4 \ 2 $};
    \node[right=40pt] at (4.75,-28) () {$\lambda_5=\big[4^0,4^2;3^1 \big]_5$};
    \draw (nodo10-1)[red]--(nodo12-1) node[midway,right] {$a_2$};
    \draw (nodo10-2)[red]--(nodo12-1) node[midway,left] {$a_2$}; 
    \draw (nodo11-1)--(nodo12-2) node[midway,right] {$A_2$};
    \draw (nodo11-2)--(nodo12-3) node[midway,right=7pt,below] {$A_2$};
    \draw (nodo11-3)--(nodo12-3) node[midway,below right] {$a_1$};
    \draw (nodo11-3)--(nodo12-2) node[midway,left=15pt] {$a_1$}; 
    \draw (nodo11-1)--(nodo12-1) node[midway,left] {$a_1$};
    \draw (nodo11-2)--(nodo12-1) node[midway,below right] {$a_1$};

    \node[rectangle, draw] at (-5,-30.2) (nodo13-1) {$[1] \ 3  \stackon{5}{\stackon{3}{[1]}} 4 \ 3 \ [2]$};
    \node[left=40pt] at (-5,-30.2) () {$\lambda_5=\big[4^1,4^1;1^2 \big]_5$};
    \node[rectangle, draw] at (-2.5,-30.2) (nodo13-2) {$3  \stackunder{\stackon{6}{3}}{[2]} 4 \ 2 $};
    \node[right=50pt] at (-3,-30.2) () {$\lambda_5=\big[3^2 \big]_5$};
    \draw (nodo12-1)[red]--(nodo13-1) node[midway,left] {$A_2$};
    \draw (nodo12-1)[red]--(nodo13-2) node[midway,right] {$A_2$}; 
    \draw (nodo12-2)--(nodo13-2) node[midway,right=5pt] {$a_1$};
    \draw (nodo12-3)--(nodo13-2) node[midway,below right] {$a_1$};

   \draw (nodo13-1)--(-5,-31.2) node[below] () {etc.};
   \draw (nodo13-2)--(-2.5,-31.2) node[below] () {etc.};
    
    \end{tikzpicture}
   }
    \caption{The (partial) Higgs branch Hasse diagram for the $D_5^{\su_8}$ SCFT. We have drawn the tensor branch curve configuration in each vertex. Blue transitions are used to identify slices in the nilpotent cone of $\mathfrak{su}(8)$, whereas red is used for the nilpotent cone of $\mathfrak{su}(6)$.}
    \label{fig:Hasse-D5-[8]}
\end{figure}

\clearpage

Having played this exercise with the $D_{5}^{\su_{4}}([1^4])$ model, we can apply the same philosophy to study more general long $D_{N}^{\su_{2k}}([1^{2k}])$ theories. The quiver subtraction algorithm correctly predicts the full nilpotent cone $\mathcal{N}(\su_{2\ell})$ for $1\le 2\ell\le 2k$ as a subset of the Hasse
diagram for each long theory. Moreover, nilpotent cones of different algebras are connected by $D_n$ slices, giving a nested inclusion relation. It is interesting to notice that the $D_N$ slice only appears once and at the bottom of the diagram where the supersymmetry enhancement appears. In Figure \ref{fig:Hasse-D_Nsu6_Long}, we explicitly considered the Higgs branch of the long $D_{N}^{\su_{6}}([1^{6}])$ theory, while taking care to highlight the different nilpotent cones appearing in the Hasse diagram, namely $\mathcal{N}(\su_{2}),\,\mathcal{N}(\su_{4}),\,\mathcal{N}(\su_{6})$, with different colors. The nested structure of the diagram is rendered explicit from the presence of D-slices interconnecting the various nilpotent subdiagrams, with the $D_N$ slice itself leading to the $(2,0)$ theory of type $D_N$.

The case of short theories can be understood as well with the same approach; in fact, the quiver subtraction algorithm, as well as the magnetic quiver for the Higgs branch, is not sensitive to the \emph{location} of the flavor symmetries in the tensor branch description. Therefore there is no qualitative distinction between the procedure for determination of the Higgs branch for long and short theories. What is different is the fact that the full nilpotent cone of the ultraviolet flavor algebra is no longer realized as a subset of the Hasse diagram; as highlighted in Figure \ref{fig:hasseeeee}. On the other hand, it is partially realized and still nested to other (partially realized) cones via D-type transverse slices. One further caveat is that there may also be additional theories between one partial nilpotent cone and another. Figures \ref{fig:Hasse-D5-[6]-partition} and \ref{fig:Hasse-D5-[8]} provide examples of Hasse diagrams for short theories, respectively $D_{5}^{\su_6}([1^6])$ and $D_{5}^{\su_8}([1^8])$. In these examples, it is clear to see that, in contrast to the long theories of Figure \ref{fig:Hasse-D_Nsu6_Long}, there are many possibilities for the RG flow that are not associated with elements of any nilpotent cone.

\subsection{An Observation on Higgsing from the Tensor Branch}\label{sec:slicesub}

We have derived the interacting non-product subdiagram of the Higgs branch Hasse diagram of any given 6d $(1,0)$ SCFT Higgsable to D-type 6d $(2,0)$ SCFTs, of the form in equation \eqref{eqn:TBDNdecorated}, via quiver subtraction on the associated magnetic quiver. For the (both long and short) theories $D_N^{\mathfrak{su}_{2k}}(O)$, which have a tensor branch description of the form in equation \eqref{eqn:TBDNdecorated}, it is straightforward to observe that whenever the tensor branch geometry takes a specific form, then there exists a Higgs branch RG flow to another SCFT where the transverse slice also takes a particular form. We elucidate this observation in this subsection.

There are precisely three kinds of transverse slices that appear in the Hasse diagram: $a_n$, $A_n$, and $D_n$. The first is the closure of the minimal nilpotent orbit of $\mathfrak{su}(n+1)$, and the latter two are orbifold singularities of AD-type, respectively. By studying the tensor branch geometries before and after such elementary transverse slices, we observe that they occur under the following circumstances.
\begin{itemize}
    \item An \textbf{$a_n$ transverse slice} exists whenever there is an $\mathfrak{su}(n+1 \geq 2)$ flavor algebra attached to a single curve supporting a non-trivial gauge algebra. After the transition, the rank of that gauge algebra is reduced by one, and the flavor algebras are fixed via anomaly cancellation. To illustrate this, we consider the following example:
    \begin{equation}\label{eqn:a_n-transition}
    5 \ \stackunder{\stackon{$10$}{$5$}}{[2]} \ 8 \ 6 \ [4] \Rightarrow \begin{cases}
        [1] \ 5 \ \stackon{$9$}{\stackon{$5$}{[1]}}  \stackunder{8}{[1]} \ 6 \ [4] & \text{Higgsing the $[2]$ flavor via an $a_1$ transition,} \\

        5 \ \stackunder{\stackon{$10$}{$5$}}{[2]} \stackunder{8}{[1]} \ 5 \ [2] & \text{Higgsing the $[4]$ flavor via an $a_3$ transition.}
    \end{cases}
\end{equation}
    That is, the Higgsable to $D_5$ tensor branch on the left has two elementary Higgsings where the transverse slice is of $a_n$ type, to the two tensor branch configurations on the right. We emphasize that we know that these two transitions exist via the construction of the magnetic quiver and the application of the quiver subtraction algorithm; however, a posteriori, we note that are observed to correspond to the presence of the $[2]$ and the $[4]$, respectively.
    \item An \textbf{$A_n$ transverse slice} occurs whenever there are two curves supporting non-trivial gauge algebras each with a single dangling fundamental flavor (that is, each has a $[1]$), and an inclusive linear chain of $n$ curves, without attached flavor algebras, between them. After the Higgsing, the ranks of the gauge algebras on each of the $n$ curves are reduced by one. Again, it is best to illustrate via an example. There exists the following elementary transition between the SCFTs associated with the tensor branches:
    \begin{align} \label{eqn:ADE-transition1}
        2  \stackunder{4}{[1]}  \stackon{7}{\stackon{4}{[1]}}  \stackunder{6}{[2]} \ 3   \qquad\xrightarrow{\,\,\,A_3\,\,\,}\qquad [1] \ 2 \ 3  \  \stackon{6}{3}  \stackunder{6}{[3]} \ 3 \,.
    \end{align}
    We can see there is a linear chain of three curves of the form
    \begin{equation}
        [1] \, 4 \, 7 \, 4 \, [1] \,,
    \end{equation}
    indicating the existence of a Higgsing to a new SCFT where the transverse slice in the foliation of the symplectic singularity is $A_3$. Turning to, for example, Figure \ref{fig:Hasse-D5-[8]}, we can see this structure realized in each instance of such a tensor branch configuration.
    \item Finally, we turn to the \textbf{$D_n$ transverse slice}. Such a transition occurs whenever we have a tensor branch configuration of the form
    \begin{equation}
        k_1 \overset{\displaystyle k_2}{k_3} k_4 \cdots \underset{\displaystyle [1]}{k_{n-1}} k_n \cdots \,,
    \end{equation}
    where $k_1, k_2, k_n \geq 1$ and $k_i \geq 2$ for $i = 3, \cdots n-1$. It is important to note that there are no dangling hypermultiplets attached to any of the depicted curves. After Higgsing, each of the $k_1, k_2, k_n$ is decreased by one, the $k_i$ for $i = 3, \cdots, n-1$ are each decreased by two, and the infrared flavor symmetry is determined from anomaly cancellation. An example is the following elementary Higgsing:
    \begin{align} 
     3 \stackunder{\stackon{6}{3}}{[1]} 5 \ 4  \ [3]  \qquad\xrightarrow{\,\,\,D_4\,\,\,}\qquad 2 \ \stackon{4}{2} \ 4 \ 4  \ [4] \,.
    \end{align}

    There is one special case of the $D_n$ transverse slice, which occurs at the bottom of the Hasse diagram, i.e., the final Higgsing before the $(2,0)$ supersymmetry enhancement occurs. This Higgsing is always between tensor branches of the form:
    \begin{equation}
        1 \overset{\displaystyle 1}{2} 2 \cdots \underset{\left[\frac{3}{2}\right]}{2} 1 \qquad \xrightarrow{\,\,\,D_N\,\,\,} \qquad 0 \overset{\displaystyle 0}{0} 0 \cdots 0 0 \,,
    \end{equation}
    where there are $N$ compact curves in the configuration. In this case, there is a $[3/2]$ instead of a $[1]$, but otherwise, the standard rule for the $D_n$ transverse slice applies.
\end{itemize}

From this experimental reverse engineering of the transverse slices in terms of the tensor branch configurations, we can produce a catalogue of ``slices'' that need to be subtracted from the tensor branch to generate the Higgs branch Hasse diagrams for $D_N^{\mathfrak{su}_{2k}}(O)$. To wit, we denote these tensor branch slices as
\begin{align}\label{eqn:slicesub}
    a_n &:= \ 1 - [n+1] \,, \\
    A_n &:= \ [1]- \underbrace{1 - \cdots -1}_{n \text{ curves}}-[1] \,,\\
    D_n &:= \ 1-\stackon{2}{\stackon{$|$}{1}}-\underbrace{2 - \cdots - 2}_{n-5 \text{ curves}} - \stackunder{2}{\stackunder{$|$}{[1]}} -1  \,.
\end{align}
In particular, if the tensor branch configuration contains one of these subgraphs, then there exists an elementary Higgs branch RG flow, with transverse slice as given, to a new SCFT where the ranks of the gauge algebras on the tensor branch are obtained via subtracting the gauge ranks of the subgraph. We refer to this as performing ``slice subtraction'' directly on the 6d tensor branch. 

Working out the Higgsing pattern for $D^{\su_{2k}}_{N}\left( O \right)$ theories and drawing the associated Hasse diagram is a straightforward computation with the slice subtraction algorithm, and it can be checked from the previously drawn Hasse diagrams in Figure \ref{fig:Hasse-D_Nsu6_Long} for long theories and in Figures \ref{fig:Hasse-D5-[6]-partition} and \ref{fig:Hasse-D5-[8]} for short theories. 

We emphasize once again that this is an experimental observation, based on the transverse slices as worked out from the quiver subtraction algorithm as applied to the magnetic quiver for the Higgs branch. The slice subtraction pattern holds for the $D^{\su_{2k}}_{N}\left( O \right)$ SCFTs, but it does not a priori hold beyond that regime. In the case of the family of the 6d $(1,0)$ SCFTs known as conformal matter, there is a similar set of slice subtraction rules, which can be understood microscopically in terms of giving vacuum expectation values to specific Higgs branch chiral ring operators directly in 6d. For conformal matter, this analysis appears in \cite{DKL}, and the generalization to the microscopics of the slice subtraction algorithm for Higgsable to D-type $(2,0)$ SCFTs can be determined. 

\section{Geometric Approach to Product Higgsing}\label{sec:product}

In the previous section, one of the approaches to understanding the Higgs branch of $D_N^{\mathfrak{su}_{2k}}$ was via a magnetic quiver for the Higgs branch. That is, we applied the quiver subtraction algorithm to the magnetic quiver for the Higgs branch worked out in Section \ref{sec:Magnetic_quiver}. Such a procedure allows us to extract both the symplectic leaves and the transverse slices between them; however, due to the incompleteness of the quiver subtraction algorithm for the unitary-orthosymplectic quivers that we are considering (see, e.g., equation \eqref{quiver:Magnetic_Dn_inf}), we do not see the structure of the full Higgs branch in this way, only the leaves/slices corresponding to interacting non-product 6d $(1,0)$ SCFTs. In particular, the subtlety arises from the $\mathfrak{so}(2N)$ gauge node with the antisymmetric hypermultiplet. 

We can partially overcome this limitation if instead of the structure of the Higgs branch, i.e., knowing both the leaves and the transverse slices between them, we only focus on the possible Higgsing results, i.e., only the leaves. The geometry itself automatically encodes the Higgsing pattern of an SCFT, since the former is realized as complex structure deformation of the elliptically fibered Calabi-Yau threefold $Y$ on which F-theory is compactified. Thus we can study such complex structure deformations to extract the Hasse diagram of a 6d SCFT. One drawback of this approach is that the information about the transverse slice corresponding to any particular Higgsing is obscured.

The second drawback is that, given a non-compact elliptically-fibered Calabi--Yau threefold engineering a 6d $(1,0)$ SCFT, it is generally challenging to study the space of complex structure deformations. In particular, it is not straightforward to explore the complex structure moduli space of the geometry engineering the tensor branch SQFT, which has the advantage of a smooth base and only minimal singularities in the fiber, to learn about the complex structure deformations of the SCFT geometry. Therefore, in this section, we review the complex structure deformation approach as applied to the geometries engineering the 6d $(2,0)$ SCFTs. From the magnetic quiver of those theories that are Higgsable to 6d $(2,0)$ SCFTs of type $\Gamma$, we have seen that the ``end'' of the Hasse diagram corresponds to the 6d $(2,0)$ theory of type $\Gamma$ itself. In fact, these $(2,0)$ theories themselves possess a non-trivial Higgs branch, which must be further studied; this continuation of the Higgs branch Hasse diagram is what we explore here. Recall that the $(2,0)$ theories are realized geometrically as a trivial fibration over a base space of the form $\mathbb{C}^2/\Gamma$, with $\Gamma \subset SU(2)$ a finite subgroup.

\begin{table}[t]
    \centering
    \begin{threeparttable}
    \begin{tabular}{cc}
    \toprule
    $\Gamma$ & Possible Deformations  \\ \midrule
     $A_n$ & $A_{n-k}\times A_{k-1}$ \\
     $D_n$ & $D_{n-k}\times A_{k-1}$ \\
    $E_6$ & $D_{5}, A_{5}, A_{4} \times A_{1},A_{2} \times A_{2} \times A_{1}$\\
    $E_7$ & $E_{6}, D_{6}, A_{6}, D_{5} \times A_{1}, A_{5} \times A_{1}, A_{4} \times A_{2}, A_{3} \times A_{2} \times A_{1}$\\
     $E_8$ & $E_{7}, D_{7}, A_{7}, E_{6} \times A_1, A_{6} \times A_{1}, D_{5} \times A_{2}, A_{4} \times A_{3}, A_{4} \times A_{2} \times A_{1}$ \\ \bottomrule
    \end{tabular} 
    \end{threeparttable}
    \caption{Complex structure deformations of a $\mathbb{C}^2/\Gamma$ singularity. Such deformations can be applied recursively. Here, we implicitly allowed the trivial $A_0$ and $D_0$ cases.}\label{tbl:deletions}
\end{table}

The deformation space of a $\mathbb{C}^2/\Gamma$ orbifold singularity was worked out in \cite{katz1992gorenstein}, following \cite{MR715649,MR597833,MR422270,MR713260}, amongst others. We first consider a simple example: the $A_{N-1} = \mathbb{C}^2/\mathbb{Z}_N$ orbifold can be written as the zero-locus of the polynomial
\begin{equation}
    uv = z^N \,,
\end{equation}
in $\mathbb{C}^3$, where the singular point is at the origin. There exists a complex structure deformation which modifies this hypersurface equation as follows:
\begin{equation}
    uv = z^N \quad \longrightarrow \quad uv = (z - t_1)^{N_1}(z - t_2)^{N_2} \,,
\end{equation}
where $N_1 + N_2 = N$. Now, there is locally a $\mathbb{C}^2/\mathbb{Z}_{N_1}$ singularity at $u = v = z - t_1 = 0$ and a $\mathbb{C}^2/\mathbb{Z}_{N_2}$ singularity at $u = v = z - t_2 = 0$. The space of all such deformations, together with the foliation structure of the symplectic singularity, is simply $\operatorname{Sym}^{N}(\mathbb{C}^2)$, as discussed previously in \cite{Bourget:2022ehw}. The general result can be summarized as follows. By abuse of notation, let $\Gamma$ denote the Dynkin diagram of the Lie algebra associated with the finite group $\Gamma$ via the McKay correspondence. Let $\Gamma_0$ denote a subgraph of $\Gamma$. The graph
\begin{equation}
    \Gamma' = \Gamma - \Gamma_0 \,\,=\,\, \bigsqcup_{i=1}^n \, \Gamma_i' \,,
\end{equation}
consists of a disjoint union of Dynkin diagrams $\Gamma_i'$. There exists a deformation of $\mathbb{C}^2/\Gamma$ to a space with $n$ isolated singular points locally of the form
\begin{equation}
    \mathbb{C}^2/\Gamma_i' \,,
\end{equation}
respectively. Again, we have abused notation and used $\Gamma_i'$ to denote both the Dynkin diagrams and the associated finite subgroups of $SU(2)$.

 \begin{figure}[!b]
     \centering
     \resizebox{0.98\textwidth}{!}{
     \begin{tikzpicture}
         \node[] at (0,0) (0) {\footnotesize {$\varnothing$}};
         \node[] [below=1cm of 0] (2) {\footnotesize {$2$}};
         \node[] [below right=1cm and 1cm of 2] (22) {\footnotesize {$22$}};
         \node[] [below left=1cm and 1cm of 2] (2u2) {\footnotesize {$2 \sqcup 2$}};
         \node[] [below right=1cm and 1cm of 22] (222) {\footnotesize {$222$}};
         \node[] [below left=1cm and 1cm of 22] (22u2) {\footnotesize {$22 \sqcup 2$}};
         \node[] [below left=1cm and 1cm of 2u2] (2u2u2) {\footnotesize {$2\sqcup 2 \sqcup 2$}};
        \node[] [below right=1cm and 1cm of 222] (2^222) {\footnotesize {$2\stackon{$2$}{$2$}2$}};
         \node[] [below left=1cm and 1cm of 222] (2222) {\footnotesize {$2222$}};
         \node[] [below left=1cm and 1cm of 22u2] (222u2) {\footnotesize {$222 \sqcup 2$}};
        \node[] [below left=1cm and 1cm of 2u2u2] (22u2u2) {\footnotesize {$22\sqcup 2 \sqcup 2$}};
        \node[] [below left=1cm and 1cm of 2222] (2^2222) {\footnotesize {$2\stackon{$2$}{$2$}22$}};

        \draw (0)--(2) node[midway,right] {\footnotesize $A_1$};
        \draw (2)--(2u2) node[midway,above left] {\footnotesize $A_1$};
        \draw[blue] (2)--(22) node[midway,above right] {\footnotesize $m$};
        \draw (2u2)--(2u2u2) node[midway,above left] {\footnotesize $A_1$};
        \draw[blue] (22)--(222) node[midway,above right] {\footnotesize $m$};
        \draw[blue] (2u2)--(22u2) node[midway,above right] {\footnotesize $m$};
        \draw (22)--(22u2) node[midway,above left] {\footnotesize $A_1$};
        \draw[blue] (2u2u2)--(222u2) node[midway,above right] {\footnotesize $m$};
        \draw[blue] (22u2)--(2222) node[midway,above right] {\footnotesize $m$};
        \draw[blue] (222)--(2222) node[midway,above left] {\footnotesize $m$};
        \draw[blue] (2u2u2)--(22u2u2) node[midway,above left] {\footnotesize $m$};
        \draw (22u2)--(222u2) node[midway,above left] {\footnotesize $A_1$};
        \draw[dashed] (222)--(2^222);
        \draw[dashed] (22u2u2)--(2^2222);
        \draw[dashed] (222u2)--(2^2222);
        \draw[dashed] (2222)--(2^2222);
        \draw[dashed] (2^222)--(2^2222);
     \end{tikzpicture}}
     \caption{Hasse diagram of the 6d $\mathcal{N}=(2,0)$ SCFT of type $D_5$. Known slices are labeled and colored according to their nature (black for $A_1$ slices and blue for $m$ ones), whereas unknown slices are depicted with dashed lines.}
     \label{fig:Hasse-D5(2,0)}
 \end{figure}
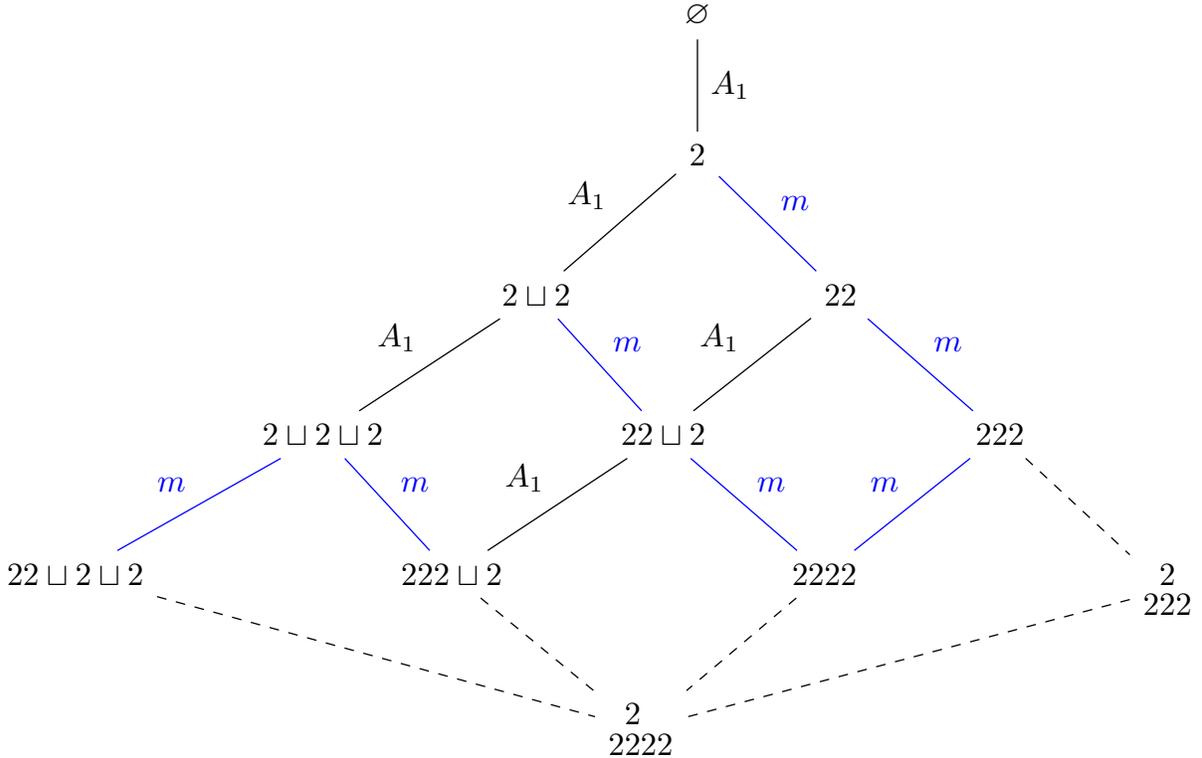

It is clear that a partial ordering exists on such deformations given via inclusion of subgraphs. Let $\Gamma'$ denote the (possibly disconnected) Dynkin diagram after an arbitrary number of deformations, then the elementary transitions under this partial ordering occur when the subgraph removed from $\Gamma'$ consists only of a single vertex, that is, removing an $A_1$ Dynkin diagram. Thus, the Hasse diagram of deformations of $\mathbb{C}^2/\Gamma$ under this partial ordering can be straightforwardly determined once the single vertex deletions of each simple Dynkin diagram are known. For convenience, we list these in Table \ref{tbl:deletions}.\footnote{We thank Amihay Hanany for pointing out that this procedure determines the Levi subgroups of the algebra $\mathfrak{g}$ associated with $\Gamma$.} This provides the symplectic leaves of the Hasse diagram for the Higgs branch of the 6d $(2,0)$ SCFTs of type $\Gamma$, however, it does not, in general, provide the data of the transverse slice, even though it is easy to see that each of these slices is one-dimensional. 

\vspace{1.5cm}

\begin{landscape}
\pagestyle{empty}
    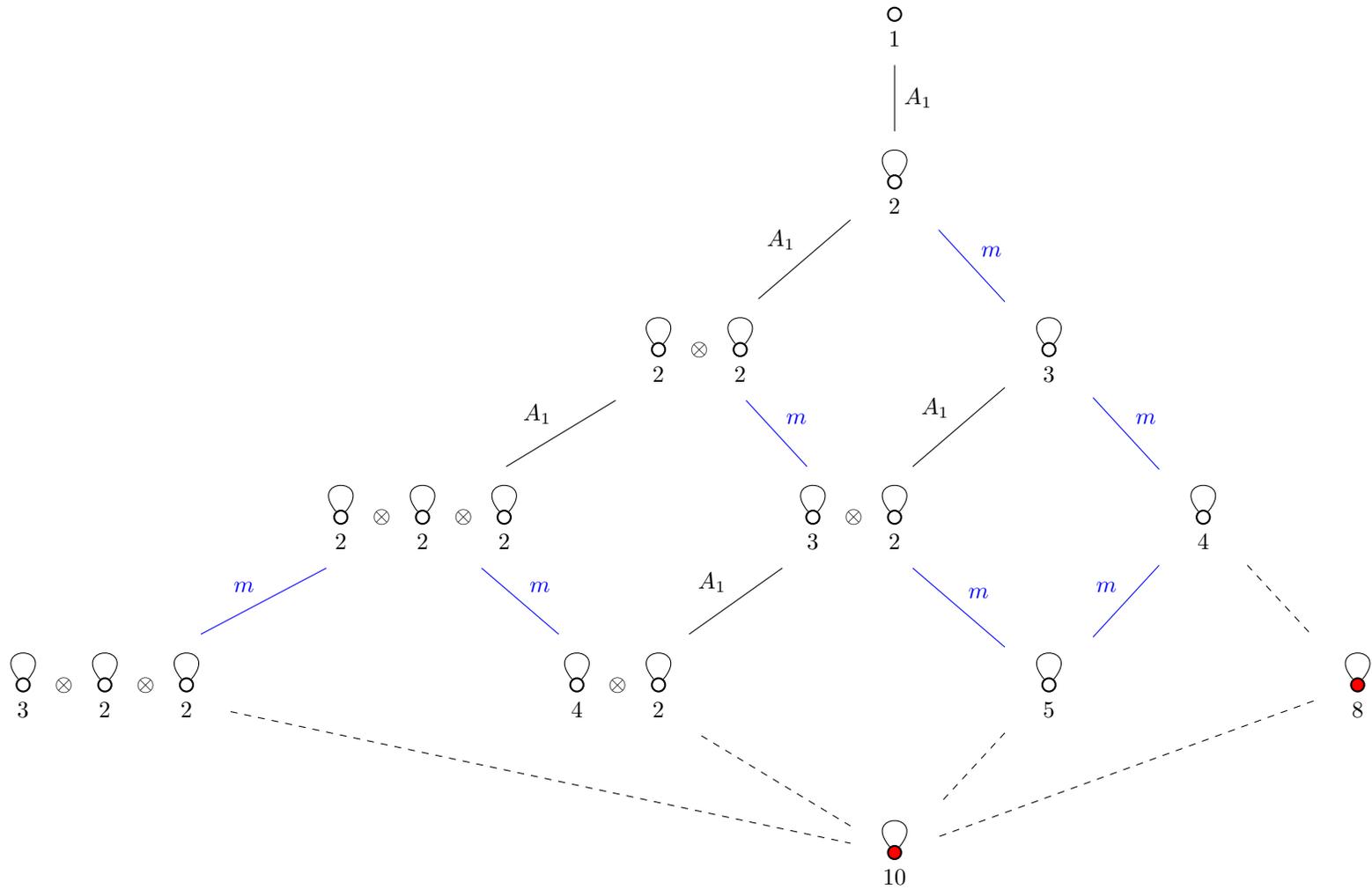
\begin{figure}[t]
     \centering
     \begin{tikzpicture}
         \node[] at (0,0) (0) {\begin{tikzpicture}
             \node[node,draw, label=below:{\footnotesize $1$}] at (0,0) {};
         \end{tikzpicture}};
         \node[] [below=1cm of 0] (2) {\begin{tikzpicture}
             \node[node,draw, label=below:{\footnotesize $2$}] at (0,0) (2a) {};
             \draw (2a) to[out=130, in=410, looseness=12] (2a);
         \end{tikzpicture}};
         \node[] [below right=1cm and 1cm of 2] (22) {\begin{tikzpicture}
             \node[node,draw, label=below:{\footnotesize $3$}] at (0,0) (2a) {};
             \draw (2a) to[out=130, in=410, looseness=12] (2a);
         \end{tikzpicture}};
         \node[] [below left=1cm and 1cm of 2] (2u2) {\begin{tikzpicture}
             \node[node,draw, label=below:{\footnotesize $2$}] at (0,0) (2a) {};
             \node[tnode] [right=0.2cm of 2a] (oplus) {\footnotesize{$\otimes$}};
             \node[node,draw, label=below:{\footnotesize $2$}] [right=0.2cm of oplus] (2b) {};
             \draw (2a) to[out=130, in=410, looseness=12] (2a);
             \draw (2b) to[out=130, in=410, looseness=12] (2b);
         \end{tikzpicture}};
         \node[] [below right=1cm and 1cm of 22] (222) {\begin{tikzpicture}
             \node[node,draw, label=below:{\footnotesize $4$}] at (0,0) (2a) {};
             \draw (2a) to[out=130, in=410, looseness=12] (2a);
         \end{tikzpicture}};
         \node[] [below left=1cm and 1cm of 22] (22u2) {\begin{tikzpicture}
             \node[node,draw, label=below:{\footnotesize $3$}] at (0,0) (22a) {};
             \node[tnode] [right=0.2cm of 22a] (oplus) {\footnotesize{$\otimes$}};
             \node[node,draw, label=below:{\footnotesize $2$}] [right=0.2cm of oplus] (2b) {};
             \draw (22a) to[out=130, in=410, looseness=12] (22a);
             \draw (2b) to[out=130, in=410, looseness=12] (2b);
         \end{tikzpicture}};
         \node[] [below left=1cm and 1cm of 2u2] (2u2u2) {\begin{tikzpicture}
             \node[node,draw, label=below:{\footnotesize $2$}] at (0,0) (2a) {};
             \node[tnode] [right=0.2cm of 2a] (oplus) {\footnotesize{$\otimes$}};
             \node[node,draw, label=below:{\footnotesize $2$}] [right=0.2cm of oplus] (2b) {};
             \node[tnode] [right=0.2cm of 2b] (oplus1) {\footnotesize{$\otimes$}};
             \node[node,draw, label=below:{\footnotesize $2$}] [right=0.2cm of oplus1] (2c) {};
             \draw (2a) to[out=130, in=410, looseness=12] (2a);
             \draw (2b) to[out=130, in=410, looseness=12] (2b);
             \draw (2c) to[out=130, in=410, looseness=12] (2c);
         \end{tikzpicture}};
        \node[] [below right=1cm and 1cm of 222] (2^222) {\begin{tikzpicture}
             \node[node,draw,fill=red, label=below:{\footnotesize $8$}] at (0,0) (2a) {};
             \draw (2a) to[out=130, in=410, looseness=12] (2a);
         \end{tikzpicture}};
         \node[] [below left=1cm and 1cm of 222] (2222) {\begin{tikzpicture}
             \node[node,draw, label=below:{\footnotesize $5$}] at (0,0) (2a) {};
             \draw (2a) to[out=130, in=410, looseness=12] (2a);
         \end{tikzpicture}};
         \node[] [below left=1cm and 1cm of 22u2] (222u2) {\begin{tikzpicture}
             \node[node,draw, label=below:{\footnotesize $4$}] at (0,0) (222a) {};
             \node[tnode] [right=0.2cm of 222a] (oplus) {\footnotesize{$\otimes$}};
             \node[node,draw, label=below:{\footnotesize $2$}] [right=0.2cm of oplus] (2b) {};
             \draw (222a) to[out=130, in=410, looseness=12] (222a);
             \draw (2b) to[out=130, in=410, looseness=12] (2b);
         \end{tikzpicture}};
        \node[] [below left=1cm and 1cm of 2u2u2] (22u2u2) {\begin{tikzpicture}
             \node[node,draw, label=below:{\footnotesize $3$}] at (0,0) (22a) {};
             \node[tnode] [right=0.2cm of 22a] (oplus) {\footnotesize{$\otimes$}};
             \node[node,draw, label=below:{\footnotesize $2$}] [right=0.2cm of oplus] (2b) {};
             \node[tnode] [right=0.2cm of 2b] (oplus1) {\footnotesize{$\otimes$}};
             \node[node,draw, label=below:{\footnotesize $2$}] [right=0.2cm of oplus1] (2c) {};
             \draw (22a) to[out=130, in=410, looseness=12] (22a);
             \draw (2b) to[out=130, in=410, looseness=12] (2b);
             \draw (2c) to[out=130, in=410, looseness=12] (2c);
         \end{tikzpicture}};
        \node[] [below left=1cm and 1cm of 2222] (2^2222) {\begin{tikzpicture}
           \node[node,draw,fill=red, label=below:{\footnotesize $10$}] at (0,0) (q) {};
           \draw (q) to[out=130, in=410, looseness=12] (q);
        \end{tikzpicture}};

        \draw (0)--(2) node[midway,right] {\footnotesize $A_1$};
        \draw (2)--(2u2) node[midway,above left] {\footnotesize $A_1$};
        \draw[blue] (2)--(22) node[midway,above right] {\footnotesize $m$};
        \draw (2u2)--(2u2u2) node[midway,above left] {\footnotesize $A_1$};
        \draw[blue] (22)--(222) node[midway,above right] {\footnotesize $m$};
        \draw[blue] (2u2)--(22u2) node[midway,above right] {\footnotesize $m$};
        \draw (22)--(22u2) node[midway,above left] {\footnotesize $A_1$};
        \draw[blue] (2u2u2)--(222u2) node[midway,above right] {\footnotesize $m$};
        \draw[blue] (22u2)--(2222) node[midway,above right] {\footnotesize $m$};
        \draw[blue] (222)--(2222) node[midway,above left] {\footnotesize $m$};
        \draw[blue] (2u2u2)--(22u2u2) node[midway,above left] {\footnotesize $m$};
        \draw (22u2)--(222u2) node[midway,above left] {\footnotesize $A_1$};
        \draw[dashed] (222)--(2^222);
        \draw[dashed] (22u2u2)--(2^2222);
        \draw[dashed] (222u2)--(2^2222);
        \draw[dashed] (2222)--(2^2222);
        \draw[dashed] (2^222)--(2^2222);
     \end{tikzpicture}
     \caption{Hasse diagram of the 6d $\mathcal{N}=(2,0)$ SCFT of type $D_5$ with the magnetic quiver for the Higgs branch of the theory at each vertex depicted. Known slices are labeled and colored according to their nature (black for $A_1$ slices and blue for $m$), whereas unknown slices are depicted via dashed lines.}
     \label{fig:Hasse-D5(2,0)-MQ}
 \end{figure}
\end{landscape}

In the first case in Table \ref{tbl:deletions}, where an A-type orbifold splits into two, we do have an understanding of the transverse slice \cite{Bourget:2022ehw}. The slice is $A_1$ if the orbifold splits into two copies of the same Dynkin diagram, otherwise it is $m$:\footnote{For the definition of the one-dimensional $m$ slice, see \cite{MR3570131}.}
\begin{equation}
    \begin{aligned}
     &A_{2k+1} \quad & &\xrightarrow{\,\,A_1\,\,} \quad & &A_k \times A_k \,, \\
    &A_{k} \quad & &\xrightarrow{\,\,m\,\,} \quad & &A_{k-1-p} \times A_p \,,\quad \quad p\ge0  \,.
    \end{aligned}
\end{equation}
While it is unknown for the generic case, the geometry allows us to draw Hasse diagrams, without the slice data, for cases where brane constructions are not viable or only partially understood. As an example, we have depicted the Hasse diagram of the complex structure deformations of the $\mathbb{C}^2/D_5$ singularity in Figure \ref{fig:Hasse-D5(2,0)}; we have labelled each vertex by the resolution of the deformed singularity. This is then the Hasse diagram for the Higgs branch of the 6d $(2,0)$ SCFT of type $D_5$. Even more, since we have a brane construction from each of the theories appearing in the $D_5$-type Hasse diagram of Figure \ref{fig:Hasse-D5(2,0)}, and in general for every theory in the $D_n$-type Hasse diagram, we can associate a magnetic quiver to each of the theories as done in Figure \ref{fig:Hasse-D5(2,0)-MQ} and take it as a future starting point for a more systematic understanding of subtraction in the context of orthosymplectic quivers.

\section{Higgs Branch of Higgsable to E-type \texorpdfstring{\boldmath{$(2,0)$}}{(2,0)} SCFTs}\label{sec:Etype}

In addition to the Higgsable to D-type $(2,0)$ SCFTs, $D_N^{\mathfrak{su}_{2k}}(O)$, in Section \ref{sec:thetheories} we also introduced the SCFTs $E_N(k_1, \cdots, k_N)$ which have supersymmetry-enhancing Higgs branch RG flows to the E-type $(2,0)$ SCFTs. While we could determine the (interacting, non-product subdiagram of the) Higgs branch Hasse diagram from the geometric perspective, following equation \eqref{eqn:EPO}, there is no known brane description of these theories and thus no magnetic quiver for the Higgs branches. As a result, we lack a method to learn about the structure of the Higgs branch as a foliation of a symplectic singularity; this is, we are missing the information on the transverse slices. Specifically, given an elementary Higgsing:
\begin{equation}
    E_N(k_1, \cdots, k_N) \quad \xrightarrow{\,\,\,S\,\,\,} \quad E_N(k_1', \cdots, k_N') \,, 
\end{equation}
what can we learn about the transverse slice $S$?

In Section \ref{sec:slicesub}, we proposed an empirical ``slice subtraction algorithm'' that allowed for the determination of the transverse slice directly via studying the structure of the tensor branch field theory before Higgsing. This algorithm is conjectural and based on the matching with the transverse slices as determined from the magnetic quiver for the $D_N^{\mathfrak{su}_{2k}}(O)$ SCFTs. However, we may suppose that the algorithm applies more broadly, in particular to the $E_N(k_1, \cdots, k_N)$ SCFTs, and test whether it provides a transverse slice consistent with the known properties of the 6d $(1,0)$ theories before and after the flow.

The $a_n$ slice subtraction involves reducing the rank of a single gauge algebra by one. It is straightforward to check the change in the Higgs branch dimension using equation \eqref{eqn:dimHE}. Let $i$ denote the index of the compact curve to which the $\mathfrak{su}(n+1)$ flavor algebra is attached, and we find that the difference in the Higgs branch dimension is:
\begin{equation}
    \delta \operatorname{dim}(\mathcal{H}) = \sum_{j=1}^N A^{ij}k_j - \frac{1}{2} A^{ii} = m_i - 1 = n \,.
\end{equation}
This is the dimension of the closure of the minimal nilpotent orbit of $\mathfrak{su}(n+1)$, as expected for an $a_n$ transverse slice. Similar arguments can be made to show that the change in the Higgs branch dimension whenever there is an $A_n$ or $D_n$ transition according to equation \eqref{eqn:slicesub} is one. This provides a consistency check that the transverse slices that we have associated via slice subtraction are consistent with the expectation from the Higgs branch dimension.

\begin{figure}[p]
    \centering
    \resizebox{!}{\textheight-3.5cm}{
    \begin{tikzpicture}
    
    \node[rectangle, draw] at (0,0) (nodo1) {$[1] \ 6 \ 11 \  \stackon{16}{8} \ 13 \ 10 \ 7 \ 4 \ [1]$ , $\text{dim} \ \mathcal{H}=13$};
    \node[rectangle, draw] at (0,-4) (nodo2a) {$5 \ 10 \  \stackon{15}{\stackon{8}{[1]}} \ 12 \ 9 \ 6 \ 3$, $\text{dim} \ \mathcal{H}=12$};
    \node[rectangle, draw] at (8,-4) (nodo2b) {$4 \ 8 \  \stackon{12}{6} \ 10 \ 8 \ 6 \ 4 \ [2]$, $\text{dim} \ \mathcal{H}=12$};
    \draw[blue] (nodo1)--(nodo2a) node[midway,right] {$A_7$};

    \node[rectangle, draw] at (0,-8) (nodo3) {$4 \ 8 \  \stackon{12}{6} \ 10 \ 8 \stackunder{6}{[1]}  3 $, $\text{dim} \ \mathcal{H}=11$};
    \draw[red] (nodo2a)--(nodo3) node[midway,right] {$E_6$};
    \draw[blue] (nodo2b)--(nodo3) node[midway,below right ] {$A_1$};

    \node[rectangle, draw] at (0,-12) (nodo4) {$[1] \ 4 \ 7 \  \stackon{10}{5} \ 8 \ 6 \ 4 \ 2 $, $\text{dim} \ \mathcal{H}=10$};
    \draw[blue] (nodo3)--(nodo4) node[midway,right] {$D_7$};

    \node[rectangle, draw] at (0,-16) (nodo5) {$ 2 \ 4 \  \stackon{6}{3} \ 4 \ 3 \ 2 \ [1] $, $\text{dim} \ \mathcal{H}=9$};
    \draw[red] (nodo4)--(nodo5) node[midway,right] {$E_7$};

    \node[rectangle, draw] at (0,-20) (nodo6) {$ 0 \ 0 \ \stackon{0}{0} \ 0 \ 0 \ 0 $, $\text{dim} \ \mathcal{H}=8$};
    \draw[red] (nodo5)--(nodo6) node[midway,right] {$E_8$};
 
    \end{tikzpicture}
    }
    \caption{The bottom of the (intersecting, non-product part of the) Hasse diagram for the $E_8(k_1, \cdots, k_8)$ SCFTs; we depict only configurations with $k_5 \leq 16$. At each vertex we write the tensor branch description of the SCFT as well as the dimension of the Higgs branch. Vertices are connected according to equation \eqref{eqn:EPO}, and the transverse slice labelling each link is conjectured from the slice subtraction algorithm in equations \eqref{eqn:slicesub} and \eqref{eqn:slicesub2}.}
    \label{fig:Hasse-E8}
\end{figure}
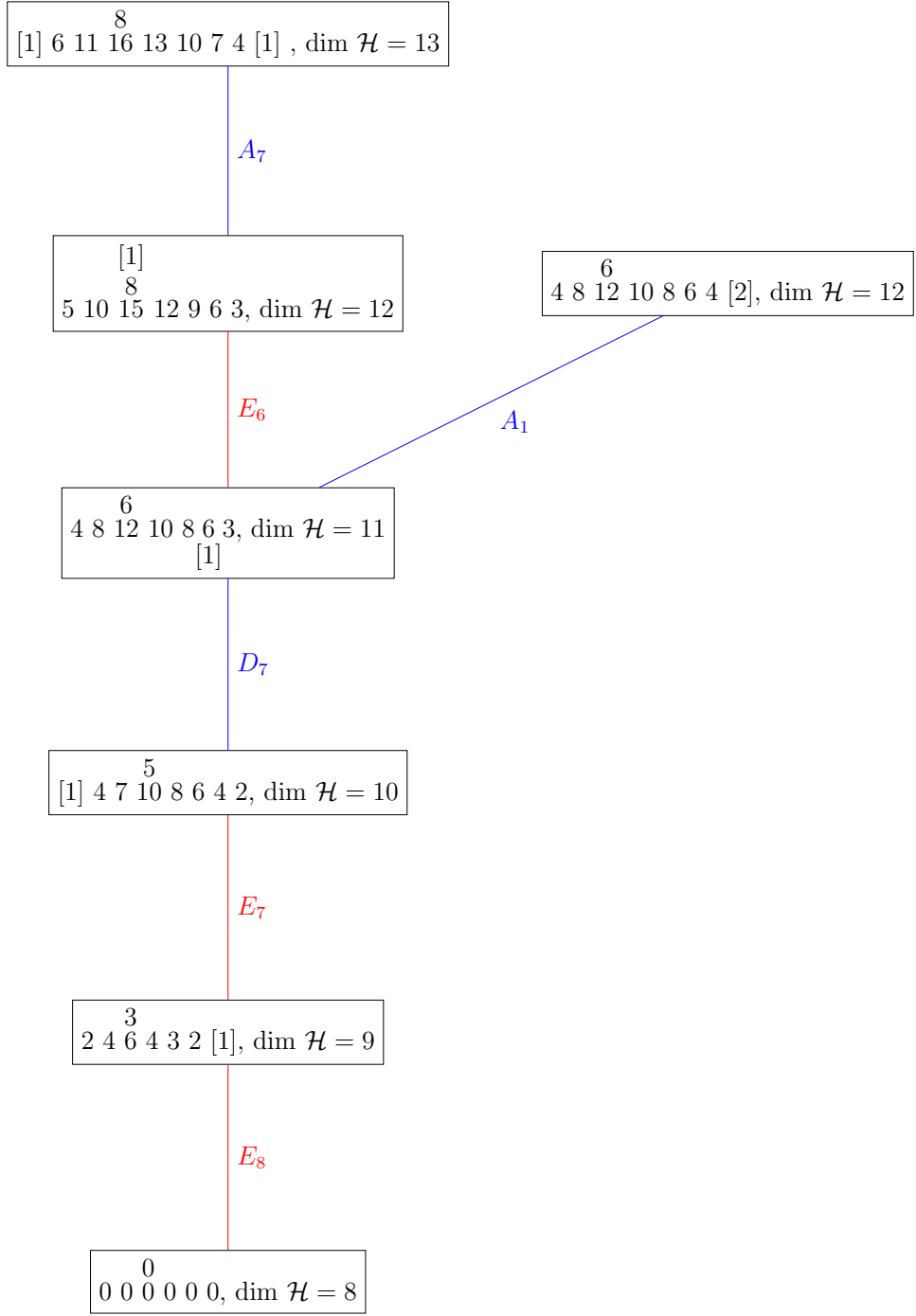

\begin{figure}[p]
    \centering
    \includegraphics[page=2,height=0.90\textheight]{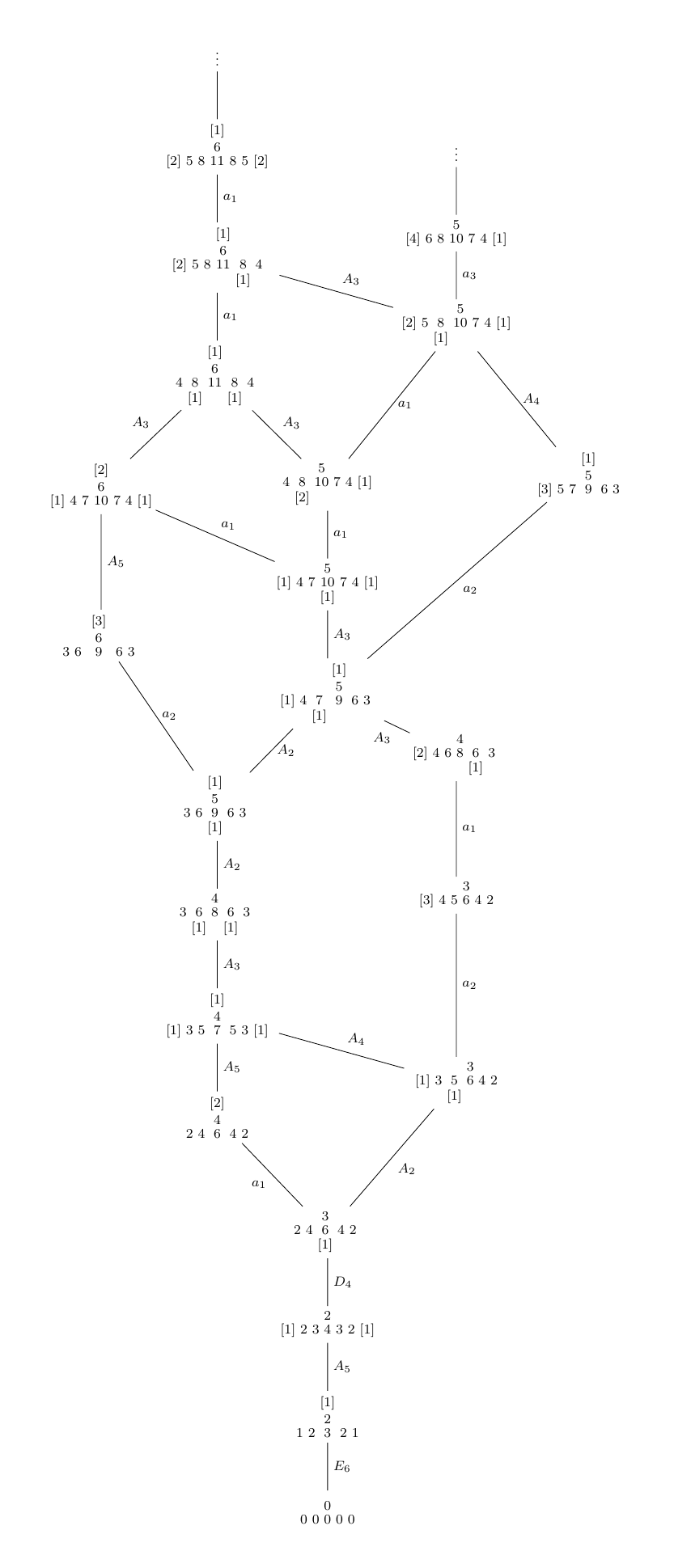}
    \caption{Portion of the bottom part of the Higgs branch Hasse diagram obtained via slice subtraction on a Higgsable to $E_6$ $(2,0)$ theory: $E_6(k_1, \cdots, k_6)$. This figure is the same as Figure \ref{fig:hasseE6}, but with the addition of the transverse slice labelling.}
    \label{fig:hasseE6take2}
\end{figure}

However, we can see that there are a small number of transitions in the Higgs branch Hasse diagram of the $E_N(k_1, \cdots, k_N)$ SCFTs where there is no transverse slice associated via the slice subtraction rules. In each of these cases, the transverse slices must be one-dimensional due to the known change in the Higgs branch dimension. These are best illustrated by the lower part of the $E_8(k_1, \cdots, k_8)$ Hasse diagram, which we depict in Figure \ref{fig:Hasse-E8}. Blue transitions denote elementary Higgsings where the transverse slice follows from the slice subtraction rules in equation \eqref{eqn:slicesub}, whereas the red transitions do not. However, these red transitions have a very evocative form; they appear to involve the subtraction of an $E_N$ type Dynkin diagram, weighted by the Dynkin labels, from the gauge ranks of the tensor branch configuration. This suggests that we should propose the following slice subtraction rules, in addition to those in equation \eqref{eqn:slicesub}:
\begin{equation}
    \begin{gathered}\label{eqn:slicesub2}
    E_6 :=
        \ 1-2-\stackon{3}{\stackon{$|$}{\stackon{2}{\stackon{$|$}{[1]}}}}-2 -1 \,, \qquad 
    E_7 := \ [1]-2-3-\stackon{4}{\stackon{$|$}{2}}-3-2-1 \,, \\[0.3cm]
    E_8 := \ 2-4-\stackon{6}{\stackon{$|$}{3}}-5-4-3-2-[1] \,.
\end{gathered}
\end{equation}

Interestingly, the slice subtraction rules in equations \eqref{eqn:slicesub} and \eqref{eqn:slicesub2} are sufficient to provide a transverse slice, with the correct change in the Higgs branch dimension, to each link in the Hasse diagram of interacting non-product theories on the Higgs branch of $E_N(k_1, \cdots, k_N)$. For example, see Figure \ref{fig:hasseE6take2} where we have taken the Hasse diagram from the F-theory geometric engineering (for $N = 6$) that was drawn in Figure \ref{fig:hasseE6} and added the transverse slices obtained using the slice subtraction algorithm. Of course, the change in the Higgs branch dimension is a relatively crude invariant, and further study is necessary to verify that this algorithm is producing the correct transverse slices for this class of Higgsable to E-type $(2,0)$ SCFTs.

\begin{figure}[t!]
    \centering
    \begin{subfigure}[b]{0.4\textwidth}
        \includegraphics[page=5,width=\textwidth]{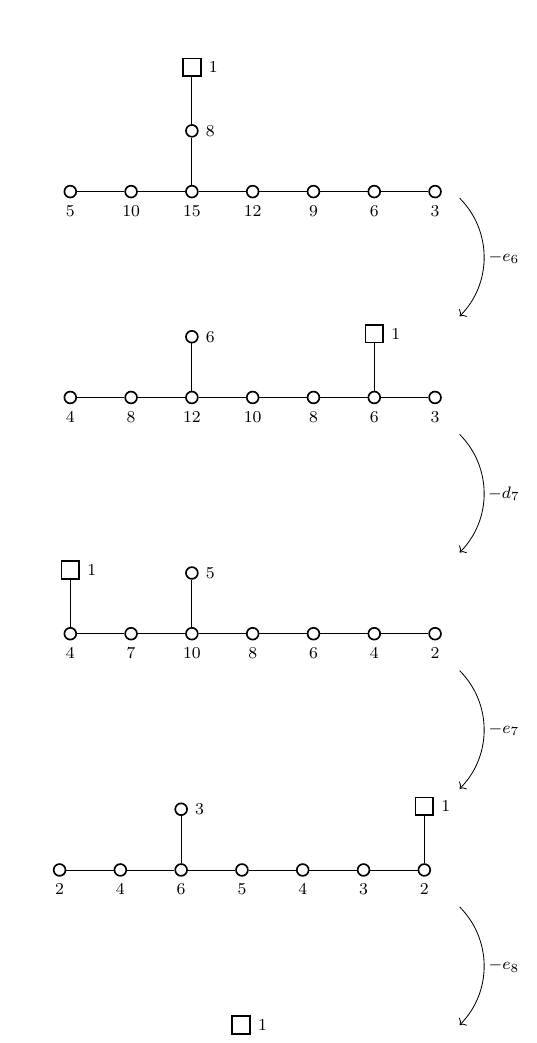}
    \end{subfigure}
        \begin{subfigure}[b]{0.4\textwidth}
        \includegraphics[page=6,width=\textwidth]{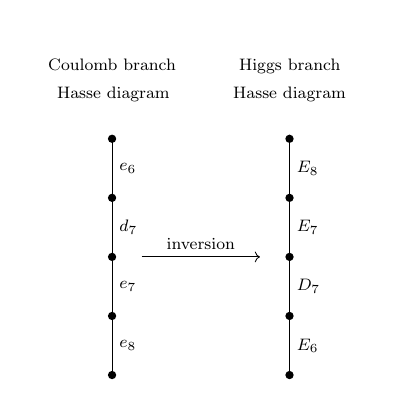}\vspace{3.5cm}
    \end{subfigure}
    \caption{Quiver subtraction technique applied to the 3d reduction of the tensor branch SQFT of the $E_8(5,10,15,12,9,6,3,8)$ SCFT together with its Coulomb branch Hasse diagram and its Higgs branch Hasse diagram derived via inversion. The Hasse diagrams are ordered according to the foliation induced by the underlying symplectic singularity, which is converse to the ordering used in other figures of this paper that decrease in Higgs branch dimension.}
    \label{fig:inversion}
\end{figure}

Further evidence for these assignments of transverse slices arises from the 3d reduction of the tensor branch SQFT, which we briefly summarize here. For an arbitrary $E_6(k_1, \cdots, k_6)$ SCFT, it is believed that the $T^3$ reduction of the tensor branch theory gives rise to the following unitary 3d $\mathcal{N}=4$ quiver:
\begin{equation}
    \underset{[m_1]}{k_1} \,  \underset{[m_2]}{k_2}  \overset{\displaystyle \overset{[m_6]}{k_6}}{\underset{[m_3]}{k_3}}  \underset{[m_4]}{k_4}  \underset{[m_5]}{k_5} \qquad\xrightarrow{\,\,\,T^3\,\,\,}\qquad \begin{gathered}
        \begin{tikzpicture}
            \node[node,draw,label=above:{\footnotesize $k_1$}] at (0,0) (1) {};
            \node[node,draw,label=above:{\footnotesize $k_2$}] [right=0.7cm of 1] (2) {};
            \node[node,draw,label={[xshift=-8pt,yshift=-20pt]\footnotesize $k_3$}] [right=0.7cm of 2] (3) {};
            \node[node,draw,label=above:{\footnotesize $k_4$}] [right=0.7cm of 3] (4) {};
            \node[node,draw,label=above:{\footnotesize $k_5$}] [right=0.7cm of 4] (5) {};
            \node[node,draw,label=right:{\footnotesize $k_6$}] [above=0.7cm of 3] (6) {};
            
            \node[fnode,draw,label=below:{\footnotesize $m_1$}] [below=0.5 of 1] (1f) {};
            \node[fnode,draw,label=below:{\footnotesize $m_2$}] [below=0.5 of 2] (2f) {};
            \node[fnode,draw,label=below:{\footnotesize $m_3$}] [below=0.5 of 3] (3f) {};
            \node[fnode,draw,label=below:{\footnotesize $m_4$}] [below=0.5 of 4] (4f) {};
            \node[fnode,draw,label=below:{\footnotesize $m_5$}] [below=0.5 of 5] (5f) {};
            \node[fnode,draw,label=above:{\footnotesize $m_6$}] [above=0.5 of 6] (6f) {};

            \draw (1)--(2)--(3)--(4)--(5);
            \draw (3f)--(3)--(6)--(6f);
            \draw (1)--(1f);
            \draw (2)--(2f);
            \draw (4)--(4f);
            \draw (5)--(5f);
        \end{tikzpicture}
    \end{gathered}\,,
\end{equation}
and similarly for the $N=7$ and $N=8$ cases. Since these 3d reductions have only unitary gauge/flavor nodes and bifundamental matter, we can apply the technique of \emph{inversion} \cite{Grimminger:2020dmg} to understand their Higgs branches,\footnote{More precisely, if the Coulomb branch Hasse diagram only has transverse slices which are closures of minimal nilpotent orbits or Kleinian singularities, then we can use inversion to derive the Higgs branch Hasse diagram.} which are isomorphic to the Higgs branches of the 6d tensor branch SQFTs. Inversion notes that the Coulomb branch Hasse diagram and the Higgs branch Hasse diagram are identical as graphs, and the only difference is the labelling of the edges via transverse slices. If a transition in the Coulomb branch is labelled by the closure of a minimal nilpotent orbit of a simple and simply-laced Lie algebra $\mathfrak{g}$, $\overline{\operatorname{min.} \mathfrak{g}}$, then the corresponding transition in the Higgs branch is labelled by the Kleinian singularity $\mathbb{C}^2/\Gamma_\mathfrak{g}$, where $\Gamma_\mathfrak{g}$ is the finite subgroup of $SU(2)$ related to $\mathfrak{g}$ via the McKay correspondence, and vice versa. To wit, 
\begin{equation}
        \overline{\operatorname{min.} \mathfrak{g}} \qquad \longleftrightarrow \qquad \mathbb{C}^2/\Gamma_\mathfrak{g} \,.
\end{equation}
Therefore, we can use quiver subtraction to determine the Coulomb branch Hasse diagram of the 3d reduction and use inversion to determine the transverse slices in the Higgs branch Hasse diagram. When performing this process, we note that the slices obtained in this way are the same as the slices obtained when following the slice subtraction algorithm. As an example, we carry this out in Figure \ref{fig:inversion} for a Higgsable to $E_8$-type theory, where the resulting Higgs branch Hasse diagram can be compared to that from slice subtraction in Figure \ref{fig:Hasse-E8}. While this is a statement about the Higgs branch of the tensor branch theories, as opposed to the SCFT Higgs branches, we generally (though not always) expect that operators of the Higgs branch chiral ring on the generic point of the tensor branch survive the contraction to the SCFT point, and thus can still trigger Higgsing.

We would also like to understand such a Higgsing microscopically in six dimensions. For example: when considering the $E_8$ transverse slice that appears in Figure \ref{fig:Hasse-E8}, it is natural to ask: what is the operator in the Higgs branch chiral ring of the UV 6d $(1,0)$ SCFT to which a vacuum expectation value is given to trigger this renormalization group flow? First, one can analyze the 1/2-BPS Higgs branch operators belonging to the tensor branch SQFT, and attempt to understand how the $E_8$ transition is triggered between the tensor branch SQFTs in the UV and the IR. Then, it is necessary to make an argument that the relevant operators survive in the SCFT limit where all of the compact curves are shrunk to zero volume. Some of these questions are answered in the conformal matter theories (or Higgsable to A-type $(2,0)$ SCFTs) in \cite{DKL}. We leave the general case for future work.

\section{6d \texorpdfstring{\boldmath{$(2,0)$}}{(1,0)} LSTs with SUSY-enhancing RG Flows}\label{sec:LSTs}

Thus far, we have considered 6d $(1,0)$ SCFTs which have supersymmetry-enhancing renormalization group flows. There exists a closely related class of six-dimensional theories known as little string theories (LSTs). These are non-gravitational theories, but they differ from their SCFTs cousins as they are non-local and have an intrinsic length scale $M_\text{string}$. While the UV-completion of such theories is not a standard quantum field theory, below the scale set by $M_\text{string}$, we can treat the theory as a QFT with a cutoff \cite{Aharony:1999ks,Seiberg:1997zk}. This non-local nature allows LSTs to experience T-duality;  in \cite{Lawrie:2023uiu} it was recognized that a robust feature for the exploration of the T-duality landscape of LSTs is the magnetic quiver of the Higgs branch.

As we have mentioned in Section \ref{sec:thetheories}, 6d $(1,0)$ LSTs have a similar geometric construction in F-theory to the 6d $(1,0)$ SCFTs  \cite{Bhardwaj:2015oru}. In this construction, the 6d $(2,0)$ LSTs arise when the curve configuration consists of an affine ADE Dynkin diagram formed out of $(-2)$-curves, with no reducible singular fibers supported over the compact curves. Thus, in this section, we consider 6d $(1,0)$ LSTs, and the magnetic quivers for their Higgs branches, which have Higgs branch renormalization group flows that evince SUSY-enhancement. This class of 6d $(1,0)$ LSTs has been considered, for example, in \cite{Bhardwaj:2015oru,Hohenegger:2016eqy,Hohenegger:2016yuv,Kim:2017xan,Bastian:2017ing,Bastian:2017ary,Haghighat:2018dwe,Wei:2022hjx,Filoche:2023yfm}. 

First, we revisit the 6d $(1,0)$ LSTs that are Higgsable to the 6d $(2,0)$ LSTs of A-type. These theories consist of a ring of conformal matter theories fused together. The LSTs that result from fusing $N$ copies of $(\mathfrak{g}, \mathfrak{g})$ conformal matter, where $\mathfrak{g}$ is an ADE Lie algebra, in this way are denoted $\widehat{A}_{N-1}^\mathfrak{g}$. We have written their tensor branch curve configurations in Table \ref{tbl:someLSTs},\footnote{A note on the notation of \cite{Bhardwaj:2015oru}, when we write $// \cdots //$, we are indicating that the leftmost and rightmost curves in $\cdots$ intersect, with intersection number one.} together with a variety of their well-studied properties. In fact, for Higgsable to $(2,0)$ LSTs of types $A_1$ and $A_2$ there are additional configurations; these occur when the endpoint configuration, obtained by repetitively blowing-down all $(-1)$-curves, consists of either two smooth rational curves of self-intersection $(-2)$ intersecting at the single point of multiplicity two, or of three smooth rational curves of self-intersection $(-2)$ all intersecting at a single point. These endpoints are referred to as $III$ and $IV$, as these intersection patterns match those of the Kodaira--Neron fibers of the same labelling. Since we are principally concerned with DE-type SUSY enhancement in this paper, we do not enumerate these exceptional Higgsable to A-type $(2,0)$ LSTs.

Let us now consider 6d $(1,0)$ LSTs Higgsable to D-type $(2,0)$ LSTs. The generic class takes the form:\footnote{Recall that a 6d $(1,0)$ LST is associated with a tensor branch curve configuration together with a choice of contraction map. Such contraction maps are in general not unique \cite{Aspinwall:1997ye}, and thus we should consider each curve configuration as being associated with a family of LSTs parametrized by possible contraction maps.}
\begin{equation}
    \overset{\mathfrak{su}_{k_1}}{2}\,\overset{\displaystyle \overset{\mathfrak{su}_{k_2}}{2}}{\overset{\mathfrak{su}_{k_3}}{2}}\underbrace{\,\overset{\mathfrak{su}_{k_4}}{2}\,\cdots\,\overset{\mathfrak{su}_{k_{N-2}}}{2}\,}_{N-5}\overset{\displaystyle \overset{\mathfrak{su}_{k_{N+1}}}{2}}{\overset{\mathfrak{su}_{k_{N-1}}}{2}}\overset{\mathfrak{su}_{k_N}}{2} \,.
\end{equation}
Anomaly cancellation fixes that the $k_i$ must satisfy the condition that
\begin{equation}\label{eqn:affineDN}
    A^{ij}k_j = m_i \geq 0 \,,
\end{equation}
where $A^{ij}$ is the affine Cartan matrix of $D_N$. In fact, almost all solutions are parametrized by a single positive integer $K$, and take the form
\begin{equation}\label{eqn:dora}
    \overset{\mathfrak{su}_{K}}{2}\,
    \overset{\displaystyle 
        \overset{\mathfrak{su}_{K}}{2}}{\overset{\mathfrak{su}_{2K}}{2}}
    \underbrace{\,\overset{\mathfrak{su}_{2K}}{2}\,\cdots\,\overset{\mathfrak{su}_{2K}}{2}\,}_{N-5}
    \overset{\displaystyle 
        \overset{\mathfrak{su}_{K}}{2}}{\overset{\mathfrak{su}_{2K}}{2}}
    \overset{\mathfrak{su}_{K}}{2} \,.
\end{equation}
Therefore, we refer to this class of LSTs as
\begin{equation}
    \widehat{D}_N^{\mathfrak{su}_{2K}} \,.
\end{equation}
Here, we have removed those solutions of equation \eqref{eqn:affineDN} which are incompatible with the F-theory geometry, as discussed around equation \eqref{eqn:screwy}.

We can also consider the LSTs that arise as the affinization of the tensor branch configuration in equation \eqref{eqn:DNE6}
\begin{equation}\label{eqn:DNhatE6}
    \overset{\mathfrak{su}_3}{3}1 \overset{\displaystyle \overset{\mathfrak{su}_3}{3}}{\overset{\displaystyle 1}{\overset{\mathfrak{e}_6}{6}}} 1\overset{\mathfrak{su}_3}{3}1\overbrace{\overset{\mathfrak{e}_6}{6} 1\overset{\mathfrak{su}_3}{3}1 \cdots \overset{\mathfrak{e}_6}{6} 1\overset{\mathfrak{su}_3}{3}1}^{N-5}\overset{\displaystyle \overset{\mathfrak{su}_3}{3}}{\overset{\displaystyle 1}{\overset{\mathfrak{e}_6}{6}}}1\overset{\mathfrak{su}_3}{3}  \,.
\end{equation}
Finally, we note that there are two additional 6d $(1,0)$ LSTs Higgsable to 6d $(2,0)$ D-type LSTs, analogous to those appearing in equation \eqref{eqn:DtypeEXCEPT}. These have tensor branch configurations of the following forms
\begin{equation}\label{eqn:DhatEXCEPT}
    2 \overset{\displaystyle 2}{\overset{\mathfrak{su}_{2}}{2}}\underbrace{\,\overset{\mathfrak{su}_{3}}{2} \cdots \overset{\mathfrak{su}_{3}}{2} \,}_{N-5}\overset{\displaystyle 2}{\overset{\mathfrak{su}_{2}}{2}}2 \qquad \text{ and } \qquad 
    \overset{\mathfrak{su}_2}{2}\overset{\displaystyle\overset{\mathfrak{su}_2}{2}}{\overset{\mathfrak{so}_7}{3}}1\underbrace{\overset{\mathfrak{so}_8}{4}1\cdots\overset{\mathfrak{so}_8}{4}1\,}_{N-5}\overset{\displaystyle\overset{\mathfrak{su}_2}{2}}{\overset{\mathfrak{so}_7}{3}}\overset{\mathfrak{su}_2}{2}
    \,.
\end{equation}
In equation \eqref{eqn:dora}, equation \eqref{eqn:DNhatE6}, and the left of equation \eqref{eqn:DhatEXCEPT},\footnote{For the configuration on the right in equation \eqref{eqn:DhatEXCEPT}, when we take the $N=4$ limit we have the configuration:
\begin{equation}
    \overset{\mathfrak{su}_2}{2}
    \underset{\displaystyle \overset{\mathfrak{su}_2}{2}}{\overset{
      \displaystyle \overset{\mathfrak{su}_2}{2}}{
      \overset{\mathfrak{g}_2}{2}
    }}
    \overset{\mathfrak{su}_2}{2} \,.
\end{equation}} we can take $N=4$, in which case the two curves which intersect three other curves are identified. Note: for particularly small values of $N$, there are additional 6d $(1,0)$ LSTs that are Higgsable to the $(2,0)$ LSTs of D-type, however, we do not consider them further in this work. 

A similar analysis can be performed for the 6d $(1,0)$ LSTs that are Higgsable to the 6d $(2,0)$ LSTs of E-type. For $E_6$ the only possibilities are
\begin{equation}
    \begin{gathered}\overset{\mathfrak{su}_{K}}{2} 
    \overset{\mathfrak{su}_{2K}}{2}
    \overset{
      \displaystyle \overset{\mathfrak{su}_{K}}{2}
    }{
    \overset{
      \displaystyle \overset{\mathfrak{su}_{2K}}{2}
    }{
      \overset{\mathfrak{su}_{3K}}{2}
    }}
    \overset{\mathfrak{su}_{2K}}{2}
    \overset{\mathfrak{su}_{K}}{2} \,, \qquad\qquad 2
    \overset{\mathfrak{su}_2}{2}
    \overset{\displaystyle 2}{\overset{
      \displaystyle \overset{\mathfrak{su}_2}{2}}{
      \overset{\mathfrak{g}_2}{2}
    }}
    \overset{\mathfrak{su}_2}{2}
    2 \,, \qquad\qquad
    \overset{\mathfrak{su}_2}{2}
    \overset{\mathfrak{so}_7}{3}
    \overset{\mathfrak{su}_2}{2}1
    \overset{
      \displaystyle \overset{
        \displaystyle \overset{
          \displaystyle \overset{
            \displaystyle \overset{\mathfrak{su}_2}{2}
          }{
            \overset{\mathfrak{so}_7}{3}}}{\overset{\mathfrak{su}_2}{2}}
        }{
          1
      }
    }{\overset{\mathfrak{e}_7}{8}}
    1\overset{\mathfrak{su}_2}{2}
    \overset{\mathfrak{so}_7}{3}
    \overset{\mathfrak{su}_2}{2} \,.
  \end{gathered}
\end{equation}
For $E_7$ there is still one exceptional theory, in addition to the standard family; altogether:
\begin{equation}
    \begin{gathered}
    \overset{\mathfrak{su}_{K}}{2} 
    \overset{\mathfrak{su}_{2K}}{2}
    \overset{\mathfrak{su}_{3K}}{2}
    \overset{
      \displaystyle \overset{\mathfrak{su}_{2K}}{2}
    }{
      \overset{\mathfrak{su}_{4K}}{2}
    }
    \overset{\mathfrak{su}_{3K}}{2}
    \overset{\mathfrak{su}_{2K}}{2}
    \overset{\mathfrak{su}_{K}}{2} \,,
    \qquad\qquad 
        2\overset{\mathfrak{su}_2}{2}\overset{\mathfrak{g}_2}{3}1\overset{\displaystyle \overset{\displaystyle \overset{\mathfrak{su}_3}{3}}{1}}{\overset{\mathfrak{f}_4}{5}}1\overset{\mathfrak{g}_2}{3}\overset{\mathfrak{su}_2}{2}2 \,,
    \end{gathered}
\end{equation}
whereas for $E_8$ we have simply:
\begin{equation}
    \begin{gathered}
    \overset{\mathfrak{su}_{K}}{2} 
    \overset{\mathfrak{su}_{2K}}{2}
    \overset{\mathfrak{su}_{3K}}{2}
    \overset{\mathfrak{su}_{4K}}{2}
    \overset{\mathfrak{su}_{5K}}{2}
    \overset{
      \displaystyle \overset{\mathfrak{su}_{3K}}{2}
    }{
      \overset{\mathfrak{su}_{6K}}{2}
    }
    \overset{\mathfrak{su}_{4K}}{2}
    \overset{\mathfrak{su}_{2K}}{2}
 \,.
 \end{gathered}
\end{equation}

Each of the 6d $(1,0)$ LSTs which have SUSY-enhancing Higgs branch RG flows to the 6d $(2,0)$ LSTs that we have just found are written in Tables \ref{tbl:someLSTs} and \ref{tbl:someLSTs2}.

\begin{table}[p]
    \centering
    \begin{threeparttable}
    \begin{tabular}{cccc}
        \toprule
         LST & Curve Configuration & $\operatorname{dim}(\mathcal{C})$ & $\kappa_R$  \\\midrule
         $\widehat{A}_{N-1}^{\mathfrak{su}_{K}}$ & $\begin{gathered}
            \big/\big/\underbrace{
    \overset{\mathfrak{su}_{K}}{2}
    \overset{\mathfrak{su}_{K}}{2}
    \cdots 
    \overset{\mathfrak{su}_{K}}{2}
    \overset{\mathfrak{su}_{K}}{2}
    }_{N \text{ $(-2)$-curves}}\big/\big/
         \end{gathered}$  & $NK-1$ & $NK$ \\
         $\widehat{A}_{N-1}^{\mathfrak{so}_{2K}}$ & $\begin{gathered}
            \big/\big/ \underbrace{\,\overset{\mathfrak{so}_{2K}}{4}\overset{\mathfrak{sp}_{K-4}}{1} \cdots \overset{\mathfrak{so}_{2K}}{4}\overset{\mathfrak{sp}_{K-4}}{1}\,}_{N \text{ $(-4)$-curves}} \big/\big/
         \end{gathered}$  & $2N(K-1)-1$& $4NK-8N$ \\
         $\widehat{A}_{N-1}^{\mathfrak{e}_{6}}$ & $\begin{gathered}
    \big/\big/ \underbrace{\,\overset{\mathfrak{e}_{6}}{6}1\overset{\mathfrak{su}_{3}}{3}1 \cdots \overset{\mathfrak{e}_{6}}{6}1\overset{\mathfrak{su}_{3}}{3}1\,}_{N \text{ $(-6)$-curves}} \big/\big/
\end{gathered}$  & $12N-1$ & $24N$ \\ 
        $\widehat{A}_{N-1}^{\mathfrak{e}_{7}}$ & $\begin{gathered}
            \big/\big/ \underbrace{\,\overset{\mathfrak{e}_{7}}{8}1\overset{\mathfrak{su}_{2}}{2}\overset{\mathfrak{so}_{7}}{3}\overset{\mathfrak{su}_{2}}{2}1 \cdots \overset{\mathfrak{e}_{7}}{8}1\overset{\mathfrak{su}_{2}}{2}\overset{\mathfrak{so}_{7}}{3}\overset{\mathfrak{su}_{2}}{2}1\,}_{N \text{ $(-8)$-curves}} \big/\big/
         \end{gathered}$  & $18N-1$& $48N$ \\
         $\widehat{A}_{N-1}^{\mathfrak{e}_{8}}$ & $\begin{gathered}
            \big/\big/ \underbrace{\,\overset{\mathfrak{e}_{8}}{(12)}12\overset{\mathfrak{su}_{2}}{2}\overset{\mathfrak{g}_{2}}{3}1\overset{\mathfrak{f}_{4}}{5}1\overset{\mathfrak{g}_{2}}{3}\overset{\mathfrak{su}_{2}}{2}21 \cdots \overset{\mathfrak{e}_{8}}{(12)}12\overset{\mathfrak{su}_{2}}{2}\overset{\mathfrak{g}_{2}}{3}1\overset{\mathfrak{f}_{4}}{5}1\overset{\mathfrak{g}_{2}}{3}\overset{\mathfrak{su}_{2}}{2}21 \,}_{N \text{ $(-12)$-curves}} \big/\big/
         \end{gathered}$  & $30N-1$ & $120N$ \\
         $\widehat{D}_N^{\mathfrak{su}_{2K}}$ & $\begin{gathered}
             \overset{\mathfrak{su}_{K}}{2}\,
    \overset{\displaystyle 
        \overset{\mathfrak{su}_{K}}{2}}{\overset{\mathfrak{su}_{2K}}{2}}
    \underbrace{\,\overset{\mathfrak{su}_{2K}}{2}\,\cdots\,\overset{\mathfrak{su}_{2K}}{2}\,}_{N-5 \text{ $(-2)$-curves}}
    \overset{\displaystyle 
        \overset{\mathfrak{su}_{K}}{2}}{\overset{\mathfrak{su}_{2K}}{2}}
    \overset{\mathfrak{su}_{K}}{2}
         \end{gathered}$ & $2K(N-1)-1$ & $4NK-8K$ \\ 
        $\widehat{D}_N^{\mathfrak{su}_{3}}$ & 
        $\begin{gathered}
            2 \overset{\displaystyle 2}{\overset{\mathfrak{su}_{2}}{2}}\underbrace{\,\overset{\mathfrak{su}_{3}}{2} \cdots \overset{\mathfrak{su}_{3}}{2} \,}_{N-5 \text{ $(-2)$-curves}}\overset{\displaystyle 2}{\overset{\mathfrak{su}_{2}}{2}}2
        \end{gathered}$ & $3N-8$& $6N-18$ \\
        $\widehat{D}_N^{\mathfrak{so}_{8}}$ & 
        $\begin{gathered}
            \overset{\mathfrak{su}_2}{2}\overset{\displaystyle\overset{\mathfrak{su}_2}{2}}{\overset{\mathfrak{so}_7}{3}}1\underbrace{\overset{\mathfrak{so}_8}{4}1\cdots\overset{\mathfrak{so}_8}{4}1\,}_{N-5 \text{ $(-4)$-curves}}\overset{\displaystyle\overset{\mathfrak{su}_2}{2}}{\overset{\mathfrak{so}_7}{3}}\overset{\mathfrak{su}_2}{2}
        \end{gathered}$ & $6N-14$ & $16N-48$ \\
        $\widehat{D}_N^{\mathfrak{e}_{6}}$ & 
        $\begin{gathered}
            \overset{\mathfrak{su}_3}{3}1 \overset{\displaystyle \overset{\mathfrak{su}_3}{3}}{\overset{\displaystyle 1}{\overset{\mathfrak{e}_6}{6}}} 1\overset{\mathfrak{su}_3}{3}1\overbrace{\overset{\mathfrak{e}_6}{6} 1\overset{\mathfrak{su}_3}{3}1 \cdots \overset{\mathfrak{e}_6}{6} 1\overset{\mathfrak{su}_3}{3}1}^{N-5 \text{ $(-6)$-curves}}\overset{\displaystyle \overset{\mathfrak{su}_3}{3}}{\overset{\displaystyle 1}{\overset{\mathfrak{e}_6}{6}}}1\overset{\mathfrak{su}_3}{3} 
        \end{gathered}$ & $12N-26$ & $48N-144$ \\\bottomrule
    \end{tabular}
    \end{threeparttable}
    \caption{We list 6d $(1,0)$ LSTs that have SUSY-enhancing Higgs branch renormalization group flows to the $(2,0)$ LSTs. For their properties, we list the dimension of the Coulomb branch of the $S^1$ compactification, and the generalized symmetry structure constant $\kappa_R$.}
    \label{tbl:someLSTs}
\end{table}

\begin{table}[t]
    \centering
    \begin{threeparttable}
    \begin{tabular}{cccc}
    \toprule
         LST & Curve Configuration & $\operatorname{dim}(\mathcal{C})$ & $\kappa_R$  \\\midrule
        $\widehat{E}_6^{\mathfrak{su}_{3K}}$ &
        $\begin{gathered}
            \overset{\mathfrak{su}_{K}}{2} 
    \overset{\mathfrak{su}_{2K}}{2}
    \overset{
      \displaystyle \overset{\mathfrak{su}_{K}}{2}
    }{
    \overset{
      \displaystyle \overset{\mathfrak{su}_{2K}}{2}
    }{
      \overset{\mathfrak{su}_{3K}}{2}
    }}
    \overset{\mathfrak{su}_{2K}}{2}
    \overset{\mathfrak{su}_{K}}{2}
        \end{gathered}$ & $12K-1$ & $24K$ \\
        $\widehat{E}_6^{\mathfrak{g}_{2}}$ &
        $\begin{gathered}
            2
    \overset{\mathfrak{su}_2}{2}
    \overset{\displaystyle 2}{\overset{
      \displaystyle \overset{\mathfrak{su}_2}{2}}{
      \overset{\mathfrak{g}_2}{2}
    }}
    \overset{\mathfrak{su}_2}{2}
    2
        \end{gathered}$ & $11$ & $27$ \\


        $\widehat{E}_6^{\mathfrak{e}_{7}}$ &
        $\begin{gathered}
            \overset{\mathfrak{su}_2}{2}
    \overset{\mathfrak{so}_7}{3}
    \overset{\mathfrak{su}_2}{2}1
    \overset{
      \displaystyle \overset{
        \displaystyle \overset{
          \displaystyle \overset{
            \displaystyle \overset{\mathfrak{su}_2}{2}
          }{
            \overset{\mathfrak{so}_7}{3}}}{\overset{\mathfrak{su}_2}{2}}
        }{
          1
      }
    }{\overset{\mathfrak{e}_7}{8}}
    1\overset{\mathfrak{su}_2}{2}
    \overset{\mathfrak{so}_7}{3}
    \overset{\mathfrak{su}_2}{2}
        \end{gathered}$ & $34$ & $144$ \\
        $\widehat{E}_7^{\mathfrak{su}_{4K}}$ &
        $\begin{gathered}
            \overset{\mathfrak{su}_{K}}{2} 
    \overset{\mathfrak{su}_{2K}}{2}
    \overset{\mathfrak{su}_{3K}}{2}
    \overset{
      \displaystyle \overset{\mathfrak{su}_{2K}}{2}
    }{
      \overset{\mathfrak{su}_{4K}}{2}
    }
    \overset{\mathfrak{su}_{3K}}{2}
    \overset{\mathfrak{su}_{2K}}{2}
    \overset{\mathfrak{su}_{K}}{2}
        \end{gathered}$ & $18K-1$ & $48K$ \\
        $\widehat{E}_7^{\mathfrak{f}_{4}}$ &
        $\begin{gathered}
            2\overset{\mathfrak{su}_2}{2}\overset{\mathfrak{g}_2}{3}1\overset{\displaystyle \overset{\displaystyle \overset{\mathfrak{su}_3}{3}}{1}}{\overset{\mathfrak{f}_4}{5}}1\overset{\mathfrak{g}_2}{3}\overset{\mathfrak{su}_2}{2}2
        \end{gathered}$ & $22$ & $96$ \\
        $\widehat{E}_8^{\mathfrak{su}_{6K}}$ &
        $\begin{gathered}
            \overset{\mathfrak{su}_{K}}{2} 
    \overset{\mathfrak{su}_{2K}}{2}
    \overset{\mathfrak{su}_{3K}}{2}
    \overset{\mathfrak{su}_{4K}}{2}
    \overset{\mathfrak{su}_{5K}}{2}
    \overset{
      \displaystyle \overset{\mathfrak{su}_{3K}}{2}
    }{
      \overset{\mathfrak{su}_{6K}}{2}
    }
    \overset{\mathfrak{su}_{4K}}{2}
    \overset{\mathfrak{su}_{2K}}{2}
        \end{gathered}$ & $30K-1$ & $120K$ \\\bottomrule
    \end{tabular}
    \end{threeparttable}
    \caption{The continuation of Table \ref{tbl:someLSTs}; see the caption there for the details.}
    \label{tbl:someLSTs2}
\end{table}

We now turn to a discussion of the physical properties of little string theories. LSTs possess a $\mathfrak{u}(1)^{(1)}_{LST}$ one-form symmetry. The symmetry structure of LSTs typically involves combined transformations involving the background field of the one-form symmetry, together with the background fields of the zero-form symmetries: $\mathfrak{su}(2)_R$, $\mathfrak{p}$, and $\mathfrak{f}$ \cite{DelZotto:2020sop,Cordova:2020tij}. Here, $\mathfrak{su}(2)_R$ is the R-symmetry, $\mathfrak{p}$ denotes the Poincare symmetry, and $\mathfrak{f}$ encodes any additional flavor symmetry. This symmetry structure, sometimes referred to as Green--Schwarz symmetry,\footnote{See \cite{Kang:2023uvm}, for a careful analysis of such continuous Green--Schwarz symmetries, highlighting several major subtleties.} can be written as
\begin{equation}
    \left(\mathfrak{su}(2)_R \oplus \mathfrak{p} \oplus \mathfrak{f} \right)^{(0)} \times_{\kappa_R ,\kappa_P , \kappa_{F} } \ \mathfrak{u}(1)_{LST}^{(1)} \,.
\end{equation}
The quantities $\kappa_R$, $\kappa_P$, and $\kappa_{F}$ capture how the R-symmetry, Poincare symmetry, and flavor symmetries, respectively, mix with the one-form symmetry; they are referred to as the generalized symmetry structure constants. For each of the LSTs that have SUSY-enhancing Higgs branch RG flows, we have $\kappa_P = 0$ and there generically is no non-Abelian flavor symmetry;\footnote{Even in the cases of $(2,0)$ LSTs realized as a cluster of undecorated $(-2)$-curves intersecting as an affine Dynkin diagram where an $SU(2)$ flavor symmetry can be identified, we choose not to consider $\kappa_F$ for simplicity.} therefore, we only consider $\kappa_R$ from this point onwards. One of the key results of \cite{DelZotto:2020sop,DelZotto:2023ahf} is that the structure constants can be determined from the F-theory geometry; in particular:
\begin{equation}
    \kappa_R = \sum_i \ell_i h_{\mathfrak{g}_i}^\vee \,.
\end{equation}
Here, the sum runs over the compact curves at the generic point of the tensor branch, $\ell_i$ is the normalized zero-eigenvector associated with the zero-eigenvalue of the intersection matrix:
\begin{equation}
    A^{ij} \ell_j = 0 \qquad \text{ such that } \qquad \ell_j > 0 \,\,\text{ and }\,\, \operatorname{gcd}(\ell_1, \ell_2,  \cdots) = 1 \,,
\end{equation}
and $h_{\mathfrak{g}_i}^\vee$ is the dual Coxeter number of the algebra $\mathfrak{g}_i$ supported over the $i$th curve.\footnote{If the singular fiber is irreducible over the curve $C_i$, then we define $h_{\mathfrak{g}_i}^\vee = 1$.} 

Furthermore, we can consider moduli spaces of supersymmetric vacua associated with the LSTs. We can consider the dimension of the Coulomb branch of the five-dimensional theory that is obtained via compactification of the LST on an $S^1$. This quantity can be obtained from the configuration at the generic point of the tensor branch as follows:
\begin{equation}
    \operatorname{dim}(\mathcal{C}) = \sum_i \big(1 + \operatorname{rank}(\mathfrak{g}_i)\big) - 1 \,.
\end{equation}
The sum runs over the compact curves $C_i$ supporting algebras $\mathfrak{g}_i$. 

For each of the Higgsable to $(2,0)$ LSTs that we study in this paper, we compute these two physical quantities, $\kappa_R$ and $\operatorname{dim}(\mathcal{C})$,
and we list them in Tables \ref{tbl:someLSTs} and \ref{tbl:someLSTs2}. We remind the reader that these quantities depend only on the tensor branch curve configuration, and not the choice of contraction map, and thus we can write them down without specifying the latter.

One of the exciting features of LSTs that follows from their non-local nature is T-duality. Two LSTs are said to be T-dual if they give rise to the same theory after circle compactification. More specifically, if $\mathcal{T}_1$ and $\mathcal{T}_2$ are 6d LSTs, then we can consider the $S^1$ compactifications, with radii $R_1$ and $R_2$, where, along the circles, we can turn on Wilson lines valued in the (continuous) flavor symmetry of the LSTs. We denote these compactified theories as
\begin{equation}
    \mathcal{T}_1\langle S^1_{R_1}, \operatorname{WL}_1\rangle \qquad \text{ and } \qquad \mathcal{T}_2\langle S^1_{R_2}, \operatorname{WL}_2\rangle \,,
\end{equation}
where $\operatorname{WL}_i$ abstractly specifies which Wilson lines are turned on. If these two 5d theories are identical, at some point on the Coulomb branch and for some choices of $\operatorname{WL}_1$ and $\operatorname{WL}_2$, then $\mathcal{T}_1$ and $\mathcal{T}_2$ are said to be T-dual. 

To identify a T-dual pair, we should see that both $\mathcal{T}_1$ and $\mathcal{T}_2$ are engineered in string theory from the same compactification space, up to inequivalent fibration structures on that space; this approach was pioneered in \cite{Aspinwall:1997ye}. Alternatively, one can identify putative T-dual pairs by determining invariants of the theories $\mathcal{T}_1$ and $\mathcal{T}_2$ which are known to either be unchanged or change predictably under the $S^1$-compactification with Wilson lines. Some examples of such T-duality invariant properties are \cite{DelZotto:2020sop}
\begin{equation}\label{eqn:numericalinv}
    \kappa_P \,, \qquad \kappa_R \,, \qquad \kappa_F \,, \qquad \operatorname{dim}(\mathcal{C}) \,, \qquad \operatorname{rank}(\mathfrak{f}) \,,
\end{equation}
where the latter is the rank of the 6d flavor algebra. This provides a necessary, but not sufficient, condition for T-duality. In general, the 6d Higgs branch does not match across T-duality, as the choice of Wilson lines modifies the 5d Higgs branch from the 6d Higgs branch. However, in cases where there is no flavor symmetry in 6d, there are no Wilson lines to turn on, and the 5d Higgs branch is identical to the 6d Higgs branch, and thus any T-dual pair $(\mathcal{T}_1, \mathcal{T}_2)$ must satisfy
\begin{equation}
    \mathcal{H}_{\mathcal{T}_1} = \mathcal{H}_{\mathcal{T}_2} \,,
\end{equation}
where $\mathcal{H}_\mathcal{T}$ denotes the Higgs branch of theory $\mathcal{T}$. For some of the Higgsable to DE-type $(2,0)$ LSTs that we consider in this paper, there is no 6d flavor symmetry, and thus the Higgs branches of any T-dual pair must match.

In Tables \ref{tbl:someLSTs} and \ref{tbl:someLSTs2}, we have written the putative T-dual for each 6d $(1,0)$ LSTs evincing DE-type SUSY enhancement by identifying another LSTs where the numerical invariants in equation \eqref{eqn:numericalinv}
match. Importantly, each of these quantities is independent of the choice of contraction map, and thus they can be determined directly from the curve configuration; the Higgs branch itself \emph{is} sensitive to the choice of contraction map, and we discuss this particular subtlety anon.

Since we have enumerated a collection of LSTs that realize Higgs branch RG flows with SUSY enhancement in Tables \ref{tbl:someLSTs} and \ref{tbl:someLSTs2}, together with their T-duality invariant properties, we can identify pairs of LSTs for which $\operatorname{dim}(\mathcal{C})$ and $\kappa_R$ (as well as $\kappa_P$, $\kappa_F$, and $\operatorname{rank}(\mathfrak{f})$) match. We find that the putative T-duals for each of the five classes of LSTs whose base geometry is a ring of $(-2)$-curves are as follows:
\begin{equation}
    \begin{aligned}
        \widehat{A}_{N-1}^{\mathfrak{su}_K} \quad &\longleftrightarrow \quad \widehat{A}_{K-1}^{\mathfrak{su}_N} \,, \\
        \widehat{A}_{N-1}^{\mathfrak{so}_{2K}} \quad &\longleftrightarrow \quad \widehat{D}_{K}^{\mathfrak{su}_{2N}} \,, \\
        \widehat{A}_{N-1}^{\mathfrak{e}_{6}} \quad &\longleftrightarrow \quad \widehat{E}_{6}^{\mathfrak{su}_{3N}} \,, \\
        \widehat{A}_{N-1}^{\mathfrak{e}_{7}} \quad &\longleftrightarrow \quad \widehat{E}_{7}^{\mathfrak{su}_{4N}} \,, \\
        \widehat{A}_{N-1}^{\mathfrak{e}_{8}} \quad &\longleftrightarrow \quad \widehat{E}_{8}^{\mathfrak{su}_{6N}} \,.
    \end{aligned}
\end{equation}
We note that this is expected from fiber-base duality, as was already pointed out in \cite{Bhardwaj:2015oru}; see also \cite{Haghighat:2018dwe} for a careful analysis of the second row. Next, we can consider a possible T-dual for the $\widehat{D}_N^{\mathfrak{e}_6}$ LST. It is easy to see that there is no other theory in Tables \ref{tbl:someLSTs} or \ref{tbl:someLSTs2} for which the numerical invariants match. Instead, consider the LSTs associated with the following tensor branch curve configuration:
\begin{equation}\label{eqn:funkyA2LST}
    \overset{\mathfrak{so}_{2N}}{4} 
    \overset{\mathfrak{sp}_{2N-8}}{1}
    \overset{
      \displaystyle \overset{\mathfrak{so}_{2N}}{4}
    }{
    \overset{
      \displaystyle \overset{\mathfrak{sp}_{2N-8}}{1}
    }{
      \overset{\mathfrak{so}_{6N-16}}{4}
    }}
    \overset{\mathfrak{sp}_{2N-8}}{1}
    \overset{\mathfrak{so}_{2N}}{4} \,.
\end{equation}
The endpoint configuration for this theory is type $IV$; that is, it is three $(-2)$-curves that intersect simultaneously at one point. We label the associated LSTs as $IV^{\mathfrak{so}_{6N-16}}$. We can determine the T-duality invariant numerical quantities:
\begin{equation}
    \operatorname{dim}(\mathcal{C}) = 12N - 26 \,, \qquad \kappa_R = 48N - 144 \,.
\end{equation}
Therefore, as all the numerical invariants we can compute agree, we have the putative T-duality:
\begin{equation}
    \widehat{D}_N^{\mathfrak{e}_6} \quad \longleftrightarrow \quad IV^{\mathfrak{so}_{6N-16}} \,.
\end{equation}
While a similar discussion for the putative LSTs T-dual to the $\widehat{D}_N^{\mathfrak{su}_3}$, $\widehat{D}_N^{\mathfrak{so}_8}$, $\widehat{E}_6^{\mathfrak{g}_2}$, $\widehat{E}_6^{\mathfrak{e}_7}$, and $\widehat{E}_7^{\mathfrak{f}_4}$ theories can be carried out, it is beyond the scope of this paper and we leave it for future work.\footnote{A detailed analysis of T-duality for non-heterotic LSTs has recently appeared in \cite{FELLAS}.} Here, following the theme of this paper, we discuss the Higgs branches of some of these LSTs. 

There are two approaches that we can take to determine the Higgs branch of the LSTs under discussion. We would like to capture a Higgs branch by providing a magnetic quiver, that is, a 3d $\mathcal{N}=4$ Lagrangian quiver such that the Coulomb branch is isomorphic to the Higgs branch of the theory we are interested in. The first approach is to engineer the LSTs via a brane system in Type IIA (or Type I') string theory, and then pass to the magnetic phase; this is the approach we have taken for SCFTs throughout this paper. 

Alternatively, we can take advantage of the fact that LSTs satisfy the tensor decoupling condition \cite{Bhardwaj:2015oru}, that is, if one takes the volume of any compact curve in the curve configuration of an LST to infinity, then one obtains either an SCFT or a product of SCFTs. In reverse, we can think of an LST as arising from the fusion of a non-Abelian flavor algebra of a product of SCFTs. If we know magnetic quivers for the Higgs branches of the fused SCFTs, then we can determine the magnetic quiver for the LST via Coulomb gauging of the Coulomb symmetry of the SCFT magnetic quivers.

We first consider the LSTs that we denote via $\widehat{A}_{N-1}^{\mathfrak{su}_K}$; the tensor branch curve configuration of such LSTs takes the form of a ring of $(-2)$-curves, each supporting an $\mathfrak{su}(K)$ algebra. This LST can be obtained from the rank $N$ $(\mathfrak{su}(K), \mathfrak{su}(K))$ conformal matter theory, and fusing the two $\mathfrak{su}(K)$ flavor factors. We depict this as
\begin{equation}
    [\mathfrak{su}(K)]\,\underbrace{\,\overset{\mathfrak{su}_K}{2} \cdots \overset{\mathfrak{su}_K}{2}\,}_{N-1}\,[\mathfrak{su}(K)] \,\, \xrightarrow{\,\text{fusion }\,} \,\, \big/\big/ \underbrace{\,\overset{\mathfrak{su}_K}{2} \cdots \overset{\mathfrak{su}_K}{2}\,}_{N} \big/\big/ \,.
\end{equation}
The magnetic quiver for the Higgs branch of rank $N$ $(\mathfrak{su}(K), \mathfrak{su}(K))$ conformal matter is \cite{Hanany:2018vph}
\begin{equation}
    \begin{gathered}
        \begin{tikzpicture}
      \node[node, label=below:{\footnotesize $1$}] (Z1)  {};
      \node[node, label=below:{\footnotesize $2$}] (Z2) [right=6mm of Z1] {};
      \node[tnode] (dots) [right=6mm of Z2] {\footnotesize $\cdots$};
      \node[node, label=below:{\footnotesize $K-1$}] (Zk-1) [right=6mm of dots] {};
      \node[node, label=below:{\footnotesize $K$}] (Zk) [right=6mm of Zk-1] {};
      \node[node, label=right:{\footnotesize $N$}] (Fk) [above=6mm of Zk] {};
      \node[node, label=below:{\footnotesize $K-1$}] (Zk1) [right=6mm of Zk] {};
      \node[tnode] (dots1) [right=6mm of Zk1] {\footnotesize $\cdots$};
      \node[node, label=below:{\footnotesize $2$}] (Z21) [right=6mm of dots1] {};
      \node[node, label=below:{\footnotesize $1$}] (Z11) [right=6mm of Z21] {};
      \draw (Z1.east) -- (Z2.west);
      \draw (Z2.east) -- (dots.west);
      \draw (dots.east) -- (Zk-1.west);
      \draw (Zk-1.east) -- (Zk.west);
      \draw (Zk.north) -- (Fk.south);
      \draw (Zk.east) -- (Zk1.west);
      \draw (Zk1.east) -- (dots1.west);
      \draw (dots1.east) -- (Z21.west);
      \draw (Z21.east) -- (Z11.west);
      \draw (Fk) to[out=130, in=410, looseness=12] (Fk);

        \def\radius{0.2cm}
        \def\linethickness{8pt}
        \begin{pgfonlayer}{background}
            \foreach \coord in {Z1,Z2,dots,Zk-1,Zk1,dots1,Z21,Z11} {
            \fill[blue!30] (\coord) circle (\radius);
            }     
            \foreach \start/\ending in {Z1/Z2, Z2/dots.east, dots.east/Zk-1} {
            \draw[line width=\linethickness, draw=blue!30, fill=blue!30, fill opacity=0.3] (\start) -- (\ending);}
            \foreach \start/\ending in {Zk1/dots1.west, dots1.west/Z21, Z21/Z11} {
            \draw[line width=\linethickness, draw=blue!30, fill=blue!30, fill opacity=0.3] (\start) -- (\ending);
             }
        \end{pgfonlayer};
    \end{tikzpicture}
    \end{gathered} \,,
\end{equation}
where the $\mathfrak{su}(K) \oplus \mathfrak{su}(K)$ Coulomb symmetry arises from the balanced nodes highlighted in blue. The behavior of the 3d quiver under Coulomb gauging follows from \cite{Benini:2010uu}, and thus we find that the Higgs branch of the LST $\widehat{A}_{N-1}^{\mathfrak{su}_K}$ is simply the Coulomb branch of the following quiver:
\begin{equation}
  \begin{gathered}
    \begin{tikzpicture}
    \node[node, label=right:{\footnotesize $K$}] (Zk) [right=6mm of Zk-1] {};
      \node[node, label=right:{\footnotesize $N$}] (Fk) [above=6mm of Zk] {};
      \draw (Zk.north) -- (Fk.south);
      \draw (Fk) to[out=130, in=410, looseness=12] (Fk);
      \draw (Zk) to[out=-50, in=230, looseness=12] (Zk);
      
    \end{tikzpicture}
  \end{gathered}\,.
\end{equation}
As we can see, this quiver is identical under $K \leftrightarrow N$, which is consistent with the T-duality\footnote{Notice that this theory actually enjoys the triality \cite{Bastian:2017ary}: $\widehat{A}_{N-1}^{\mathfrak{su}_K}  \leftrightarrow  \widehat{A}_{K-1}^{\mathfrak{su}_N} \leftrightarrow  \widehat{A}_{\ell-1}^{\mathfrak{su}_{\frac{NK}{\ell}}}$ with $\ell=\mathrm{gcd}(N,K)$, not seen from the magnetic quiver.} 
\begin{equation}
    \widehat{A}_{N-1}^{\mathfrak{su}_K} \qquad \leftrightarrow \qquad \widehat{A}_{K-1}^{\mathfrak{su}_N} \,.
\end{equation}

Next, we consider the LSTs $\widehat{A}_{N-1}^{\mathfrak{so}_{2K}}$; these theories can be obtained via the diagonal fusion of the $\mathfrak{so}(2K) \,\oplus\, \mathfrak{so}(2K)$ flavor algebra of rank $N$ $(\mathfrak{so}(2K), \mathfrak{so}(2K))$ conformal matter. Pictorially, we have
\begin{equation}
    [\mathfrak{so}(2K)]\, 
    \underbrace{\,\overset{\mathfrak{sp}_{K-4}}{1}\overset{\mathfrak{so}_{2K}}{4} \cdots \overset{\mathfrak{sp}_{K-4}}{1}\overset{\mathfrak{so}_{2K}}{4} \overset{\mathfrak{sp}_{K-4}}{1}\,}_{N-1 \text{ $(-4)$-curves}}
    \,[\mathfrak{so}(2K)] 
    \,\, \xrightarrow{\,\text{fusion }\,} \,\, \big/\big/
        \underbrace{\,\overset{\mathfrak{sp}_{K-4}}{1}\overset{\mathfrak{so}_{2K}}{4} \cdots \overset{\mathfrak{sp}_{K-4}}{1}\overset{\mathfrak{so}_{2K}}{4}\,}_{N \text{ $(-4)$-curves}}
    \big/\big/
    \,.
\end{equation}
The magnetic quiver for the conformal matter theory is known \cite{Hanany:2018vph}; it is
\begin{equation}
    \begin{gathered}
        \begin{tikzpicture}
      \node[node, fill=red, label=below:{\footnotesize $2$}] (Z1)  {};
      \node[node,fill=blue, label=below:{\footnotesize $2$}] (Z2) [right=7mm of Z1] {};
      \node[tnode] (dots) [right=7mm of Z2] {\footnotesize $\cdots$};
      \node[node,fill=blue, label=below:{\footnotesize $2K-2$}] (Zk-1) [right=7mm of dots] {};
      \node[node,fill=red,  label=below:{\footnotesize $2K$}] (Zk) [right=7mm of Zk-1] {};
      \node[node,fill=blue, label=right:{\footnotesize $2N$}] (Fk) [above=6mm of Zk] {};
      \node[node,fill=blue, label=below:{\footnotesize $2K-2$}] (Zk1) [right=7mm of Zk] {};
      \node[tnode] (dots1) [right=6mm of Zk1] {\footnotesize $\cdots$};
      \node[node,fill=blue,  label=below:{\footnotesize $2$}] (Z21) [right=7mm of dots1] {};
      \node[node,fill=red,  label=below:{\footnotesize $2$}] (Z11) [right=7mm of Z21] {};
      \draw (Z1.east) -- (Z2.west);
      \draw (Z2.east) -- (dots.west);
      \draw (dots.east) -- (Zk-1.west);
      \draw (Zk-1.east) -- (Zk.west);
      \draw (Zk.north) -- (Fk.south);
      \draw (Zk.east) -- (Zk1.west);
      \draw (Zk1.east) -- (dots1.west);
      \draw (dots1.east) -- (Z21.west);
      \draw (Z21.east) -- (Z11.west);
      \draw (Fk) to[out=130, in=410, looseness=12] (Fk);

        \def\radius{0.2cm}
        \def\linethickness{8pt}
        \begin{pgfonlayer}{background}
            \foreach \coord in {Z1,Z2,dots,Zk-1,Zk1,dots1,Z21,Z11} {
            \fill[blue!30] (\coord) circle (\radius);
            }     
            \foreach \start/\ending in {Z1/Z2, Z2/dots.east, dots.east/Zk-1} {
            \draw[line width=\linethickness, draw=blue!30, fill=blue!30, fill opacity=0.3] (\start) -- (\ending);}
            \foreach \start/\ending in {Zk1/dots1.west, dots1.west/Z21, Z21/Z11} {
            \draw[line width=\linethickness, draw=blue!30, fill=blue!30, fill opacity=0.3] (\start) -- (\ending);
             }
        \end{pgfonlayer};
    \end{tikzpicture}
    \end{gathered} \,,
\end{equation}
where, again, we have highlighted the balanced nodes giving rise to the $\mathfrak{so}(2K) \, \oplus \, \mathfrak{so}(2K)$ Coulomb symmetry.\footnote{The loop on the $\mathfrak{usp}(2N)$ gauge node denotes an antisymmetric hypermultiplet.} The Coulomb gauging is described in \cite{Benini:2010uu}, and thus we find that the magnetic quiver for the LST $\widehat{A}_{N-1}^{\mathfrak{so}_{2K}}$ is
\begin{equation}\label{eqn:AD_HB}
    \begin{gathered}
        \begin{tikzpicture}
    \node[node,fill=red, label=right:{\footnotesize $2K$}] (Zk) [right=6mm of Zk-1] {};
      \node[node,fill=blue, label=right:{\footnotesize $2N$}] (Fk) [above=6mm of Zk] {};
      \draw (Zk.north) -- (Fk.south);
      \draw (Fk) to[out=130, in=410, looseness=12] (Fk);
      \draw (Zk) to[out=-50, in=230, looseness=12] (Zk);
      
    \end{tikzpicture}
    \end{gathered} \,.
\end{equation}

We are now ready to turn to the LSTs $\widehat{D}_{N}^{\mathfrak{su}_{2K}}$. Such LSTs can be obtained via the diagonal fusion of the $\mathfrak{su}(2K)$ flavor symmetry of the SCFTs $D_{N'}^{\mathfrak{su}_{2K}}$ and $D_{N''}^{\mathfrak{su}_{2K}}$ where $N' + N'' = N$. The magnetic quivers of the Higgsable to D-type $(2,0)$ SCFTs have been determined in this paper; therefore, for the LSTs $\widehat{D}_{N}^{\mathfrak{su}_{2K}}$, we have
\begin{equation}
    \begin{gathered}
        \begin{tikzpicture}
      \node[node,  label=below:{\footnotesize $1$}] (A1) {};
      \node[node, label=below:{\footnotesize $2$}] (A2) [left=10mm of A1] {};
      \node[tnode] (T1) [left=5mm of A2] {$\cdots$};
      \node[node, label=below:{\footnotesize $2K-2$}] (A3) [left=5mm of T1] {};
      \node[node, label=below:{\footnotesize $2K-1$}] (A4) [left=10mm of A3] {};
    \node[node, label=below:{\footnotesize $2K$},fill=blue] (Usp) [left=10mm of A4] {};
      \node[node, label=right:{\footnotesize $2N'$},fill=red] (Nu) [above =6mm of Usp] {};

            \node[node,  label=below:{\footnotesize $1$}] [right=8mm of A1] (A1l) {};
      \node[node, label=below:{\footnotesize $2$}] (A2l) [right=10mm of A1l] {};
      \node[tnode] (T1l) [right=5mm of A2l] {$\cdots$};
      \node[node, label=below:{\footnotesize $2K-2$}] (A3l) [right=5mm of T1l] {};
      \node[node, label=below:{\footnotesize $2K-1$}] (A4l) [right=10mm of A3l] {};
    \node[node, label=below:{\footnotesize $2K$},fill=blue] (Uspl) [right=10mm of A4l] {};
      \node[node, label=right:{\footnotesize $2N''$},fill=red] (Nul) [above =6mm of Uspl] {};

\node[tnode] (oplus) [right=1mm of A1] {$\oplus$};
    
      \draw (A1.west) -- (A2.east);
      \draw (A2.west) -- (T1.east);
      \draw (T1.west) -- (A3.east);
      \draw (A3.west) -- (A4.east);
      \draw (A4.west) -- (Usp.east);    
      \draw (Usp.north) -- (Nu.south);
    \draw (Nu) to[out=130, in=410, looseness=12] (Nu);  

          \draw (A1l.east) -- (A2l.west);
      \draw (A2l.east) -- (T1l.west);
      \draw (T1l.east) -- (A3l.west);
      \draw (A3l.east) -- (A4l.west);
      \draw (A4l.east) -- (Uspl.west);    
      \draw (Uspl.north) -- (Nul.south);
    \draw (Nul) to[out=130, in=410, looseness=12] (Nul);  

\def\radius{0.2cm}
        \def\linethickness{8pt}
        \begin{pgfonlayer}{background}
            \foreach \coord in {A1,A2,T1,A3,A4,A1l,A2l,T1l,A3l,A4l} {
            \fill[blue!30] (\coord) circle (\radius);
            }     
            \foreach \start/\ending in {A1/A2, A2/T1.east, T1.east/A3, A3/A4} {
            \draw[line width=\linethickness, draw=blue!30, fill=blue!30, fill opacity=0.3] (\start) -- (\ending);}
            \foreach \start/\ending in {A1l/A2l, A2l/T1l.west, T1l.west/A3l, A3l/A4l} {
            \draw[line width=\linethickness, draw=blue!30, fill=blue!30, fill opacity=0.3] (\start) -- (\ending);
             }
        \end{pgfonlayer};

    \end{tikzpicture}
    \end{gathered} \,,
\end{equation}
where the $\oplus$ indicates Coulomb gauging along the Coulomb symmetries that arise from the blue-highlighted balanced nodes. When Coulomb gauging $\mathfrak{su}(2K)$ symmetries that are connected to $\mathfrak{usp}(2K)$ gauge nodes, the post-gauging $\mathfrak{usp}(2K)$ node picks up an antisymmetric hypermultiplet due to the decomposition of the adjoint representation of $\mathfrak{su}(2K)$, as discussed in \cite{Benini:2010uu,Beratto:2020wmn,Kang:2022zsl}. Therefore, the result of the Coulomb gauging is
\begin{equation}\label{eqn:DA_HB}
    \begin{gathered}
        \begin{tikzpicture}
    \node[node,fill=blue, label=right:{\footnotesize $2K$}] (Zk) [right=6mm of Zk-1] {};
      \node[node,fill=red, label=right:{\footnotesize $2N''$}] (Fk) [above right=6mm and 3mm of Zk] {};\node[node,fill=red, label=left:{\footnotesize $2N'$}] (Fk1) [above left=6mm and 3mm of Zk] {};
      \draw (Zk) -- (Fk);
      \draw (Zk) -- (Fk1);
      \draw (Fk) to[out=130, in=410, looseness=12] (Fk);
      \draw (Fk1) to[out=130, in=410, looseness=12] (Fk1);
      \draw (Zk) to[out=-50, in=230, looseness=12] (Zk);
    \end{tikzpicture}
        \,.
    \end{gathered}
\end{equation}
We can see that there are different Higgs branches for each of the combinations of $(N', N'') \geq (3, 3)$ such that $N' + N'' = N$ and $N' \leq N''$. This reflects the fact that to prescribe an LST we must give both a tensor branch curve configuration and a contraction map; the contraction map is not unique, and the choice can be captured by the splitting of $N$ into $N'$ and $N''$. We require that both $N'$ and $N''$ are $\geq 3$ as we have engineered this LST via fusion of $D_{N'}^{\mathfrak{su}_{2K}}$ and $D_{N''}^{\mathfrak{su}_{2K}}$ SCFTs, and we have only defined such theories when the number of $(-2)$-curves in the tensor branch configuration is at least three. The Higgs branch given by equation \eqref{eqn:DA_HB} does not match, under interchange of $N$ and $K$, the Higgs branch of the putative T-dual as given in equation \eqref{eqn:AD_HB} for any such choice of $N'$ and $N''$. However, the fusions that we have written here are not necessarily all of the distinct contraction maps for the tensor branch configuration in equation \eqref{eqn:dora}; with a sensible analytic continuation, we could believe that formally $N'=0$ and $N''=N$ captures a valid contraction map, and does in fact result in the same Higgs branch across the T-duality. A more careful matching and verification of these Higgs branches under T-duality would be an interesting subject for future research.

As a final point in this section, we highlight a family of LSTs which arise from the fusion of a $D_N^{\mathfrak{su}_{2K}}$ SCFT together with a rank $N'$ $(\mathfrak{e}_8, \mathfrak{su}(2K))$ orbi-instanton SCFT. The latter is the SCFT that arises on a stack of $N'$ M5-branes probing a $\mathbb{C}^2/\mathbb{Z}_{2K}$ orbifold singularity, and contained inside of an M9-brane \cite{DelZotto:2014hpa}. We can depict this fusion as follows:
\begin{equation}
    \begin{gathered}
        \begin{aligned}
        &\underbrace{\,\overset{\mathfrak{su}_{K}}{2} 
    \overset{
      \displaystyle \overset{\mathfrak{su}_{K}}{2}
    }{
      \overset{\mathfrak{su}_{2K}}{2}
    }
    \overset{\mathfrak{su}_{2K}}{2}
    \cdots 
    \overset{\mathfrak{su}_{2K}}{2}}_{N \text{ $(-2)$-curves}}\, [\mathfrak{su}(2K)] \quad\oplus\quad
[\mathfrak{su}(2K)]  \,       \underbrace{\,      \overset{\mathfrak{su}_{2K}}{2}
    \cdots 
    \overset{\mathfrak{su}_{2K}}{2}}_{N' -1 \text{ $(-2)$-curves}} 
    \overset{\mathfrak{su}_{2K-1}}{2} \cdots
    {
      \overset{\mathfrak{su}_{2}}{2}
    } 2 \, 1
\\
    &\qquad\qquad \longrightarrow \qquad \underbrace{\,\overset{\mathfrak{su}_{K}}{2} 
    \overset{
      \displaystyle \overset{\mathfrak{su}_{K}}{2}
    }{
      \overset{\mathfrak{su}_{2K}}{2}
    }
    \overset{\mathfrak{su}_{2K}}{2}
    \cdots 
\overset{\mathfrak{su}_{2K}}{2}
    \cdots 
    \overset{\mathfrak{su}_{2NK}}{2}}_{N + N' \text{ $(-2)$-curves}} 
    \overset{\mathfrak{su}_{2K-1}}{2} \cdots
    {
      \overset{\mathfrak{su}_{2}}{2}
    } 2 \, 1 \,,
        \end{aligned}
    \end{gathered} 
\end{equation}
where the $\oplus$ denotes the fusion of two $\mathfrak{su}(2K)$ flavor symmetries. We note that these LSTs have $\kappa_P = 2$, and thus they cannot be T-dual to the Higgsable to $(2,0)$ LSTs we have discussed above; however, even if these theories are not purely heterotic, they may be T-dual to the heterotic LSTs that have recently been explored in, for example, \cite{Lawrie:2023uiu,Mansi:2023faa,Ahmed:2023lhj,DelZotto:2023nrb,DelZotto:2022xrh,Bhardwaj:2022ekc,DelZotto:2022ohj} via matching the generalized symmetry structure constants. The magnetic quiver for the orbi-instanton theory and its Higgsed products is well-known \cite{Cabrera:2019izd,Fazzi:2023ulb} (see \cite{Lawrie:2023uiu} for a recent summary). Therefore the procedure of Coulomb gauging again reveals the magnetic quiver for the Higgs branch of this family of LSTs; we find that it is the Coulomb branch of the following quiver:
\begin{equation}\label{eqn:MQ_D1-type}
\begin{gathered}
\begin{aligned}
    &\begin{tikzpicture}
      \node[node,  label=below:{\footnotesize $1$}] (A1) {};
      \node[node, label=below:{\footnotesize $2$}] (A2) [left=10mm of A1] {};
      \node[tnode] (T1) [left=5mm of A2] {$\cdots$};
      \node[node, label=below:{\footnotesize $2K-2$}] (A3) [left=5mm of T1] {};
      \node[node, label=below:{\footnotesize $2K-1$}] (A4) [left=10mm of A3] {};
    \node[node, label=below:{\footnotesize $2K$},fill=blue] (Usp) [left=10mm of A4] {};
      \node[node, label=right:{\footnotesize $2N$},fill=red] (Nu) [above =6mm of Usp] {};

            \draw (A1.west) -- (A2.east);
      \draw (A2.west) -- (T1.east);
      \draw (T1.west) -- (A3.east);
      \draw (A3.west) -- (A4.east);
      \draw (A4.west) -- (Usp.east);    
      \draw (Usp.north) -- (Nu.south);
    \draw (Nu) to[out=130, in=410, looseness=12] (Nu);  
    \def\radius{0.2cm}
        \def\linethickness{8pt}
        \begin{pgfonlayer}{background}
            \foreach \coord in {A1,A2,T1,A3,A4} {
            \fill[blue!30] (\coord) circle (\radius);
            }     
            \foreach \start/\ending in {A1/A2, A2/T1.east, T1.east/A3, A3/A4} {
            \draw[line width=\linethickness, draw=blue!30, fill=blue!30, fill opacity=0.3] (\start) -- (\ending);}
        \end{pgfonlayer};
    \end{tikzpicture}
    \\
    &\qquad\qquad \begin{tikzpicture}
      \node[node, label=below:{\footnotesize $1$}] [right=6mm of A1] (Z1)  {};
      \node[node, label=below:{\footnotesize $2$}] (Z2) [right=6mm of Z1] {};
      \node[tnode] (adots) [right=6mm of Z2] {\footnotesize $\cdots$};
      \node[node, label=below:{\footnotesize $2K-1$}] (Zk-1) [right=6mm of adots] {};
      \node[node, label=below:{\footnotesize $2K$}] (Zk) [right=7mm of Zk-1] {};
      \node[node, label={[below right=5pt and -9pt]:\rotatebox{-30}{\footnotesize $(2K+N')$}}] (A1a)  [right=6mm of Zk] {};
      \node[node, label={[below right=5pt and -9pt]:\rotatebox{-30}{\footnotesize $2(2K+N')$}}] (A2a) [right=6mm of A1a] {};
      \node[node, label={[below right=5pt and -9pt]:\rotatebox{-30}{\footnotesize $3(2K+N')$}}] (A3a) [right=6mm of A2a] {};
      \node[node, label={[below right=5pt and -9pt]:\rotatebox{-30}{\footnotesize $4(2K+N')$}}] (A4a) [right=6mm of A3a] {};
      \node[node, label={[below right=5pt and -9pt]:\rotatebox{-30}{\footnotesize $5(2K+N')$}}] (A5a) [right=6mm of A4a] {};
      \node[node, label={[below right=5pt and -9pt]:\rotatebox{-30}{\footnotesize $6(2K+N')$}}] (N3a) [right=6mm of A5a] {};
      \node[node, label={[below right=5pt and -9pt]:\rotatebox{-30}{\footnotesize $4(2K+N')$}}] (B4a) [right=6mm of N3a] {};
      \node[node, label={[below right=5pt and -9pt]:\rotatebox{-30}{\footnotesize $2(2K+N')$}}] (B2a) [right=6mm of B4a] {};
      \node[node, label=right:{\footnotesize $3(2K+N')$}] (Nua) [above=7mm of N3a] {};
      \draw (Z1.east) -- (Z2.west);
      \draw (Z2.east) -- (adots.west);
      \draw (adots.east) -- (Zk-1.west);
      \draw (Zk-1.east) -- (Zk.west);
      \draw (Zk.east) -- (A1a.west);
      \draw (A1a.east) -- (A2a.west);
      \draw (A2a.east) -- (A3a.west);
      \draw (A3a.east) -- (A4a.west);
      \draw (A4a.east) -- (A5a.west);
      \draw (A5a.east) -- (N3a.west);
      \draw (N3a.east) -- (B4a.west);
      \draw (B4a.east) -- (B2a.west);
      \draw (N3a.north) -- (Nua.south); 

\node[tnode] (oplus) [left=2mm of Z1] {$\oplus$};
    
\def\radius{0.2cm}
        \def\linethickness{8pt}
        \begin{pgfonlayer}{background}
            \foreach \coord in {Z1,Z2,adots,Zk-1} {
            \fill[blue!30] (\coord) circle (\radius);
            }     
            \foreach \start/\ending in {Z1/Z2, Z2/adots.east, adots.east/Zk-1} {
            \draw[line width=\linethickness, draw=blue!30, fill=blue!30, fill opacity=0.3] (\start) -- (\ending);
             }
        \end{pgfonlayer};

    \end{tikzpicture} \,,
  \end{aligned}  
  \end{gathered} 
\end{equation}
where the $\oplus$ now indicates the Coulomb gauging of the two highlighted $\mathfrak{su}(2K)$ Coulomb symmetries.

From the magnetic quivers for the Higgs branches of the LSTs, we can determine the foliation of the symplectic singularity capturing the fixed points under Higgs branch RG flow via the technique of quiver subtraction, as we have done for Higgsable to D-type SCFTs throughout this paper. We leave a careful analysis of the structure of the Higgs branches of LSTs, and the implications with regards to T-duality, for future work.

\section{Discussion}\label{sec:disc}

In this paper, we have studied the renormalization group flows along the Higgs branch of 6d $(1,0)$ SCFTs which present supersymmetry enhancement along some locus inside of the Higgs branch. In particular, we were interested in the lesser-studied 6d $(1,0)$ SCFTs that have an enhancement to the DE-type $(2,0)$ SCFTs. From the geometric engineering perspective, almost all such theories are constructed in terms of elliptically-fibered Calabi--Yau threefolds where the base contains a collection of $(-2)$-curves intersecting according to the associated DE-type Dynkin diagram. The singular fibers over these curves were of special unitary type, and interacting non-product SCFTs on the Higgs branch were obtained by (consistently) reducing the ranks of these special unitary algebras. Thus, this geometric picture allows the derivation (of the interacting non-product subdiagram) of the Higgs branch Hasse diagram, but without the information of the transverse slices in the foliation of the symplectic singularity. 

Focusing on the Higgsable to D-type $(2,0)$ theories, a Type IIA realisation is explicitly given via a system of D6-D8-NS5-branes in the presence of an $\mathrm{ON}^-$-plane. This brane construction of the theory is crucial since it allows, with due extensions, the application of the magnetic quiver technique to study the Higgs branch of the six-dimensional theory. In fact, from a joint approach intertwining tensor branch geometries and magnetic quivers, it has been rendered possible to describe not only the various leaves in the Higgs branch of $D_N^{\mathfrak{su}_{2k}}(O)$ theories but also the transverse slices between them. The lessons learnt from the D-type case can then be conjectured to extend to Higgsable to E-type $(2,0)$ models, for which a controlled Type IIA brane system is not available. Therefore, the proposed new slice subtraction method allows the construction of the Hasse diagram by simply looking at the tensor branch curve configuration and subtracting the Dynkin labels of the appropriate Dynkin diagram from the ranks of the gauge algebras.

While most of this paper was concerned with the 6d $(1,0)$ SCFTs that have a Higgs branch renormalization group flow to the 6d $(2,0)$ SCFTs; the same techniques can be used to study the Higgs branch of little string theories. We considered the natural ``affinized extension'' of the SCFTs considered in this paper, that is, the 6d $(1,0)$ LSTs that are Higgsable to the 6d $(2,0)$ LSTs. We derive the magnetic quivers for the Higgs branches and determine the structure constants of the generalized symmetries, which led to some interesting observations about T-dualities of LSTs.

\paragraph{\textbf{\uline{Quiver subtraction for (unitary-)orthosymplectic quivers:}}}

As we have highlighted throughout Section \ref{sec:Higgsbranch}, when constructing the Hasse diagram from the unitary-orthosymplectic magnetic quiver there are several subtleties. One important caveat was that the unitary-orthosymplectic quiver subtraction algorithm that we motivated from the brane dynamics perspective is incomplete; subtractions involving the special orthogonal gauge node with antisymmetric matter were not defined. Thus, the Hasse diagram as derived via the subtraction algorithm is only a subdiagram of the Higgs branch Hasse diagram of the 6d SCFT. This subtlety is deeply connected to the orthosymplectic analogue of the notion of decoration \cite{Bourget:2022tmw}, which has yet to be understood. Relatedly, the 6d $(2,0)$ SCFT that lives at the ``end'' of the Higgs branch RG flow still has a non-trivial Higgs branch, indicating that there exist further Higgsings which we have not captured; in Section \ref{sec:product}, we understood some of these Higgsings from a geometric perspective.

\begin{figure}[t]
    \centering
    \begin{tikzpicture}
        \node[] at (0,0) (0) {\footnotesize $\varnothing$};
        \node[] [below=1cm of 0] (2) {\footnotesize $2$};
        \node[] [below left=1cm and 0.9cm of 2] (2u2) {\footnotesize $2\sqcup2$};
        \node[] [below right=1cm and 1cm of 2] (22) {\footnotesize $22$};
        \node[] [below=2.5cm of 2] (222) {\footnotesize $222$};
        \node[] [below=1cm of 222] (22su22) {\footnotesize $2 \stackon{$2$}{$\su_2$} 2$};

        \draw (0)--(2) node[midway, right] () {\footnotesize $A_1$};
        \draw (2)--(2u2) node[midway, above left] () {\footnotesize $A_1$};
        \draw (2)--(22) node[midway, above right] () {\footnotesize $m$};
        \draw (2u2)--(222) node[midway, below left] () {\footnotesize $A_1$};
        \draw (22)--(222) node[midway, below right] () {\footnotesize $m$};
        \draw (222)--(22su22) node[midway, right] () {\footnotesize $\mathcal{Y}(4)\cong a_2$};
    \end{tikzpicture} $\qquad$
    \begin{tikzpicture}
    \node[tnode] at (0,0) (label) {\footnotesize $A$-type or $d_4/\mathfrak{S}_4(1)$:};
    \node[node, label=below:{\footnotesize $2$}] (Zk) [right=6mm of label] {};
      \node[node, label=below:{\footnotesize $4$}] (Fk) [right=6mm of Zk] {};
      \draw (Zk.east) -- (Fk.west);
      \draw (Fk) to[out=130, in=410, looseness=12] (Fk);
      
        \node[tnode] at (0,-2) (label) {$d_4/\mathfrak{S}_4(2)$:};
        \node[node, label=below:{\footnotesize $1$}] (Z1) [right=6mm of label] {};
    \node[node, label=below:{\footnotesize $2$}] (Zk) [right=6mm of Z1] {};
      \node[node, label=below:{\footnotesize $3$}] (Fk) [right=6mm of Zk] {};
      \draw (Z1.east) -- (Zk.west);
      \draw[transform canvas={xshift=0pt,yshift=1pt}] (Zk.east)-- (Fk.west);
      \draw[transform canvas={xshift=0pt,yshift=-1pt}] (Zk.east)-- (Fk.west);
      \draw (3.1,-2.1)--(3,-2)--(3.1,-1.9);
      \draw (Fk) to[out=130, in=410, looseness=12] (Fk);

              \node[tnode] at (0,-4) (label) {$D$-type:};
        \node[node, label=below:{\footnotesize $1$}] (Z1) [right=6mm of label] {};
    \node[node,fill=blue, label=below:{\footnotesize $2$}] (Zk) [right=6mm of Z1] {};
      \node[node,fill=red, label=below:{\footnotesize $6$}] (Fk) [right=6mm of Zk] {};
      \draw (Z1.east) -- (Zk.west);
      \draw(Zk.east)-- (Fk.west);
      \draw (Fk) to[out=130, in=410, looseness=12] (Fk);
    \end{tikzpicture}
    
    \caption{Hasse diagram for the $D_3^{\mathfrak{su}_2}([1^2])$ theory, along with the different magnetic quiver realizations of the Higgs branch.}
    \label{fig:something_Hasse}
\end{figure}
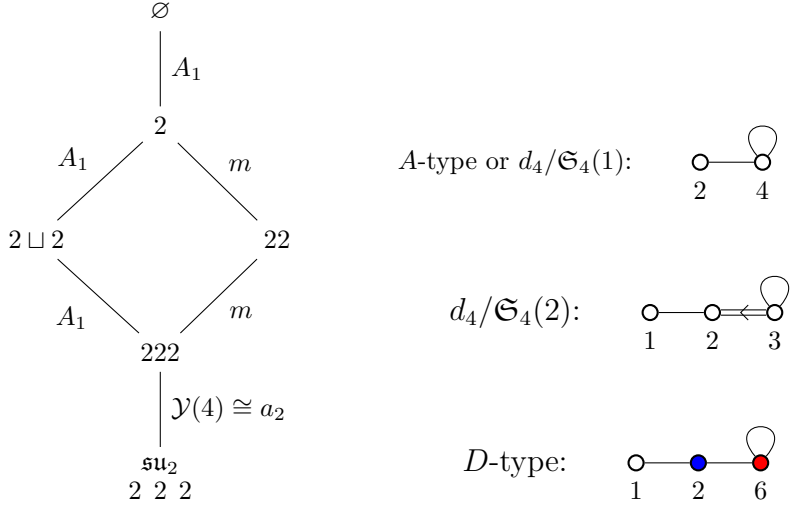

An extreme example of this subtlety with decoration in the unitary-orthosymplectic case can be seen in the $D_3^{\mathfrak{su}_2}([1^2])$ SCFT. In addition to the unitary-orthosymplectic magnetic quiver for the Higgs branch that has been the principle object of study in this paper, there exist two unitary 3d $\mathcal{N}=4$ quivers which have the same Coulomb branch Hilbert series as the unitary-orthosymplectic quiver \cite{Hanany:2023uzn}. These three quivers are depicted on the right in Figure \ref{fig:something_Hasse}. Whereas in the unitary case, the extraction of the Hasse diagram (shown on the left in Figure \ref{fig:something_Hasse}) from the quiver is fully under control and reproduces that expected from the 6d geometry, for the unitary-orthosymplectic quiver the lack of a notion of decoration hinders an analogous procedure, and only a subdiagram of the full Hasse diagram can be produced.\footnote{In fact, in this $D_3^{\mathfrak{su}_2}([1^2])$ example, the subdiagram observable from the unitary-orthosymplectic quiver consists only of a single vertex; in more general examples it is a more involved subdiagram.} This plurality of descriptions, from the geometry and alternative unitary quivers, should provide an important hint at the nature of decoration for orthosymplectic quivers.

\paragraph{\textbf{\uline{Non-invertible symmetries across the Higgs branch:}}} The 6d $(2,0)$ SCFTs of type $D_{2n}$ have non-invertible global symmetries \cite{Lawrie:2023tdz}, realized via the duality defect construction \cite{Choi:2021kmx,Kaidi:2021xfk}. This occurs because the intermediate defect group, $\mathbb{Z}_2 \oplus \mathbb{Z}_2$, admits distinct polarizations, related via gauging of two-form symmetries, and there exists a Green--Schwarz automorphism/duality \cite{Apruzzi:2017iqe} which acts on the charge lattice of string-like defects. When combined, these lead to non-invertible duality defects; see \cite{Lawrie:2023tdz,Gukov:2020btk} for details and references. For long 6d $(1,0)$ SCFTs $D_N^{\mathfrak{su}_{2k}}(O)$, the $\mathbb{Z}_2$ Green--Schwarz automorphism still exists, and thus the non-invertible symmetries are also realized in such theories; that is, at any interacting non-product fixed point on the Higgs branch of these Higgsable to D-type $(2,0)$ SCFTs, the non-invertible symmetry is preserved. In contrast, for short quivers, certain choices of generalized partition can break the $\mathbb{Z}_2$ Green--Schwarz automorphism, thereby destroying the duality defect construction of the non-invertible symmetries. Thus, for short quivers, the non-invertible symmetries appear to be alternatively broken and emergent along various loci in the Higgs branch. It would be interesting to make a detailed study of the microscopic behavior of such non-invertible symmetries along Higgs branch renormalization group flows.

\paragraph{\textbf{\uline{Higgs branches and 4d reductions:}}} 

While we have studied the Higgs branches of 6d $(1,0)$ SCFTs in this paper, it is well-known that compactifying on a $d$-dimensional torus $T^d$ leads to a lower-dimensional field theory which has the same Higgs branch as the original 6d SCFT. Such lower-dimensional theories may have dual constructions, and this may lead to different perspectives on the Higgs branch. For example, a 6d $(1,0)$ SCFT compactified on a $T^2$ may have a dual description via the class $\mathcal{S}$ construction \cite{Gaiotto:2009hg,Gaiotto:2009we}. That is, starting from a 6d $(2,0)$ SCFT and compactifying on a punctured Riemann surface. Class $\mathcal{S}$ theories have known magnetic quivers, see e.g., \cite{Benini:2010uu}, and thus any class $\mathcal{S}$ duals can lead to distinct magnetic quivers for the 6d Higgs branch, and patching together these different descriptions may lead to an understanding of aspects of the Higgs branch that are obscured in one formulation. This plurality of origins was vital for studying the Higgs branch of 6d conformal matter in \cite{Baume:2021qho,DKL}.

Furthermore, when compactifying on a torus, one is free to turn on a non-trivial Stiefel--Whitney twist, which leads to classes of novel 4d $\mathcal{N}=2$ SCFTs related to S-folds \cite{Heckman:2020svr,Giacomelli:2020jel,Giacomelli:2020gee,Heckman:2022suy,Apruzzi:2020pmv,Giacomelli:2024dbd}; these theories often have an interesting Higgs branch structure that can be related to the magnetic quiver for the 6d $(1,0)$ Higgs branch via folding-type operations \cite{Bourget:2020mez,Bourget:2021csg}. Determining the structure of the Higgs branch of such SCFTs originating in the 6d $(1,0)$ SCFTs Higgsable to D-type would be a natural extension of the current work. In another direction, the 6d $(1,0)$ SCFTs Higgsable to the A-type $(2,0)$ SCFT were important for proving subtle dualities amongst class $\mathcal{S}$ theories in \cite{Distler:2022kjb}; it is natural to ask if there are analogous dualities that can also be studied from the perspective of the Higgs branch in the D-type sector.

\paragraph{\textbf{\uline{Complex structure deformations of \boldmath{$\mathrm{CY}_3$}:}}} From the geometric engineering perspective, the Higgs branch of 6d $(1,0)$ SCFTs is encoded in the complex structure deformations of the associated Calabi--Yau threefold. A systematic study of such deformation spaces has not been carried out (however, see \cite{Anderson:2013rka,Anderson:2017rpr}), even at the level of determining when two Calabi--Yau threefolds associated with 6d $(1,0)$ SCFTs are connected, let alone the extraction of the transverse slice between them. Given that the approach via the magnetic quiver for the Higgs branch, for theories which admit a brane engineering description, can be utilized to determine the structure of the Higgs branch, it should be possible to draw a clearer connection between the quiver subtraction algorithm for 3d $\mathcal{N}=4$ quivers and complex structure deformations of singular elliptically-fibered Calabi--Yau threefolds. For specific geometries engineering little string theories, steps have been taken in this direction in \cite{Mansi:2023faa}, where all the deformations have been tracked and associated with different transverse slices in the Hasse diagram for LSTs where the tensor branch geometry involves a single curve of self-intersection $0$. We hope to return to such questions in the future.

\subsection*{Acknowledgements}

We thank Florent Baume, Jacques Distler, Julius Grimminger, Monica Jinwoo Kang, and Deshuo Liu for interesting discussions. 
The authors acknowledge support from DESY (Hamburg, Germany), a member of the Helmholtz Association HGF. This work was partially supported by the Deutsche Forschungsgemeinschaft under Germany's Excellence Strategy - EXC 2121 ``Quantum Universe'' - 390833306 and the Collaborative Research Center - SFB 1624 ``Higher Structures, Moduli Spaces, and Integrability'' - 506632645.

\bibliography{bibliography}{}

\providecommand{\href}[2]{#2}\begingroup\raggedright\begin{thebibliography}{100}

\bibitem{Aharony:1999ks}
O.~Aharony, ``{A brief review of `little string theories'},''
  \href{http://dx.doi.org/10.1088/0264-9381/17/5/302}{{\em Class. Quant. Grav.}
  {\bfseries 17} (2000) 929--938},
  \href{http://arxiv.org/abs/hep-th/9911147}{{\ttfamily arXiv:hep-th/9911147}}.

\bibitem{Ahmed:2023lhj}
H.~Ahmed, P.-K. Oehlmann, and F.~Ruehle, ``{T-Duality and Flavor Symmetries in
  Little String Theories},'' \href{http://arxiv.org/abs/2311.02168}{{\ttfamily
  arXiv:2311.02168 [hep-th]}}.

\bibitem{Anderson:2013rka}
L.~B. Anderson, J.~J. Heckman, and S.~Katz, ``{T-Branes and Geometry},''
  \href{http://dx.doi.org/10.1007/JHEP05(2014)080}{{\em JHEP} {\bfseries 05}
  (2014) 080}, \href{http://arxiv.org/abs/1310.1931}{{\ttfamily arXiv:1310.1931
  [hep-th]}}.

\bibitem{Anderson:2017rpr}
L.~B. Anderson, J.~J. Heckman, S.~Katz, and L.~P. Schaposnik, ``{T-Branes at
  the Limits of Geometry},''
  \href{http://dx.doi.org/10.1007/JHEP10(2017)058}{{\em JHEP} {\bfseries 10}
  (2017) 058}, \href{http://arxiv.org/abs/1702.06137}{{\ttfamily
  arXiv:1702.06137 [hep-th]}}.

\bibitem{Apruzzi:2020pmv}
F.~Apruzzi, S.~Giacomelli, and S.~Sch\"afer-Nameki, ``{4d $\mathcal{N}=2$
  S-folds},'' \href{http://dx.doi.org/10.1103/PhysRevD.101.106008}{{\em Phys.
  Rev. D} {\bfseries 101} no.~10, (2020) 106008},
  \href{http://arxiv.org/abs/2001.00533}{{\ttfamily arXiv:2001.00533
  [hep-th]}}.

\bibitem{Apruzzi:2017iqe}
F.~Apruzzi, J.~J. Heckman, and T.~Rudelius, ``{Green-Schwarz Automorphisms and
  6D SCFTs},'' \href{http://dx.doi.org/10.1007/JHEP02(2018)157}{{\em JHEP}
  {\bfseries 02} (2018) 157}, \href{http://arxiv.org/abs/1707.06242}{{\ttfamily
  arXiv:1707.06242 [hep-th]}}.

\bibitem{Argyres:2022mnu}
P.~C. Argyres, J.~J. Heckman, K.~Intriligator, and M.~Martone, ``{Snowmass
  White Paper on SCFTs},'' \href{http://arxiv.org/abs/2202.07683}{{\ttfamily
  arXiv:2202.07683 [hep-th]}}.

\bibitem{Aspinwall:1997ye}
P.~S. Aspinwall and D.~R. Morrison, ``{Point - like instantons on K3
  orbifolds},'' \href{http://dx.doi.org/10.1016/S0550-3213(97)00516-6}{{\em
  Nucl. Phys. B} {\bfseries 503} (1997) 533--564},
  \href{http://arxiv.org/abs/hep-th/9705104}{{\ttfamily arXiv:hep-th/9705104}}.

\bibitem{Bastian:2017ing}
B.~Bastian, S.~Hohenegger, A.~Iqbal, and S.-J. Rey, ``{Dual little strings and
  their partition functions},''
  \href{http://dx.doi.org/10.1103/PhysRevD.97.106004}{{\em Phys. Rev. D}
  {\bfseries 97} no.~10, (2018) 106004},
  \href{http://arxiv.org/abs/1710.02455}{{\ttfamily arXiv:1710.02455
  [hep-th]}}.

\bibitem{Bastian:2017ary}
B.~Bastian, S.~Hohenegger, A.~Iqbal, and S.-J. Rey, ``{Triality in Little
  String Theories},'' \href{http://dx.doi.org/10.1103/PhysRevD.97.046004}{{\em
  Phys. Rev. D} {\bfseries 97} no.~4, (2018) 046004},
  \href{http://arxiv.org/abs/1711.07921}{{\ttfamily arXiv:1711.07921
  [hep-th]}}.

\bibitem{Baume:2020ure}
F.~Baume, J.~J. Heckman, and C.~Lawrie, ``{6D SCFTs, 4D SCFTs, Conformal
  Matter, and Spin Chains},''
  \href{http://dx.doi.org/10.1016/j.nuclphysb.2021.115401}{{\em Nucl. Phys. B}
  {\bfseries 967} (2021) 115401},
  \href{http://arxiv.org/abs/2007.07262}{{\ttfamily arXiv:2007.07262
  [hep-th]}}.

\bibitem{Baume:2022cot}
F.~Baume, J.~J. Heckman, and C.~Lawrie, ``{Super-spin chains for 6D SCFTs},''
  \href{http://dx.doi.org/10.1016/j.nuclphysb.2023.116250}{{\em Nucl. Phys. B}
  {\bfseries 992} (2023) 116250},
  \href{http://arxiv.org/abs/2208.02272}{{\ttfamily arXiv:2208.02272
  [hep-th]}}.

\bibitem{Baume:2021qho}
F.~Baume, M.~J. Kang, and C.~Lawrie, ``{Two 6D origins of 4D SCFTs: Class S and
  6D (1,\,0) on a torus},''
  \href{http://dx.doi.org/10.1103/PhysRevD.106.086003}{{\em Phys. Rev. D}
  {\bfseries 106} no.~8, (2022) 086003},
  \href{http://arxiv.org/abs/2106.11990}{{\ttfamily arXiv:2106.11990
  [hep-th]}}.

\bibitem{Baume:2023onr}
F.~Baume and C.~Lawrie, ``{The Bestiary of 6d (1,0) SCFTs: Nilpotent Orbits and
  Anomalies},'' \href{http://arxiv.org/abs/2312.13347}{{\ttfamily
  arXiv:2312.13347 [hep-th]}}.

\bibitem{FELLAS}
F.~Baume, P.-K. Oehlmann, and F.~Ruehle, ``{Bounds and Dualities of Type II
  Little String Theories},'' \href{http://arxiv.org/abs/2405.03877}{{\ttfamily
  arXiv:2405.03877 [hep-th]}}.

\bibitem{Beauville_2000}
A.~Beauville, ``Symplectic singularities,''
  \href{http://dx.doi.org/10.1007/s002229900043}{{\em Inventiones Mathematicae}
  {\bfseries 139} no.~3, (Mar, 2000) 541--549}.
  \url{https://doi.org/10.1007%2Fs002229900043}.

\bibitem{Benini:2010uu}
F.~Benini, Y.~Tachikawa, and D.~Xie, ``{Mirrors of 3d Sicilian theories},''
  \href{http://dx.doi.org/10.1007/JHEP09(2010)063}{{\em JHEP} {\bfseries 09}
  (2010) 063}, \href{http://arxiv.org/abs/1007.0992}{{\ttfamily arXiv:1007.0992
  [hep-th]}}.

\bibitem{Beratto:2020wmn}
E.~Beratto, S.~Giacomelli, N.~Mekareeya, and M.~Sacchi, ``{3d mirrors of the
  circle reduction of twisted A$_{2N}$ theories of class S},''
  \href{http://dx.doi.org/10.1007/JHEP09(2020)161}{{\em JHEP} {\bfseries 09}
  (2020) 161}, \href{http://arxiv.org/abs/2007.05019}{{\ttfamily
  arXiv:2007.05019 [hep-th]}}.

\bibitem{Bergman:2020bvi}
O.~Bergman, M.~Fazzi, D.~Rodr\'\i{}guez-G\'omez, and A.~Tomasiello, ``{Charges
  and holography in 6d (1,0) theories},''
  \href{http://dx.doi.org/10.1007/JHEP05(2020)138}{{\em JHEP} {\bfseries 05}
  (2020) 138}, \href{http://arxiv.org/abs/2002.04036}{{\ttfamily
  arXiv:2002.04036 [hep-th]}}.

\bibitem{Bhardwaj:2022ekc}
L.~Bhardwaj, ``{Discovering T-dualities of little string theories},''
  \href{http://dx.doi.org/10.1007/JHEP02(2024)046}{{\em JHEP} {\bfseries 02}
  (2024) 046}, \href{http://arxiv.org/abs/2209.10548}{{\ttfamily
  arXiv:2209.10548 [hep-th]}}.

\bibitem{Bhardwaj:2015oru}
L.~Bhardwaj, M.~Del~Zotto, J.~J. Heckman, D.~R. Morrison, T.~Rudelius, and
  C.~Vafa, ``{F-theory and the Classification of Little Strings},''
  \href{http://dx.doi.org/10.1103/PhysRevD.93.086002}{{\em Phys. Rev. D}
  {\bfseries 93} no.~8, (2016) 086002},
  \href{http://arxiv.org/abs/1511.05565}{{\ttfamily arXiv:1511.05565
  [hep-th]}}. [Erratum: Phys.Rev.D 100, 029901 (2019)].

\bibitem{Bourget:2019aer}
A.~Bourget, S.~Cabrera, J.~F. Grimminger, A.~Hanany, M.~Sperling, A.~Zajac, and
  Z.~Zhong, ``{The Higgs mechanism \textemdash{} Hasse diagrams for symplectic
  singularities},'' \href{http://dx.doi.org/10.1007/JHEP01(2020)157}{{\em JHEP}
  {\bfseries 01} (2020) 157}, \href{http://arxiv.org/abs/1908.04245}{{\ttfamily
  arXiv:1908.04245 [hep-th]}}.

\bibitem{Bourget:2020mez}
A.~Bourget, S.~Giacomelli, J.~F. Grimminger, A.~Hanany, M.~Sperling, and
  Z.~Zhong, ``{S-fold magnetic quivers},''
  \href{http://dx.doi.org/10.1007/JHEP02(2021)054}{{\em JHEP} {\bfseries 02}
  (2021) 054}, \href{http://arxiv.org/abs/2010.05889}{{\ttfamily
  arXiv:2010.05889 [hep-th]}}.

\bibitem{Bourget:2022tmw}
A.~Bourget and J.~F. Grimminger, ``{Fibrations and Hasse diagrams for 6d
  SCFTs},'' \href{http://dx.doi.org/10.1007/JHEP12(2022)159}{{\em JHEP}
  {\bfseries 12} (2022) 159}, \href{http://arxiv.org/abs/2209.15016}{{\ttfamily
  arXiv:2209.15016 [hep-th]}}.

\bibitem{Bourget:2022ehw}
A.~Bourget, J.~F. Grimminger, A.~Hanany, and Z.~Zhong, ``{The Hasse diagram of
  the moduli space of instantons},''
  \href{http://dx.doi.org/10.1007/JHEP08(2022)283}{{\em JHEP} {\bfseries 08}
  (2022) 283}, \href{http://arxiv.org/abs/2202.01218}{{\ttfamily
  arXiv:2202.01218 [hep-th]}}.

\bibitem{Bourget:2021csg}
A.~Bourget, J.~F. Grimminger, M.~Martone, and G.~Zafrir, ``{Magnetic quivers
  for rank 2 theories},'' \href{http://dx.doi.org/10.1007/JHEP03(2022)208}{{\em
  JHEP} {\bfseries 03} (2022) 208},
  \href{http://arxiv.org/abs/2110.11365}{{\ttfamily arXiv:2110.11365
  [hep-th]}}.

\bibitem{Bourget:2023dkj}
A.~Bourget, M.~Sperling, and Z.~Zhong, ``{Decay and Fission of Magnetic
  Quivers},'' \href{http://arxiv.org/abs/2312.05304}{{\ttfamily
  arXiv:2312.05304 [hep-th]}}.

\bibitem{Bourget:2024mgn}
A.~Bourget, M.~Sperling, and Z.~Zhong, ``{Higgs branch RG-flows via Decay and
  Fission},'' \href{http://arxiv.org/abs/2401.08757}{{\ttfamily
  arXiv:2401.08757 [hep-th]}}.

\bibitem{Buican:2016hpb}
M.~Buican, J.~Hayling, and C.~Papageorgakis, ``{Aspects of Superconformal
  Multiplets in $D>4$},'' \href{http://dx.doi.org/10.1007/JHEP11(2016)091}{{\em
  JHEP} {\bfseries 11} (2016) 091},
\href{http://arxiv.org/abs/1606.00810}{{\ttfamily arXiv:1606.00810 [hep-th]}}.

\bibitem{Cabrera:2016vvv}
S.~Cabrera and A.~Hanany, ``{Branes and the Kraft-Procesi Transition},''
  \href{http://dx.doi.org/10.1007/JHEP11(2016)175}{{\em JHEP} {\bfseries 11}
  (2016) 175}, \href{http://arxiv.org/abs/1609.07798}{{\ttfamily
  arXiv:1609.07798 [hep-th]}}.

\bibitem{Cabrera:2017njm}
S.~Cabrera and A.~Hanany, ``{Branes and the Kraft-Procesi transition: classical
  case},'' \href{http://dx.doi.org/10.1007/JHEP04(2018)127}{{\em JHEP}
  {\bfseries 04} (2018) 127}, \href{http://arxiv.org/abs/1711.02378}{{\ttfamily
  arXiv:1711.02378 [hep-th]}}.

\bibitem{Cabrera:2018ann}
S.~Cabrera and A.~Hanany, ``{Quiver Subtractions},''
  \href{http://dx.doi.org/10.1007/JHEP09(2018)008}{{\em JHEP} {\bfseries 09}
  (2018) 008}, \href{http://arxiv.org/abs/1803.11205}{{\ttfamily
  arXiv:1803.11205 [hep-th]}}.

\bibitem{Cabrera:2019izd}
S.~Cabrera, A.~Hanany, and M.~Sperling, ``{Magnetic quivers, Higgs branches,
  and 6d $N$=(1,0) theories},''
  \href{http://dx.doi.org/10.1007/JHEP06(2019)071}{{\em JHEP} {\bfseries 06}
  (2019) 071}, \href{http://arxiv.org/abs/1904.12293}{{\ttfamily
  arXiv:1904.12293 [hep-th]}}. [Erratum: JHEP 07, 137 (2019)].

\bibitem{Cabrera:2019dob}
S.~Cabrera, A.~Hanany, and M.~Sperling, ``{Magnetic quivers, Higgs branches,
  and 6d $ \mathcal{N} $ = (1, 0) theories \textemdash{} orthogonal and
  symplectic gauge groups},''
  \href{http://dx.doi.org/10.1007/JHEP02(2020)184}{{\em JHEP} {\bfseries 02}
  (2020) 184}, \href{http://arxiv.org/abs/1912.02773}{{\ttfamily
  arXiv:1912.02773 [hep-th]}}.

\bibitem{Chen:2019njf}
J.~Chen, B.~Haghighat, S.~Liu, and M.~Sperling, ``{4d $N$=1 from 6d D-type
  $N$=(1,0)},'' \href{http://dx.doi.org/10.1007/JHEP01(2020)152}{{\em JHEP}
  {\bfseries 01} (2020) 152}, \href{http://arxiv.org/abs/1907.00536}{{\ttfamily
  arXiv:1907.00536 [hep-th]}}.

\bibitem{Choi:2021kmx}
Y.~Choi, C.~Cordova, P.-S. Hsin, H.~T. Lam, and S.-H. Shao, ``{Noninvertible
  duality defects in 3+1 dimensions},''
  \href{http://dx.doi.org/10.1103/PhysRevD.105.125016}{{\em Phys. Rev. D}
  {\bfseries 105} no.~12, (2022) 125016},
  \href{http://arxiv.org/abs/2111.01139}{{\ttfamily arXiv:2111.01139
  [hep-th]}}.

\bibitem{Cordova:2015fha}
C.~Cordova, T.~T. Dumitrescu, and K.~Intriligator, ``{Anomalies,
  renormalization group flows, and the a-theorem in six-dimensional (1, 0)
  theories},'' \href{http://dx.doi.org/10.1007/JHEP10(2016)080}{{\em JHEP}
  {\bfseries 10} (2016) 080}, \href{http://arxiv.org/abs/1506.03807}{{\ttfamily
  arXiv:1506.03807 [hep-th]}}.

\bibitem{Cordova:2016xhm}
C.~Cordova, T.~T. Dumitrescu, and K.~Intriligator, ``{Deformations of
  Superconformal Theories},''
  \href{http://dx.doi.org/10.1007/JHEP11(2016)135}{{\em JHEP} {\bfseries 11}
  (2016) 135}, \href{http://arxiv.org/abs/1602.01217}{{\ttfamily
  arXiv:1602.01217 [hep-th]}}.

\bibitem{Cordova:2016emh}
C.~Cordova, T.~T. Dumitrescu, and K.~Intriligator, ``{Multiplets of
  Superconformal Symmetry in Diverse Dimensions},''
\href{http://arxiv.org/abs/1612.00809}{{\ttfamily arXiv:1612.00809 [hep-th]}}.

\bibitem{Cordova:2020tij}
C.~Cordova, T.~T. Dumitrescu, and K.~Intriligator, ``{2-Group Global Symmetries
  and Anomalies in Six-Dimensional Quantum Field Theories},''
  \href{http://dx.doi.org/10.1007/JHEP04(2021)252}{{\em JHEP} {\bfseries 04}
  (2021) 252}, \href{http://arxiv.org/abs/2009.00138}{{\ttfamily
  arXiv:2009.00138 [hep-th]}}.

\bibitem{Cremonesi:2013lqa}
S.~Cremonesi, A.~Hanany, and A.~Zaffaroni, ``{Monopole operators and Hilbert
  series of Coulomb branches of $3d$ $\mathcal{N} = 4$ gauge theories},''
  \href{http://dx.doi.org/10.1007/JHEP01(2014)005}{{\em JHEP} {\bfseries 01}
  (2014) 005}, \href{http://arxiv.org/abs/1309.2657}{{\ttfamily arXiv:1309.2657
  [hep-th]}}.

\bibitem{DelZotto:2023nrb}
M.~Del~Zotto, M.~Fazzi, and S.~Giri, ``{The Higgs branch of heterotic ALE
  instantons},'' \href{http://arxiv.org/abs/2307.11087}{{\ttfamily
  arXiv:2307.11087 [hep-th]}}.

\bibitem{DelZotto:2014hpa}
M.~Del~Zotto, J.~J. Heckman, A.~Tomasiello, and C.~Vafa, ``{6d Conformal
  Matter},'' \href{http://dx.doi.org/10.1007/JHEP02(2015)054}{{\em JHEP}
  {\bfseries 02} (2015) 054}, \href{http://arxiv.org/abs/1407.6359}{{\ttfamily
  arXiv:1407.6359 [hep-th]}}.

\bibitem{DelZotto:2022xrh}
M.~Del~Zotto, M.~Liu, and P.-K. Oehlmann, ``{Back to Heterotic Strings on ALE
  Spaces: Part II -- Geometry of T-dual Little Strings},''
  \href{http://arxiv.org/abs/2212.05311}{{\ttfamily arXiv:2212.05311
  [hep-th]}}.

\bibitem{DelZotto:2023ahf}
M.~Del~Zotto, M.~Liu, and P.-K. Oehlmann, ``{6D Heterotic Little String
  Theories and F-theory Geometry: An Introduction},''
  \href{http://arxiv.org/abs/2303.13502}{{\ttfamily arXiv:2303.13502
  [hep-th]}}.

\bibitem{DelZotto:2022ohj}
M.~Del~Zotto, M.~Liu, and P.-K. Oehlmann, ``{Back to heterotic strings on ALE
  spaces. Part I. Instantons, 2-groups and T-duality},''
  \href{http://dx.doi.org/10.1007/JHEP01(2023)176}{{\em JHEP} {\bfseries 01}
  (2023) 176}, \href{http://arxiv.org/abs/2209.10551}{{\ttfamily
  arXiv:2209.10551 [hep-th]}}.

\bibitem{DelZotto:2020sop}
M.~Del~Zotto and K.~Ohmori, ``{2-Group Symmetries of 6D Little String Theories
  and T-Duality},'' \href{http://dx.doi.org/10.1007/s00023-021-01018-3}{{\em
  Annales Henri Poincare} {\bfseries 22} no.~7, (2021) 2451--2474},
  \href{http://arxiv.org/abs/2009.03489}{{\ttfamily arXiv:2009.03489
  [hep-th]}}.

\bibitem{Distler:2022kjb}
J.~Distler, G.~Elliot, M.~J. Kang, and C.~Lawrie, ``{Isomorphisms of 4D N=2
  SCFTs from 6D},'' \href{http://dx.doi.org/10.1103/PhysRevD.107.106005}{{\em
  Phys. Rev. D} {\bfseries 107} no.~10, (2023) 106005},
  \href{http://arxiv.org/abs/2212.11983}{{\ttfamily arXiv:2212.11983
  [hep-th]}}.

\bibitem{DKL}
J.~Distler, M.~J. Kang, and C.~Lawrie. {\it IN PREPARATION}.

\bibitem{Distler:2022yse}
J.~Distler, M.~J. Kang, and C.~Lawrie, ``{Distinguishing 6D (1, 0) SCFTs: An
  extension to the geometric construction},''
  \href{http://dx.doi.org/10.1103/PhysRevD.106.066011}{{\em Phys. Rev. D}
  {\bfseries 106} no.~6, (2022) 066011},
  \href{http://arxiv.org/abs/2203.08829}{{\ttfamily arXiv:2203.08829
  [hep-th]}}.

\bibitem{Fazzi:2023ulb}
M.~Fazzi, S.~Giri, and P.~Levy, ``{Proving the 6d a-theorem with the double
  affine Grassmannian},'' \href{http://arxiv.org/abs/2312.17178}{{\ttfamily
  arXiv:2312.17178 [hep-th]}}.

\bibitem{Ferlito:2016grh}
G.~Ferlito and A.~Hanany, ``{A tale of two cones: the Higgs Branch of Sp(n)
  theories with 2n flavours},''
  \href{http://arxiv.org/abs/1609.06724}{{\ttfamily arXiv:1609.06724
  [hep-th]}}.

\bibitem{Ferlito:2017xdq}
G.~Ferlito, A.~Hanany, N.~Mekareeya, and G.~Zafrir, ``{3d Coulomb branch and 5d
  Higgs branch at infinite coupling},''
  \href{http://dx.doi.org/10.1007/JHEP07(2018)061}{{\em JHEP} {\bfseries 07}
  (2018) 061}, \href{http://arxiv.org/abs/1712.06604}{{\ttfamily
  arXiv:1712.06604 [hep-th]}}.

\bibitem{Filoche:2023yfm}
B.~Filoche, S.~Hohenegger, and T.~Kimura, ``{Non-perturbative Symmetries of
  Little Strings and Affine Quiver Algebras},''
  \href{http://dx.doi.org/10.1007/JHEP02(2024)233}{{\em JHEP} {\bfseries 02}
  (2024) 233}, \href{http://arxiv.org/abs/2311.03858}{{\ttfamily
  arXiv:2311.03858 [hep-th]}}.

\bibitem{MR3570131}
B.~Fu, D.~Juteau, P.~Levy, and E.~Sommers, ``Generic singularities of nilpotent
  orbit closures,'' \href{http://dx.doi.org/10.1016/j.aim.2016.09.010}{{\em
  Adv. Math.} {\bfseries 305} (2017) 1--77}.
  \url{https://doi.org/10.1016/j.aim.2016.09.010}.

\bibitem{Gaiotto:2009we}
D.~Gaiotto, ``{N=2 dualities},''
  \href{http://dx.doi.org/10.1007/JHEP08(2012)034}{{\em JHEP} {\bfseries 08}
  (2012) 034}, \href{http://arxiv.org/abs/0904.2715}{{\ttfamily arXiv:0904.2715
  [hep-th]}}.

\bibitem{Gaiotto:2009hg}
D.~Gaiotto, G.~W. Moore, and A.~Neitzke, ``{Wall-crossing, Hitchin systems, and
  the WKB approximation},''
  \href{http://dx.doi.org/10.1016/j.aim.2012.09.027}{{\em Adv. Math.}
  {\bfseries 234} (2013) 239--403},
  \href{http://arxiv.org/abs/0907.3987}{{\ttfamily arXiv:0907.3987 [hep-th]}}.

\bibitem{Gaiotto:2008ak}
D.~Gaiotto and E.~Witten, ``{S-Duality of Boundary Conditions In N=4 Super
  Yang-Mills Theory},''
  \href{http://dx.doi.org/10.4310/ATMP.2009.v13.n3.a5}{{\em Adv. Theor. Math.
  Phys.} {\bfseries 13} no.~3, (2009) 721--896},
  \href{http://arxiv.org/abs/0807.3720}{{\ttfamily arXiv:0807.3720 [hep-th]}}.

\bibitem{Giacomelli:2020gee}
S.~Giacomelli, M.~Martone, Y.~Tachikawa, and G.~Zafrir, ``{More on $\mathcal{N}
  =2$ S-folds},'' \href{http://dx.doi.org/10.1007/JHEP01(2021)054}{{\em JHEP}
  {\bfseries 01} (2021) 054}, \href{http://arxiv.org/abs/2010.03943}{{\ttfamily
  arXiv:2010.03943 [hep-th]}}.

\bibitem{Giacomelli:2020jel}
S.~Giacomelli, C.~Meneghelli, and W.~Peelaers, ``{New $ \mathcal{N} $ = 2
  superconformal field theories from $ \mathcal{S} $-folds},''
  \href{http://dx.doi.org/10.1007/JHEP01(2021)022}{{\em JHEP} {\bfseries 01}
  (2021) 022}, \href{http://arxiv.org/abs/2007.00647}{{\ttfamily
  arXiv:2007.00647 [hep-th]}}.

\bibitem{Giacomelli:2024dbd}
S.~Giacomelli, R.~Savelli, and G.~Zoccarato, ``{$\mathcal{N} = 2$
  Orbi-S-Folds},'' \href{http://arxiv.org/abs/2405.00101}{{\ttfamily
  arXiv:2405.00101 [hep-th]}}.

\bibitem{Gledhill:2021cbe}
K.~Gledhill and A.~Hanany, ``{Coulomb branch global symmetry and quiver
  addition},'' \href{http://dx.doi.org/10.1007/JHEP12(2021)127}{{\em JHEP}
  {\bfseries 12} (2021) 127}, \href{http://arxiv.org/abs/2109.07237}{{\ttfamily
  arXiv:2109.07237 [hep-th]}}.

\bibitem{Grimminger:2020dmg}
J.~F. Grimminger and A.~Hanany, ``{Hasse diagrams for 3d $ \mathcal{N} $ = 4
  quiver gauge theories \textemdash{} Inversion and the full moduli space},''
  \href{http://dx.doi.org/10.1007/JHEP09(2020)159}{{\em JHEP} {\bfseries 09}
  (2020) 159}, \href{http://arxiv.org/abs/2004.01675}{{\ttfamily
  arXiv:2004.01675 [hep-th]}}.

\bibitem{Gukov:2020btk}
S.~Gukov, P.-S. Hsin, and D.~Pei, ``{Generalized global symmetries of $T[M]$
  theories. Part I},'' \href{http://dx.doi.org/10.1007/JHEP04(2021)232}{{\em
  JHEP} {\bfseries 04} (2021) 232},
  \href{http://arxiv.org/abs/2010.15890}{{\ttfamily arXiv:2010.15890
  [hep-th]}}.

\bibitem{Haghighat:2018dwe}
B.~Haghighat, J.~Kim, W.~Yan, and S.-T. Yau, ``{D-type fiber-base duality},''
  \href{http://dx.doi.org/10.1007/JHEP09(2018)060}{{\em JHEP} {\bfseries 09}
  (2018) 060}, \href{http://arxiv.org/abs/1806.10335}{{\ttfamily
  arXiv:1806.10335 [hep-th]}}.

\bibitem{Hanany:2023uzn}
A.~Hanany, G.~Kumaran, C.~Li, D.~Liu, and M.~Sperling, ``{Actions on the quiver
  -- Discrete quotients on the Coulomb branch},''
  \href{http://arxiv.org/abs/2311.02773}{{\ttfamily arXiv:2311.02773
  [hep-th]}}.

\bibitem{Hanany:2018uhm}
A.~Hanany and N.~Mekareeya, ``{The small E$_{8}$ instanton and the Kraft
  Procesi transition},'' \href{http://dx.doi.org/10.1007/JHEP07(2018)098}{{\em
  JHEP} {\bfseries 07} (2018) 098},
  \href{http://arxiv.org/abs/1801.01129}{{\ttfamily arXiv:1801.01129
  [hep-th]}}.

\bibitem{Hanany:2022itc}
A.~Hanany and M.~Sperling, ``{Magnetic quivers and negatively charged
  branes},'' \href{http://dx.doi.org/10.1007/JHEP11(2022)010}{{\em JHEP}
  {\bfseries 11} (2022) 010}, \href{http://arxiv.org/abs/2208.07270}{{\ttfamily
  arXiv:2208.07270 [hep-th]}}.

\bibitem{Hanany:1996ie}
A.~Hanany and E.~Witten, ``{Type IIB superstrings, BPS monopoles, and
  three-dimensional gauge dynamics},''
  \href{http://dx.doi.org/10.1016/S0550-3213(97)00157-0}{{\em Nucl. Phys. B}
  {\bfseries 492} (1997) 152--190},
  \href{http://arxiv.org/abs/hep-th/9611230}{{\ttfamily arXiv:hep-th/9611230}}.

\bibitem{Hanany:1997gh}
A.~Hanany and A.~Zaffaroni, ``{Branes and six-dimensional supersymmetric
  theories},'' \href{http://dx.doi.org/10.1016/S0550-3213(98)00355-1}{{\em
  Nucl. Phys. B} {\bfseries 529} (1998) 180--206},
  \href{http://arxiv.org/abs/hep-th/9712145}{{\ttfamily arXiv:hep-th/9712145}}.

\bibitem{Hanany:1997sa}
A.~Hanany and A.~Zaffaroni, ``{Chiral symmetry from type IIA branes},''
  \href{http://dx.doi.org/10.1016/S0550-3213(97)00595-6}{{\em Nucl. Phys. B}
  {\bfseries 509} (1998) 145--168},
  \href{http://arxiv.org/abs/hep-th/9706047}{{\ttfamily arXiv:hep-th/9706047}}.

\bibitem{Hanany:1999sj}
A.~Hanany and A.~Zaffaroni, ``{Issues on orientifolds: On the brane
  construction of gauge theories with SO(2n) global symmetry},''
  \href{http://dx.doi.org/10.1088/1126-6708/1999/07/009}{{\em JHEP} {\bfseries
  07} (1999) 009}, \href{http://arxiv.org/abs/hep-th/9903242}{{\ttfamily
  arXiv:hep-th/9903242}}.

\bibitem{Hanany:2018vph}
A.~Hanany and G.~Zafrir, ``{Discrete Gauging in Six Dimensions},''
  \href{http://dx.doi.org/10.1007/JHEP07(2018)168}{{\em JHEP} {\bfseries 07}
  (2018) 168}, \href{http://arxiv.org/abs/1804.08857}{{\ttfamily
  arXiv:1804.08857 [hep-th]}}.

\bibitem{Hayashi:2015vhy}
H.~Hayashi, S.-S. Kim, K.~Lee, M.~Taki, and F.~Yagi, ``{More on 5d descriptions
  of 6d SCFTs},'' \href{http://dx.doi.org/10.1007/JHEP10(2016)126}{{\em JHEP}
  {\bfseries 10} (2016) 126}, \href{http://arxiv.org/abs/1512.08239}{{\ttfamily
  arXiv:1512.08239 [hep-th]}}.

\bibitem{Heckman:2022suy}
J.~J. Heckman, C.~Lawrie, L.~Lin, H.~Y. Zhang, and G.~Zoccarato, ``{6D SCFTs,
  center-flavor symmetries, and Stiefel-Whitney compactifications},''
  \href{http://dx.doi.org/10.1103/PhysRevD.106.066003}{{\em Phys. Rev. D}
  {\bfseries 106} no.~6, (2022) 066003},
  \href{http://arxiv.org/abs/2205.03411}{{\ttfamily arXiv:2205.03411
  [hep-th]}}.

\bibitem{Heckman:2020svr}
J.~J. Heckman, C.~Lawrie, T.~B. Rochais, H.~Y. Zhang, and G.~Zoccarato,
  ``{$S$-folds, string junctions, and $\mathcal{N} = 2$ SCFTs},''
  \href{http://dx.doi.org/10.1103/PhysRevD.103.086013}{{\em Phys. Rev. D}
  {\bfseries 103} no.~8, (2021) 086013},
  \href{http://arxiv.org/abs/2009.10090}{{\ttfamily arXiv:2009.10090
  [hep-th]}}.

\bibitem{Heckman:2015bfa}
J.~J. Heckman, D.~R. Morrison, T.~Rudelius, and C.~Vafa, ``{Atomic
  Classification of 6D SCFTs},''
  \href{http://dx.doi.org/10.1002/prop.201500024}{{\em Fortsch. Phys.}
  {\bfseries 63} (2015) 468--530},
  \href{http://arxiv.org/abs/1502.05405}{{\ttfamily arXiv:1502.05405
  [hep-th]}}.

\bibitem{Heckman:2015ola}
J.~J. Heckman, D.~R. Morrison, T.~Rudelius, and C.~Vafa, ``{Geometry of 6D RG
  Flows},'' \href{http://dx.doi.org/10.1007/JHEP09(2015)052}{{\em JHEP}
  {\bfseries 09} (2015) 052}, \href{http://arxiv.org/abs/1505.00009}{{\ttfamily
  arXiv:1505.00009 [hep-th]}}.

\bibitem{Heckman:2013pva}
J.~J. Heckman, D.~R. Morrison, and C.~Vafa, ``{On the Classification of 6D
  SCFTs and Generalized ADE Orbifolds},''
  \href{http://dx.doi.org/10.1007/JHEP05(2014)028}{{\em JHEP} {\bfseries 05}
  (2014) 028}, \href{http://arxiv.org/abs/1312.5746}{{\ttfamily arXiv:1312.5746
  [hep-th]}}. [Erratum: JHEP 06, 017 (2015)].

\bibitem{Heckman:2018jxk}
J.~J. Heckman and T.~Rudelius, ``{Top Down Approach to 6D SCFTs},''
  \href{http://dx.doi.org/10.1088/1751-8121/aafc81}{{\em J. Phys. A} {\bfseries
  52} no.~9, (2019) 093001}, \href{http://arxiv.org/abs/1805.06467}{{\ttfamily
  arXiv:1805.06467 [hep-th]}}.

\bibitem{Heckman:2018pqx}
J.~J. Heckman, T.~Rudelius, and A.~Tomasiello, ``{Fission, Fusion, and 6D RG
  Flows},'' \href{http://dx.doi.org/10.1007/JHEP02(2019)167}{{\em JHEP}
  {\bfseries 02} (2019) 167}, \href{http://arxiv.org/abs/1807.10274}{{\ttfamily
  arXiv:1807.10274 [hep-th]}}.

\bibitem{Hitchin:1986ea}
N.~J. Hitchin, A.~Karlhede, U.~Lindstrom, and M.~Rocek, ``{Hyperkahler Metrics
  and Supersymmetry},'' \href{http://dx.doi.org/10.1007/BF01214418}{{\em
  Commun. Math. Phys.} {\bfseries 108} (1987) 535}.

\bibitem{Hohenegger:2016eqy}
S.~Hohenegger, A.~Iqbal, and S.-J. Rey, ``{Self-Duality and Self-Similarity of
  Little String Orbifolds},''
  \href{http://dx.doi.org/10.1103/PhysRevD.94.046006}{{\em Phys. Rev. D}
  {\bfseries 94} no.~4, (2016) 046006},
  \href{http://arxiv.org/abs/1605.02591}{{\ttfamily arXiv:1605.02591
  [hep-th]}}.

\bibitem{Hohenegger:2016yuv}
S.~Hohenegger, A.~Iqbal, and S.-J. Rey, ``{Dual Little Strings from F-Theory
  and Flop Transitions},''
  \href{http://dx.doi.org/10.1007/JHEP07(2017)112}{{\em JHEP} {\bfseries 07}
  (2017) 112}, \href{http://arxiv.org/abs/1610.07916}{{\ttfamily
  arXiv:1610.07916 [hep-th]}}.

\bibitem{Intriligator:2014eaa}
K.~Intriligator, ``{6d, $ \mathcal{N}=\left(1,\;0\right) $ Coulomb branch
  anomaly matching},'' \href{http://dx.doi.org/10.1007/JHEP10(2014)162}{{\em
  JHEP} {\bfseries 10} (2014) 162},
  \href{http://arxiv.org/abs/1408.6745}{{\ttfamily arXiv:1408.6745 [hep-th]}}.

\bibitem{Kaidi:2021xfk}
J.~Kaidi, K.~Ohmori, and Y.~Zheng, ``{Kramers-Wannier-like Duality Defects in
  (3+1)D Gauge Theories},''
  \href{http://dx.doi.org/10.1103/PhysRevLett.128.111601}{{\em Phys. Rev.
  Lett.} {\bfseries 128} no.~11, (2022) 111601},
  \href{http://arxiv.org/abs/2111.01141}{{\ttfamily arXiv:2111.01141
  [hep-th]}}.

\bibitem{Kang:2023uvm}
M.~J. Kang and S.~Kang, ``{Central extensions of higher groups: Green--Schwarz
  mechanism and 2-connections},''
  \href{http://arxiv.org/abs/2311.14666}{{\ttfamily arXiv:2311.14666
  [hep-th]}}.

\bibitem{Kang:2022zsl}
M.~J. Kang, C.~Lawrie, K.-H. Lee, M.~Sacchi, and J.~Song, ``{Higgs branch,
  Coulomb branch, and Hall-Littlewood index},''
  \href{http://dx.doi.org/10.1103/PhysRevD.106.106021}{{\em Phys. Rev. D}
  {\bfseries 106} no.~10, (2022) 106021},
  \href{http://arxiv.org/abs/2207.05764}{{\ttfamily arXiv:2207.05764
  [hep-th]}}.

\bibitem{Kapustin:1998fa}
A.~Kapustin, ``{D(n) quivers from branes},''
  \href{http://dx.doi.org/10.1088/1126-6708/1998/12/015}{{\em JHEP} {\bfseries
  12} (1998) 015}, \href{http://arxiv.org/abs/hep-th/9806238}{{\ttfamily
  arXiv:hep-th/9806238}}.

\bibitem{katz1992gorenstein}
S.~Katz and D.~R. Morrison, ``Gorenstein threefold singularities with small
  resolutions via invariant theory for weyl groups,'' 1992.

\bibitem{Kim:2017xan}
J.~Kim and K.~Lee, ``{Little strings on D$_{n}$ orbifolds},''
  \href{http://dx.doi.org/10.1007/JHEP10(2017)045}{{\em JHEP} {\bfseries 10}
  (2017) 045}, \href{http://arxiv.org/abs/1702.03116}{{\ttfamily
  arXiv:1702.03116 [hep-th]}}.

\bibitem{Lawrie:2023uiu}
C.~Lawrie and L.~Mansi, ``{The Higgs Branch of Heterotic LSTs: Hasse Diagrams
  and Generalized Symmetries},''
  \href{http://arxiv.org/abs/2312.05306}{{\ttfamily arXiv:2312.05306
  [hep-th]}}.

\bibitem{Lawrie:2023tdz}
C.~Lawrie, X.~Yu, and H.~Y. Zhang, ``{Intermediate defect groups, polarization
  pairs, and noninvertible duality defects},''
  \href{http://dx.doi.org/10.1103/PhysRevD.109.026005}{{\em Phys. Rev. D}
  {\bfseries 109} no.~2, (2024) 026005},
  \href{http://arxiv.org/abs/2306.11783}{{\ttfamily arXiv:2306.11783
  [hep-th]}}.

\bibitem{Louis:2015mka}
J.~Louis and S.~L\"ust, ``{Supersymmetric AdS$_{7}$ backgrounds in half-maximal
  supergravity and marginal operators of (1, 0) SCFTs},''
  \href{http://dx.doi.org/10.1007/JHEP10(2015)120}{{\em JHEP} {\bfseries 10}
  (2015) 120}, \href{http://arxiv.org/abs/1506.08040}{{\ttfamily
  arXiv:1506.08040 [hep-th]}}.

\bibitem{Mansi:2023faa}
L.~Mansi and M.~Sperling, ``{Unravelling T-Duality: Magnetic Quivers in
  Rank-zero Little String Theories},''
  \href{http://arxiv.org/abs/2312.12510}{{\ttfamily arXiv:2312.12510
  [hep-th]}}.

\bibitem{MR604577}
J.~McKay, ``Graphs, singularities, and finite groups,'' in {\em The {S}anta
  {C}ruz {C}onference on {F}inite {G}roups ({U}niv. {C}alifornia, {S}anta
  {C}ruz, {C}alif., 1979)}, vol.~37 of {\em Proc. Sympos. Pure Math.},
  pp.~183--186.
\newblock Amer. Math. Soc., Providence, RI, 1980.

\bibitem{Morrison:2016djb}
D.~R. Morrison and T.~Rudelius, ``{F-theory and Unpaired Tensors in 6D SCFTs
  and LSTs},'' \href{http://dx.doi.org/10.1002/prop.201600069}{{\em Fortsch.
  Phys.} {\bfseries 64} (2016) 645--656},
  \href{http://arxiv.org/abs/1605.08045}{{\ttfamily arXiv:1605.08045
  [hep-th]}}.

\bibitem{Morrison:1996na}
D.~R. Morrison and C.~Vafa, ``{Compactifications of F theory on Calabi--Yau
  threefolds. 1},'' \href{http://dx.doi.org/10.1016/0550-3213(96)00242-8}{{\em
  Nucl. Phys. B} {\bfseries 473} (1996) 74--92},
  \href{http://arxiv.org/abs/hep-th/9602114}{{\ttfamily arXiv:hep-th/9602114}}.

\bibitem{Morrison:1996pp}
D.~R. Morrison and C.~Vafa, ``{Compactifications of F theory on Calabi--Yau
  threefolds. 2.},'' \href{http://dx.doi.org/10.1016/0550-3213(96)00369-0}{{\em
  Nucl. Phys. B} {\bfseries 476} (1996) 437--469},
  \href{http://arxiv.org/abs/hep-th/9603161}{{\ttfamily arXiv:hep-th/9603161}}.

\bibitem{Nahm:1977tg}
W.~Nahm, ``{Supersymmetries and their Representations},''
  \href{http://dx.doi.org/10.1016/0550-3213(78)90218-3}{{\em Nucl. Phys. B}
  {\bfseries 135} (1978) 149}.

\bibitem{Ohmori:2014pca}
K.~Ohmori, H.~Shimizu, and Y.~Tachikawa, ``{Anomaly polynomial of E-string
  theories},'' \href{http://dx.doi.org/10.1007/JHEP08(2014)002}{{\em JHEP}
  {\bfseries 08} (2014) 002}, \href{http://arxiv.org/abs/1404.3887}{{\ttfamily
  arXiv:1404.3887 [hep-th]}}.

\bibitem{Ohmori:2014kda}
K.~Ohmori, H.~Shimizu, Y.~Tachikawa, and K.~Yonekura, ``{Anomaly polynomial of
  general 6d SCFTs},'' \href{http://dx.doi.org/10.1093/ptep/ptu140}{{\em PTEP}
  {\bfseries 2014} no.~10, (2014) 103B07},
  \href{http://arxiv.org/abs/1408.5572}{{\ttfamily arXiv:1408.5572 [hep-th]}}.

\bibitem{Ohmori:2015pua}
K.~Ohmori, H.~Shimizu, Y.~Tachikawa, and K.~Yonekura, ``{6d $\mathcal{N}=(1,0)$
  theories on $T^2$ and class S theories: Part I},''
  \href{http://dx.doi.org/10.1007/JHEP07(2015)014}{{\em JHEP} {\bfseries 07}
  (2015) 014}, \href{http://arxiv.org/abs/1503.06217}{{\ttfamily
  arXiv:1503.06217 [hep-th]}}.

\bibitem{Ohmori:2015pia}
K.~Ohmori, H.~Shimizu, Y.~Tachikawa, and K.~Yonekura, ``{6d
  $\mathcal{N}=\left(1,\;0\right) $ theories on S$^{1}$ /T$^{2}$ and class S
  theories: part II},'' \href{http://dx.doi.org/10.1007/JHEP12(2015)131}{{\em
  JHEP} {\bfseries 12} (2015) 131},
  \href{http://arxiv.org/abs/1508.00915}{{\ttfamily arXiv:1508.00915
  [hep-th]}}.

\bibitem{MR713260}
H.~C. Pinkham,
  \href{http://dx.doi.org/10.1090/pspum/040.2/713260}{``Factorization of
  birational maps in dimension {$3$},''} in {\em Singularities, {P}art 2
  ({A}rcata, {C}alif., 1981)}, vol.~40 of {\em Proc. Sympos. Pure Math.},
  pp.~343--371.
\newblock Amer. Math. Soc., Providence, RI, 1983.
\newblock \url{https://doi.org/10.1090/pspum/040.2/713260}.

\bibitem{Razamat:2019mdt}
S.~S. Razamat, E.~Sabag, and G.~Zafrir, ``{From 6d flows to 4d flows},''
  \href{http://dx.doi.org/10.1007/JHEP12(2019)108}{{\em JHEP} {\bfseries 12}
  (2019) 108}, \href{http://arxiv.org/abs/1907.04870}{{\ttfamily
  arXiv:1907.04870 [hep-th]}}.

\bibitem{MR715649}
M.~Reid, \href{http://dx.doi.org/10.2969/aspm/00110131}{``Minimal models of
  canonical {$3$}-folds,''} in {\em Algebraic varieties and analytic varieties
  ({T}okyo, 1981)}, vol.~1 of {\em Adv. Stud. Pure Math.}, pp.~131--180.
\newblock North-Holland, Amsterdam, 1983.
\newblock \url{https://doi.org/10.2969/aspm/00110131}.

\bibitem{Seiberg:1997zk}
N.~Seiberg, ``{New theories in six-dimensions and matrix description of M
  theory on T**5 and T**5 / Z(2)},''
  \href{http://dx.doi.org/10.1016/S0370-2693(97)00805-8}{{\em Phys. Lett. B}
  {\bfseries 408} (1997) 98--104},
  \href{http://arxiv.org/abs/hep-th/9705221}{{\ttfamily arXiv:hep-th/9705221}}.

\bibitem{Seiberg:1996qx}
N.~Seiberg, ``{Nontrivial fixed points of the renormalization group in
  six-dimensions},''
  \href{http://dx.doi.org/10.1016/S0370-2693(96)01424-4}{{\em Phys. Lett. B}
  {\bfseries 390} (1997) 169--171},
  \href{http://arxiv.org/abs/hep-th/9609161}{{\ttfamily arXiv:hep-th/9609161}}.

\bibitem{Sperling:2021fcf}
M.~Sperling and Z.~Zhong, ``{Balanced B and D-type orthosymplectic quivers
  \textemdash{} magnetic quivers for product theories},''
  \href{http://dx.doi.org/10.1007/JHEP04(2022)145}{{\em JHEP} {\bfseries 04}
  (2022) 145}, \href{http://arxiv.org/abs/2111.00026}{{\ttfamily
  arXiv:2111.00026 [hep-th]}}.

\bibitem{Strominger:1995ac}
A.~Strominger, ``{Open p-branes},''
  \href{http://dx.doi.org/10.1016/0370-2693(96)00712-5}{{\em Phys. Lett. B}
  {\bfseries 383} (1996) 44--47},
  \href{http://arxiv.org/abs/hep-th/9512059}{{\ttfamily arXiv:hep-th/9512059}}.

\bibitem{Vafa:1996xn}
C.~Vafa, ``{Evidence for F theory},''
  \href{http://dx.doi.org/10.1016/0550-3213(96)00172-1}{{\em Nucl. Phys. B}
  {\bfseries 469} (1996) 403--418},
  \href{http://arxiv.org/abs/hep-th/9602022}{{\ttfamily arXiv:hep-th/9602022}}.

\bibitem{MR422270}
J.~M. Wahl, ``Equisingular deformations of normal surface singularities. {I},''
  \href{http://dx.doi.org/10.2307/1971049}{{\em Ann. of Math. (2)} {\bfseries
  104} no.~2, (1976) 325--356}. \url{https://doi.org/10.2307/1971049}.

\bibitem{MR597833}
J.~M. Wahl, ``Elliptic deformations of minimally elliptic singularities,''
  \href{http://dx.doi.org/10.1007/BF0322000}{{\em Math. Ann.} {\bfseries 253}
  no.~3, (1980) 241--262}. \url{https://doi.org/10.1007/BF0322000}.

\bibitem{Wei:2022hjx}
X.-Y. Wei, Y.~Sugimoto, F.~Yagi, and S.-S. Kim, ``{DE-type little strings from
  glued brane webs},'' \href{http://dx.doi.org/10.1007/JHEP05(2023)214}{{\em
  JHEP} {\bfseries 05} (2023) 214},
  \href{http://arxiv.org/abs/2212.07344}{{\ttfamily arXiv:2212.07344
  [hep-th]}}.

\bibitem{Witten:1995zh}
E.~Witten, ``{Some comments on string dynamics},'' in {\em {STRINGS 95: Future
  Perspectives in String Theory}}, pp.~501--523.
\newblock 7, 1995.
\newblock \href{http://arxiv.org/abs/hep-th/9507121}{{\ttfamily
  arXiv:hep-th/9507121}}.

\bibitem{Witten:1995gx}
E.~Witten, ``{Small instantons in string theory},''
  \href{http://dx.doi.org/10.1016/0550-3213(95)00625-7}{{\em Nucl. Phys. B}
  {\bfseries 460} (1996) 541--559},
  \href{http://arxiv.org/abs/hep-th/9511030}{{\ttfamily arXiv:hep-th/9511030}}.

\end{thebibliography}\endgroup
\bibliographystyle{sortedbutpretty}

\end{document}